\documentclass[twocolumn,prd,nofootinbib,showpacs]{revtex4}
\usepackage{graphicx}
\usepackage{color}
\usepackage{amsmath,amssymb,amsfonts,dsfont}
\usepackage{multirow}
\usepackage{ulem}

\pdfoutput=1 % if your are submitting a pdflatex (i.e. if you have
\newcommand{\lsim}{\mathrel{\mathop{\kern 0pt \rlap
  {\raise.2ex\hbox{$<$}}}
  \lower.9ex\hbox{\kern-.190em $\sim$}}}
\newcommand{\gsim}{\mathrel{\mathop{\kern 0pt \rlap
  {\raise.2ex\hbox{$>$}}}
  \lower.9ex\hbox{\kern-.190em $\sim$}}}

\def \aap  {A\&A}

\def \apjs  {ApJS}
\def \apjl  {ApJL}

\begin{document}

\title{Evidences of low-diffusion bubbles around Galactic pulsars}

\author{Mattia Di Mauro,}\email{mdimauro@slac.stanford.edu
}
\affiliation{NASA Goddard Space Flight Center, Greenbelt, MD 20771, USA}
\affiliation{Catholic University of America, Department of Physics, Washington DC 20064, USA}
\author{Silvia Manconi}\email{manconi@to.infn.it}
\affiliation{Dipartimento di Fisica, Universit\`a di Torino, via P. Giuria 1, 10125 Torino, Italy}
\affiliation{Istituto Nazionale di Fisica Nucleare, Sezione di Torino, Via P. Giuria 1, 10125 Torino, Italy}
\author{Fiorenza Donato}\email{donato@to.infn.it}
\affiliation{Dipartimento di Fisica, Universit\`a di Torino, via P. Giuria 1, 10125 Torino, Italy}
\affiliation{Istituto Nazionale di Fisica Nucleare, Sezione di Torino, Via P. Giuria 1, 10125 Torino, Italy}

%\date{\today} 
\begin{abstract}
Recently, a few-degrees extended $\gamma$-ray halo in the direction of Geminga pulsar has been detected by HAWC, Milagro and {\it Fermi}-LAT.
These observations can be interpreted with positrons ($e^+$) and electrons ($e^-$) accelerated by Geminga pulsar wind nebula (PWN), released in a Galactic environment with a low diffusion coefficient ($D_0$), and inverse Compton scattering (ICS) with the interstellar radiation fields.
We inspect here how the morphology of the ICS $\gamma$-ray flux depends on the energy, the pulsar age and distance, and the strength and extension of the low-diffusion bubble.
In particular we show that $\gamma$-ray experiments with a peak of sensitivity at TeV energies are the most promising ones to detect ICS halos.
We perform a study of the sensitivity of HAWC, HESS and the future CTA experiment finding that, with efficiencies of the order of a few \%, the first two experiments should have already detected a few tens of ICS halos while the latter will increase the number of detections by a factor of 4.
We then consider a sample of sources associated to PWNe and detected in the HESS Galactic plane survey and in the second HAWC catalog.
We use the information available in these catalogs for the $\gamma$-ray spatial morphology and flux of these sources to inspect the value of $D_0$ around them and the $e^{\pm}$ injection spectrum.
All sources are detected as extended with a $\gamma$-ray emission extended about $15-80$ pc.
Assuming that most of the $e^{\pm}$ accelerated by these sources have been released in the interstellar medium, the diffusion coefficient is $2-30 \cdot 10^{26}$ cm$^2$/s at 1 TeV, i.e. two orders of magnitude smaller than the value considered to be the average in the Galaxy. 
These observations imply that Galactic PWNe have low-diffusion bubbles with a size of at least 80 pc.

\end{abstract}

\maketitle
%\flushbottom

%%%%%%%%%%%%%%%%%%%%%%%%%%%%%%%%%%%%%%%%%%%%%%%%%%%%%%%%%%%%
\section{Introduction}
\label{sec:intro}
%%%%%%%%%%%%%%%%%%%%%%%%%%%%%%%%%%%%%%%%%%%%%%%%%%%%%%%%%%%%
A $\gamma$-ray emission at TeV energies and of a few-degrees extension size in the direction of Geminga and Monogem pulsar wind nebulae (PWNe) has been detected by HAWC \cite{Abeysekara:2017science} and Milagro \cite{2009ApJ...700L.127A}. The presence of a $\gamma$-ray halo around Geminga has been recently confirmed by \cite{DiMauro:2019yvh} with an analysis of {\it Fermi}-LAT data above 8 GeV, whose extension reaches about 15 degrees at 10 GeV.
The $\gamma$-ray halos detected around Geminga and Monogem are interpreted as photons produced by electrons ($e^-$) and $e^+$ ($e^+$) accelerated by their PWNe and inverse Compton scattering (ICS) low-energy photons of the interstellar radiation fields (ISRFs).
These observations may give us the possibility to shed light on the origin of the $e^+$ excess in cosmic rays (CR), firstly detected by Pamela \cite{Adriani:2013uda}, then by 
{\it Fermi}-LAT \cite{2012PhRvL.108a1103A} and recently, with an unprecedented precision, by AMS-02 \cite{PhysRevLett.122.041102}.
The extension of detected $\gamma$-ray halos suggests that the diffusion around these PWNe is about two orders of magnitude less intense than the value assumed to fit the latest CR data measured by AMS-02 (see, e.g. \cite{Kappl:2015bqa,Genolini:2015cta,Genolini:2019ewc}). The inferred diffusion coefficient is in fact of about $10^{27}$ cm$^2$/s at 100 GeV \cite{Abeysekara:2017science,DiMauro:2019yvh}.

The ICS halos detected around Geminga and Monogem are called by some authors ``TeV halos'', since they have been mainly detected at very-high-energy (VHE) (see, e.g., \cite{Linden:2017vvb}). 
However, we will refer to them as ``ICS halos'' because of the recent detection of the Geminga halo at {\it Fermi}-LAT energies, and because we prefer to characterize this emission with the physical process that generates it, and not with the energy at which it is detected. 
It is still unclear if these halos are generated by $e^{\pm}$ accelerated by PWNe and diffusing in the interstellar medium (ISM), or by $e^{\pm}$ propagating in a region still dominated by the PWN environment. 
Very recently, Ref.~\cite{Giacinti:2019nbu} investigated this point by using a sample of Galactic PWNe  taken from the HESS survey of the Galactic plane (HGPS) \cite{H.E.S.S.:2018zkf}. They have estimated the $e^{\pm}$ density at the location of the source VHE $\gamma$-ray emission. 
Comparing this density with the one of the ISM, they concluded that for most of these sources, except for Geminga and Monogem, the $e^{\pm}$ are probably still confined in the PWN. 
Therefore, they call these sources $e^{\pm}$ halo, rather than TeV halo. 
Their calculation is based on a series of assumptions, such as the shape of the $e^{\pm}$ injection spectrum, the energy range for accelerated $e^{\pm}$, and no time dependence considered for the spin-down luminosity. Also, the size of the ICS halos is taken directly from the HESS catalog. 
Changing some of these assumptions their results might change significantly and many of the sources in their sample could have a density of $e^{\pm}$ of the same order of the ISM.
This would imply that these cosmic particles might not be confined in the PWN. 
We will discuss in Sec.~\ref{sec:resultsD0eta} how their results would change assuming the size of the ICS emission as estimated in this paper.

%If the presence of a low-diffusion zone is confirmed also around other Galactic PWNe we should change our model for the propagation of CRs in the Galaxy.
The detection of ICS halos around Geminga and Monogem can provide key information about the acceleration mechanisms of $e^{\pm}$ from PWNe, and their propagation in the Galactic environment.
For example Ref.~\cite{DiMauro:2019yvh} used the flux and morphology of the ICS halo detected from Geminga and found that this source contributes at most $10\%$ to the $e^+$ excess. They have also found evidences for a low-diffuse bubble located around the pulsar, with a size of around 100~pc and a value of the diffusion coefficient at 1 GeV of about $2.3\times 10^{26}$ cm$^2$/s, i.e., two orders of magnitude lower than the average of the Galaxy. 
Several references (see, e.g., \cite{Hooper:2017gtd,Abeysekara:2017science,Shao-Qiang:2018zla,Tang:2018wyr,Fang:2018qco,DiMauro:2019yvh}) have studied the flux of $e^+$ from PWNe in light of the Milagro and HAWC data, and have drawn conclusions on the contribution of this source population to the $e^+$ excess. 
Reference \cite{Abeysekara:2017science} uses the low diffusion found around Geminga and Monogem PWNe to propagate particles in the entire Galaxy, and claims their contribution is negligible. 
On the other hand the authors of \cite{Hooper:2017gtd,Fang:2018qco} claim Geminga explains most of the $e^+$ data.
Finally, references \cite{Tang:2018wyr,DiMauro:2019yvh} agree on the fact that the contribution of Geminga is at the $10\%$ level.
Although, most of these papers suggest PWNe are likely the main contributors to the $e^+$ flux, they use the results based on only those two PWNe. 
Indeed, we still do not have a large enough sample of ICS halos and we have not collected evidences if such a low-diffusion bubble is present or not around a significant sample of Galactic pulsars.

In addition to Geminga and Monogem, many more ICS $\gamma$-ray halos could have been already 
detected in the direction of other Galactic pulsars by Imaging Atmospheric Cherenkov Telescopes (IACTs), HAWC, MILAGRO and {\it Fermi}-LAT.
The HAWC Collaboration has recently released the 2HWC catalog \cite{Abeysekara:2017hyn} which contains 39 sources detected close to the Galactic plane. Many of them have an extended $\gamma$-ray morphology, and are spatially close to powerful Galactic pulsars. 
The HESS Collaboration has recently published the results of a the HGPS catalog which is the most comprehensive survey of the Galactic plane in VHE $\gamma$ rays. This publication includes Galactic sky maps and the  catalog with the  properties of the 78 sources \cite{H.E.S.S.:2018zkf}.
Many of these sources  have been detected as extended, and are probably associated to PWNe.
Therefore, the 2HWC and HGPS catalogs represent two rich datasets for investigating the acceleration mechanism of $e^+$ from PWNe, and their diffusion around those sources.
Following the detection of Geminga and Monogem extended halos, the possible presence of similar objects around other Galactic pulsars has been explored in \cite{Linden:2017vvb}, by assuming a "Geminga-like" TeV halo for each considered pulsar. 
Moreover, the expected number of ICS halos detectable by current and future observatories has been estimated in \cite{Sudoh:2019lav}. As discussed in the rest of the present paper, we significantly extend the current literature by performing a complete calculation of the ICS flux for each considered source.
Furthermore, we present a novel analysis of the data provided by HGPS catalog to characterize the observed gamma-ray  emission around many Galactic pulsars, in the light of the presence a possible ICS halo.

In the first part of this paper we will inspect how the extension of the ICS halo in PWNe depends on the age and distance of the host pulsar, and on the intensity of the diffusion coefficient present around them.
The ICS halo size is a key parameter for IACTs which have a limited 	instantaneous field of view of $4-5^{\circ}$.
Then, we will show how the ICS halo size depends on the extension of the low-diffusion bubble and the pulsar proper motion.
In fact, pulsars have an average proper motion of 100 km/s \cite{Hobbs:2005yx} and, as we have shown in \cite{DiMauro:2019yvh}, this effect distorts the ICS $\gamma$-ray morphology.
In the present study, we argue that the most promising energy range for searching for ICS halos is above 100 GeV, where IACTs, and HAWC and Milagro operate.
We then use the ICS flux to predict the brightest pulsars around which HAWC should detect an ICS halo. 
Finally, we predict the number of ICS halos detectable by HESS, HAWC and in the future by the Cherenkov Telescope Array (CTA) \cite{Acharya:2017ttl}.

In the second part, we consider the PWNe already detected by IACTs. 
In particular, we use a sample of sources associated to PWNe or PWNe candidates taken from the 2HWC and HGPS catalogs. 
We use their measured size and flux to determine the diffusion 
coefficient around each source, and to estimate the minimal dimension of the low-diffusion bubble. 
%We then calculate the efficiency for the conversion of spin-down energy into $e^{\pm}$ at these VHEs.

The paper is organized as follows.
In Sec.~\ref{sec:model} we present our model for the acceleration of $e^{\pm}$ from PWNe, $e^{\pm}$ propagation in the Galaxy and the flux of $\gamma$ rays for ICS.
In Sec.~\ref{sec:sizeflux} we investigate how the ICS halos size depends on the pulsar distance, age and proper motion, 
and how it changes according to the diffusion coefficient.
In Sec.~\ref{sec:IACTs} we study the detectability of ICS halos at IACTs and rank the pulsars in ATNF catalog \cite{2005AJ....129.1993M} 
according to their expected ICS halo brightness. 
Sec.~\ref{sec:D0} contains the methodology employed for the  derivation of the diffusion coefficient around the sources, whose results are presented in Sect.
~\ref{sec:resultsD0eta}. We draw our conclusions in Sec.~\ref{sec:conclusions}.

%%%%%%%%%%%%%%%%%%%%%%%%%%%%%%%%%%%%%%%%%%%%%%%%%%%%%%%%
\section{Model for the  $e^{\pm}$ and $\gamma$-ray emission from a PWN}
\label{sec:model} 
%%%%%%%%%%%%%%%%%%%%%%%%%%%%%%%%%%%%%%%%%%%%%%%%%%%%%%%%%%%%
We recall here the basics for modeling the  $e^{\pm}$  and the consequent ICS  $\gamma$-ray  emission from PWNe. 
We follow the formalism detailed in \cite{DiMauro:2019yvh}. 

PWNe are thought to accelerate and inject $e^{\pm}$ in the ISM up to VHE (see, e.g., \cite{1996ApJ...459L..83C,Amato:2013fua,Gaensler:2006ua}).
A rapidly spinning neutron star, or pulsar, formed after a supernova explosion,  is likely the engine of this process.
 The rotation of the pulsar induces an electric field that extracts $e^-$ from the star surface. 
 These $e^-$ lose energy via curvature radiation while propagating far from the pulsar along the magnetic field lines,  and the energetic emitted photons create
 a wind of $e^{\pm}$ pairs  in the intense neutron star magnetic field. 

For the sake of completeness, we here briefly recall the basic understandings of the PWN evolution, which is then treated effectively.
According to  \cite{1996ApJ...459L..83C,Amato:2013fua,Gaensler:2006ua}, the initial phase of the PWN evolution, called free expansion phase, occurs in the first few thousands of years.
At this stage, the pulsar wind  is surrounded by
the expanding shell of the supernova remnant (SNR), which moves at a speed of about $5-10 \cdot 10^{3}$ km/s, while  the pulsar located at the center of the SNR 
has a velocity of the order of $400-500$ km/s.
The expansion velocity of the pulsar wind  increases constantly with time,  and the size $R$ of the PWN goes as $R\propto t^{1.2}$ \cite{1977ASSL...66...53C}.
During the free expansion, the pulsar wind expands very fast while the SNR ejecta interacts with the ISM creating a forward and reverse shock.
%The latter constitutes a termination shock, and its bulk energy is dissipated into a relativistically, magnetized fluid, which shines as a PWN.
The PWN reaches, at this stage, a size of about 10 pc.

After a few thousands years, the reverse shock moves inward and interacts with the outward moving PWN shock.
This interaction constitutes a termination shock, and its bulk energy is dissipated into a relativistically, magnetized fluid, which shines as a PWN.
The total energy of the SNR exceeds the one of the PWN by one or two orders of magnitude, so that the PWN can be compressed by up to a factor of 10 \cite{2009ApJ...703.2051G}.
During this process the PWNe experiences a series of contractions and expansions until a steady balance is reached.
Once the reverberations between the PWN and the SNR reverse shock have faded, the pulsar can again power a bubble steadily expanding as $R\propto t^{1.2}$ for $t<\tau_0$ and $R\propto t^{0.3}$ 
for $t>\tau_0$, where $\tau_0$ is the pulsar decay time  \cite{1984ApJ...278..630R,2001A&A...380..309V}. 
Therefore, at a time larger than $\tau_0$ the PWN size is not expected to have a strong evolution with the pulsar age.
The $e^{\pm}$ pairs produced in the pulsar magnetosphere reach the termination shock and, due to the severe energy losses, their energy is at most a few tens of GeV.
The termination shock is the place where particle acceleration eventually occurs, and a relatively large fraction (up to few tens of percent) of the wind bulk energy is converted into accelerated pairs. 
They then radiate into a photon spectrum extending from radio frequencies to TeV $\gamma$-rays, through synchrotron and ICS processes.

Given the initial velocity, the distance traveled by the pulsar from the explosion site after few tens of kyr can be comparable to or even larger than the radius of an equivalent spherical PWN around a stationary pulsar. 
The pulsar thus can abandon its original wind bubble, leaving behind itself a relic PWN, and generating a new, smaller PWN around its current position, which is called bow shock. 
Observationally, this appears as a central, possibly distorted PWN visible in radio and X-ray and powered by freshly accelerated $e^{\pm}$.
The relic PWN is powered by $e^{\pm}$ injected along its formation history. 

The PWNe considered in this paper are older than a few thousands of year. 
Therefore, these PWNe have probably already interacted with the reverse shock of the SNR.
Moreover, the $e^{\pm}$ accelerated by younger sources could be still confined inside the PWN or the SNR, while for older sources they have been probably injected from the relic and bow shock components of the PWN, and released in the ISM environment. 
In order to inspect any dependence of our results by the presence of the SNR and PWN environment, we select PWNe powered by pulsars of different ages from a few to hundreds of kyr.

We consider a model in which $e^{\pm}$ are continuously injected with a rate that follows the pulsar spin-down energy, {\it i.e.} a continuous injection scenario.
This scenario is indeed required to generate the TeV  photons  detected by Milagro and HAWC for Geminga and Monogem 
\cite{Yuksel:2008rf,Abeysekara:2017science,DiMauro:2019yvh}. A common alternative is to consider a burst like scenario, where all the particles are emitted from the source at a time equal to the age of source $T$. 
In our model, the injection spectrum $Q(E,t)$ for the accelerated $e^{\pm}$ pairs is assumed to effectively describe the particles that are produced during the acceleration process and released in the ISM, while no attempt is made to describe the dynamical evolution during the first thousands of years of the PWN, or possible modification in the spectrum of particles during the release processes.
The injection spectrum $Q(E,t)$ can be effectively described by a power law with an exponential cutoff:
 \begin{equation}
 Q(E, t)= L(t) \left( \frac{E}{E_0}\right)^{- \gamma} \exp \left(-\frac{E}{E_c} \right) ,
 \label{eq:Q_E_cont}
\end{equation}
where the magnetic dipole braking $L(t)$ (assuming a magnetic braking index of 3) is defined as:
 \begin{equation}
 L(t) = \frac{L_0}{\left( 1+ \frac{t}{\tau_0} \right)^{2} },
 \label{eq:Lt}
\end{equation}
and $\tau_0$ is the characteristic pulsar spin-down timescale. 
The cutoff energy $E_c$ is fixed to $10^3$ TeV. 
We set $\tau_0=12$~kyr if not stated otherwise, following \cite{1995A&A...294L..41A,Malyshev:2009tw,Abeysekara:2017science,DiMauro:2019yvh}.  
A smaller value of $\tau_0$, as derived for example from fits to radio, X-ray and $\gamma$-ray data to few, very young pulsars \cite{Torres:2014iua}, would have the consequence to lower the $\gamma$-ray flux at high-energy and so higher efficiency values would be found.
The injection spectrum of $e^{\pm}$ is usually measured with a broken power-law spectrum (see, e.g., \cite{Malyshev:2009tw}) with a break at energies between a few up to hundreds of GeV. The $\gamma$-ray energies we will consider are beyond a few hundreds of GeV so for $e^{\pm}$ well beyond the TeV energies. Therefore, the injection spectrum of particles from PWNe can be modeled in our care with a simple power-law and neglecting the presence of a break.

The total energy emitted by the source in $e^{\pm}$ is given by:
 \begin{equation}
 E_{{\rm tot}} =\int_0^{T} dt \int_{E_{1}}^{\infty} dE E Q(E,t),
 \label{eq:Etot}
\end{equation}
where we fix $E_1 = 0.1$ GeV \cite{Buesching:2008hr,Sushch:2013tna}.
$E_{{\rm tot}}$ is related to the pulsar total spin down energy $W_0$ with the following relation $E_{{\rm tot}}=\eta W_0$, where $\eta$ is the fraction of the pulsar's spin-down luminosity which goes into $e^{\pm}$ particles.
$W_0$ can be computed from catalogued quantities as the pulsar age $T$, 
the decay time $\tau_0$, and the spin-down luminosity $\dot{E}$:
\begin{equation}
 W_0 = \tau_0 \dot{E} \left( 1+ \frac{T}{\tau_0} \right)^2\,.
 \label{eq:W0PWN}
\end{equation}
%The \textit{actual} age $T$ and the \textit{observed} ($t_{\rm obs}$) age are related by the source distance $d$ by $T = t_{\rm obs} + d/c$. 
The normalization of the injection spectrum ($L_0$, see Eq.~\ref{eq:Q_E_cont}) is found using Eq.~\ref{eq:Etot} and assuming that $E_{{\rm tot}}=\eta W_0$.

%The spin-down luminosity $\dot{E}$, the \textit{observed} age $t_{\rm obs}$ (where $T = t_{\rm obs} + d/c$ is the \textit{actual} age) and the distance $d$ from the pulsars are taken from the ATNF catalog \cite{2005AJ....129.1993M}.
%while $\tau_0$ is the characteristic pulsar spin-down timescale. 
%We use $\tau_0=12$ kyr if not stated otherwise, following \cite{Abeysekara:2017science,DiMauro:2019yvh}.  
%Only middle-aged pulsars, with an observed age $50$~kyr$<t_{\rm obs}<10000$~kyr, are supposed to emit $e^\pm$. In younger pulsars  $e^\pm$ are believed to be confined until the expanding medium merges with the ISM, which should occur at least $40-50$~kyr after the pulsar formation \citep{1996ApJ...459L..83C, 2011ASSP...21..624B}. 
The source term in Eq.~\ref{eq:Q_E_cont} is inserted in a diffusion-loss equation to compute the $e^\pm$ number density $\mathcal{N}_e(E,\mathbf{r},t)$ per unit volume and energy, and at an observed energy $E$, a position $\mathbf{r}$, and  time $t$. 
We account for space-indipendent energy losses $b(E)$ by means of synchrotron and inverse Compton processes. 
The  interstellar photon populations
at different wavelengths have been taken from \cite{Vernetto:2016alq}. The Galactic magnetic field intensity 
has been assumed $B=3.6\; \mu$G, as resulting from the sum (in quadrature) of the regular and turbulent components \citep{2007A&A...463..993S}. For more details on the propagation model, we address to  \cite{DiMauro:2019yvh} and references therein.
In the continuous injection scenario and with a homogeneous diffusion in the Galaxy, the solution for the $e^\pm$ number density $\mathcal{N}_e(E,\mathbf{r},t)$ at an observed energy $E$, position $\mathbf{r}$ and time $t$ is given by \cite{Yuksel:2008rf}:
 \begin{eqnarray}
\label{eq:N_cont}
  \mathcal{N}_e(E,\mathbf{r},t) &=& \int_0^{t} dt'\, \frac{b(E_s)}{b(E)} \frac{1}{(\pi \lambda^2(t',t,E))^{\frac{3}{2}}} \times \nonumber \\
  && \times  \exp\left({-\frac{|\mathbf{r} -\mathbf{r_{s}} |^2}{ \lambda(t',t,E)^2}}\right)Q(E_s,t'),
\end{eqnarray}
where the integration over $t'$ accounts for  the PWN releasing $e^\pm$ continuously in time.
The  energy $E_s$ is the initial energy of $e^\pm$ that cool down to $E$ in a {\rm loss time} $\Delta \tau$:
\begin{equation}
 \Delta \tau (E, E_s) \equiv \int_{E} ^{E_s} \frac{dE'}{b(E')} = t-t_{{\rm obs}} .
\end{equation}
The $b(E)$ is the energy loss function, $\mathbf{r_{s}}$ indicates the source position, and $\lambda$ is the typical propagation scale length defined as:
\begin{equation}
\label{eq:lambda}
 \lambda^2= \lambda^2 (E, E_s) \equiv 4\int _{E} ^{E_s} dE' \frac{D(E')}{b(E')},
\end{equation} 
with $D(E)$ the diffusion coefficient.
As a matter of fact, the tipycal propagation time, defined as $\lambda^2/4D(E)$ sets grossly in the range  $10^3-10^5$~kyr, decreasing  with $E$.
The flux of $e^\pm$ at Earth is given by:
\begin{equation}
 \Phi_{e^\pm}(E) = \frac{c}{4\pi} \mathcal{N}_e(E,|\mathbf{r} -\mathbf{r_{s}} |=d,t=T).
 \label{eq:flux}
\end{equation}

Recent results  \cite{Abeysekara:2017science,DiMauro:2019yvh} suggest that the diffusion coefficient around Geminga and Monogem PWNe is $\sim 10^{26}$ cm$^2$/s  at 1 GeV, {\it i.e.}~about two orders of magnitude smaller than the value derived for the entire Galaxy through a fit to AMS-02 CR data \cite{Kappl:2015bqa,Genolini:2015cta,Genolini:2019ewc}.
A  phenomenological description  for this discrepancy proposes  a two-zone diffusion model, where the region of low diffusion is contained around the source, and delimited by an empirical radius $r_b$ \cite{Profumo:2018fmz,Tang:2018wyr}. 
The inhibition of diffusion near pulsars has been recently discussed in \cite{Evoli:2018aza}, and a possible theoretical interpretation is provided. 
This paper predicts a very strong dependence of the diffusion coefficient as a function of the pulsar age with $D_0 \sim10^{26}$ cm$^2$/s at 1 GeV for sources with $T\sim 20$ kyr and values close to the average of the Galaxy for $T>100$ kyr. 
Nevertheless, a conclusive understanding of this phenomenon is not yet achieved, and the analysis we present in this paper can give new insights on the theoretical models.

In this paper we include the phenomenological two-zone diffusion model as in \cite{Tang:2018wyr,DiMauro:2019yvh} to account for these recent observations, for which the diffusion coefficient is defined as:
\begin{eqnarray}
\label{eq:conddm}
D(E,r) =  
\left\{
\begin{array}{rl}
& D_0 (E/1{\rm \,GeV})^\delta {\rm \;for\;} 0 < r < r_b, \\
& D_2 (E/1{\rm \,GeV})^\delta {\rm \;for\;} r \geq r_b.
\end{array}
\right.
\label{eq:Diff}
\end{eqnarray}
Here $r_b$ is the size of the low-diffusion bubble while $D_0$ and $D_2$ are the diffusion coefficients inside and outside the bubbles, respectively.

%%% GAMMA-RAYS 
The $e^{\pm}$ accelerated by PWNe can produce photons whose energy covers a wide range (see, e.g., \cite{2017hsn..book.2159S} 
for a recent review).
From radio to X-ray energies, photons are produced by $e^{\pm}$ through synchrotron radiation caused by the magnetic fields.
On the other hand, at higher energies $\gamma$ rays are produced from VHE $e^{\pm}$ escaped from the PWN by the ICS off the ISRF.
We are interested here in the extended halo emission of the size of at least tens of arcminutes (i.e., around tens of parsec) generated by $e^{\pm}$ injected by PWNe in the Galactic environment, and not to the small-scale structures extended between few arcseconds to arcminutes and observed in the nebula, as for example jets and torii (see e.g. \cite{2017ApJ...835...66P}). 
%Nevertheless, the following equations hold for any  $e^+$ and $e^-$ input spectrum, target photon fields and for the synchrotron and ICS emission mechanisms.

The ICS photon flux emitted  by a PWN, at a $\gamma$-ray energy $E_{\gamma}$ and within a solid angle $\Delta \Omega$, can be computed  as \cite{1970RvMP...42..237B,Cirelli:2010xx}:
\begin{equation} \label{eq:phflux}
 \Phi_{\gamma} (E_{\gamma}, \Delta \Omega)= \int_{m_e c^2}^{\infty} dE \mathcal{M}(E, \Delta \Omega) \mathcal{P}^{\rm IC}(E, E_{\gamma})\,.
\end{equation}
The term $\mathcal{M}(E,\Delta \Omega)$ represents the spectrum of $e^+$ and $e^-$  of energy $E$ propagating in the Galaxy and from a solid angle $\Delta \Omega$:
\begin{equation}
\mathcal{M}(E,\Delta \Omega)  = \int_{\Delta \Omega} d\Omega \int_0^{\infty} d s \, \mathcal{N}_e (E,s,T).
 \label{eq:M}
\end{equation}
$\mathcal{N}_e (E,s,T)$ is the energy spectrum of $e^\pm$ taken from Eq.~\ref{eq:N_cont},  $s$ is the line of sight,
and  $\mathcal{P}^{\rm IC}(E,E_{\gamma})$ is the power of photons emitted by a single $e^\pm$ by ICS, defined  as in \cite{1970RvMP...42..237B,2010A&A...524A..51D}.
The ICS occurs off  the CMB, described by a blackbody energy density ($T_{{\rm CMB}} = 2.753$ K), the infrared light (peaked at $T_{{\rm IR}} = 3.5 \cdot 10^{-3}$ eV) and by the starlight ($T_{{\rm SL}} = 0.3$ eV) \cite{Vernetto:2016alq,2006ApJ...648L..29P,2017MNRAS.470.2539P}.
%Models for the  local ISRF are provided in \cite{Vernetto:2016alq,2006ApJ...648L..29P,2017MNRAS.470.2539P}, which all contain a careful description of the photons at all  frequencies. 
Our results are obtained for the ISRF energy density in the local Galaxy reported in \cite{Vernetto:2016alq}, but we have explicitly checked that they do not get modified by using the model in 
\cite{2006ApJ...648L..29P}. 

As shown in \cite{DiMauro:2019yvh}, the proper motion of the pulsar could alter the morphology of the $\gamma$-ray ICS halo.
The proper motion affects significantly the morphology of the $\gamma$-ray emission for pulsars older than about 100 kyr and moving with a velocity of at least 100 km s$^{-1}$.
This is particularly true for Geminga, that is a very close pulsar ($d=250$ pc), has a transverse proper motion of $v_T \approx 211$ km s$^{-1}$ \cite{Faherty:2007} and $T=340$ kyr.  We implement this effect  in Eq.~\ref{eq:N_cont} by replacing $\bf{r_s}$ with ${\bf{v_T}}t$, where $\bf{v_T}$ is the vector of the transverse velocity.
%In \cite{DiMauro:2019yvh} we have found with {\it Fermi}-LAT data a preference at about $5\sigma$ significance for the model with the pulsar velocity with respect to the one with a steady pulsar.
%We will show in the next section that this effect is not relevant for energies larger than about 100 GeV.

%\sm{REMOVE: Finally, we note that the two-zone diffusion model is used every time we compute $e^\pm$ at the Earth position. }
As for the ICS photon flux emitted by a PWN, our benchmark is the one-zone diffusion model in which, effectively, $r_b\rightarrow\infty$, and the $D_0$ corresponds to the low diffusion coefficient around the PWN. 
Using the one-zone diffusion model for the ICS is appropriate since most of the $\gamma$-ray emission is generated close to the pulsar where the low diffusion probably acts.
On the other hand, for the calculation of the $e^+$ flux at Earth the two-zone model must be considered since the size of the low-diffusion zone around the PWN is much smaller than the propagation volume from the source to the Earth.
We have already applied these choices in \cite{DiMauro:2019yvh}.
We will also discuss some examples in which the ICS photon flux is computed in a two-zone diffusion model.

%%%%%%%%%%%%%%%%%%%%%%%%%%%%%%%%%%%%%%%%%
\section{Angular size of the $\gamma$-ray ICS halos}
\label{sec:sizeflux}
%%%%%%%%%%%%%%%%%%%%%%%%%%%%%%%%%%%%%%%%%
In this section we study the size of ICS halos, defined through the $\gamma$-ray flux, for different values of $E_\gamma$, 
and as a function of the strength ($D_0$) and size ($r_b$) of the low-diffusion bubble, the age and distance of the host pulsar, and of its proper motion, in order to motivate the selection of  pulsars used in Sec. \ref{sec:sample}.

%-------------------------------
%--SIZE OF EXTENSION THETA_68
\begin{figure*}
\includegraphics[width=0.49\textwidth]{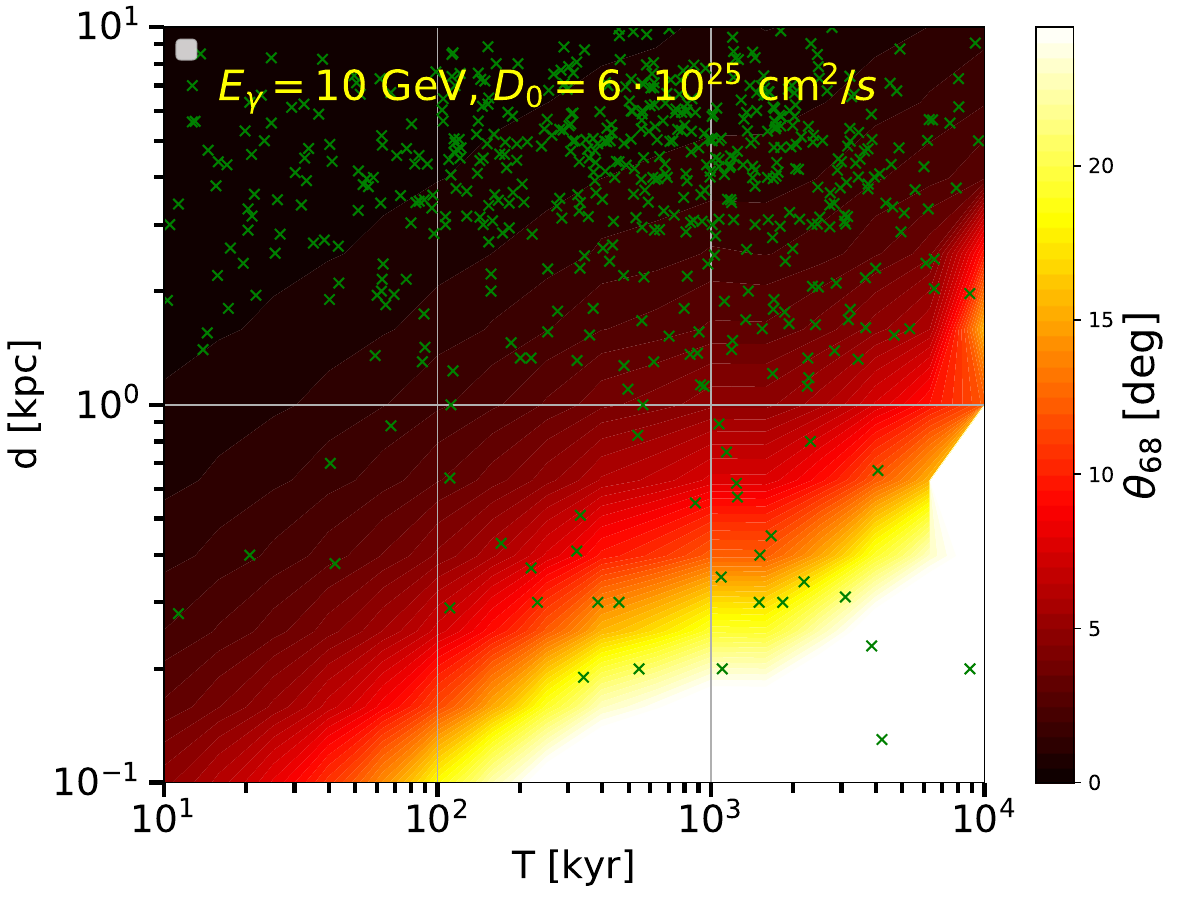}
\includegraphics[width=0.49\textwidth]{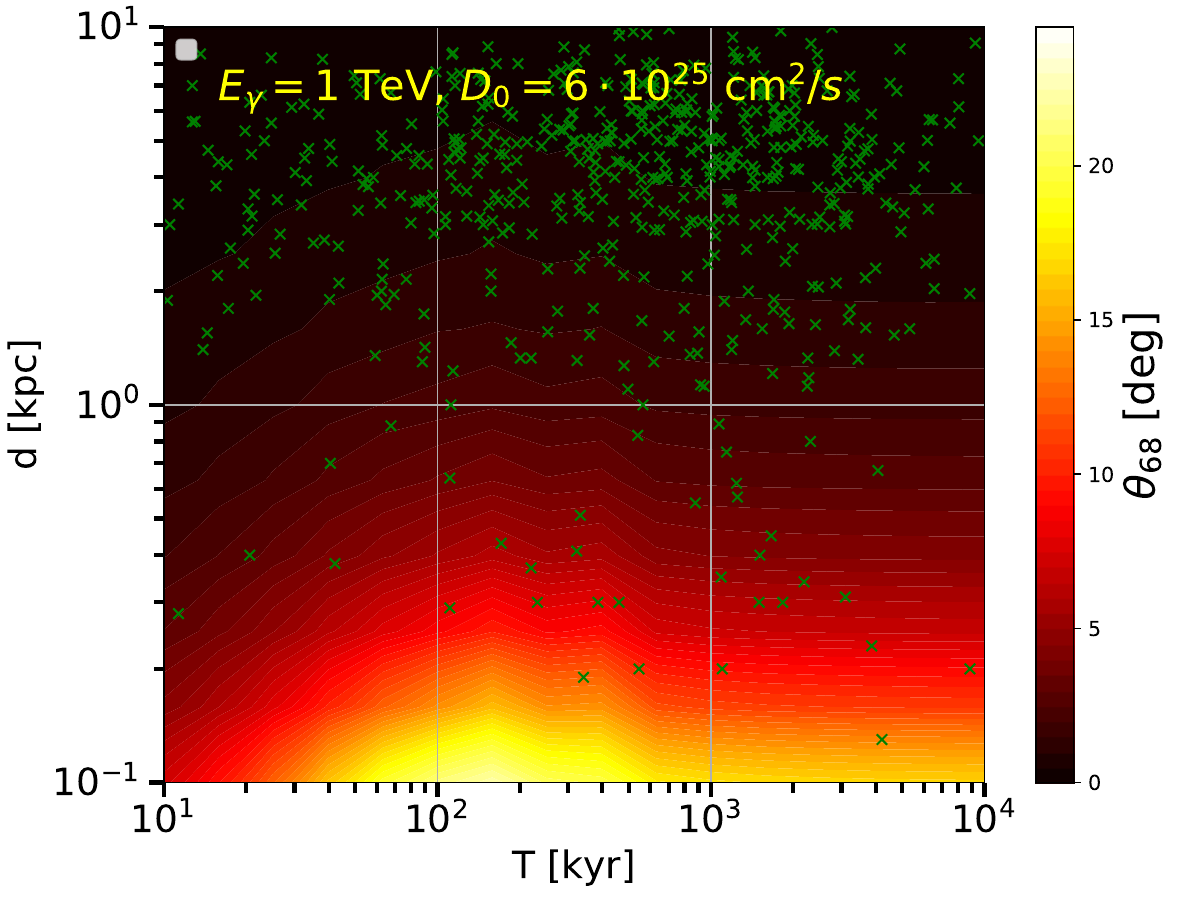}
\includegraphics[width=0.49\textwidth]{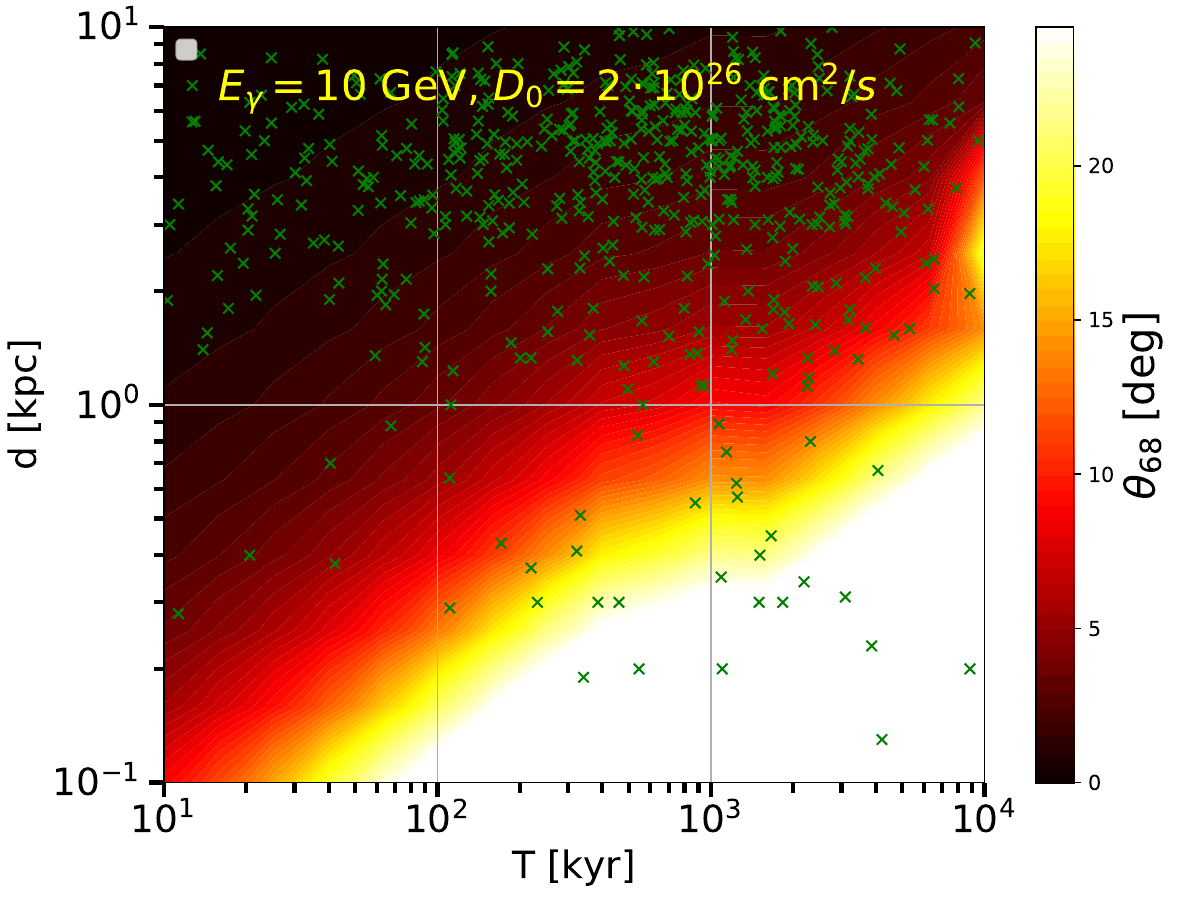}
\includegraphics[width=0.49\textwidth]{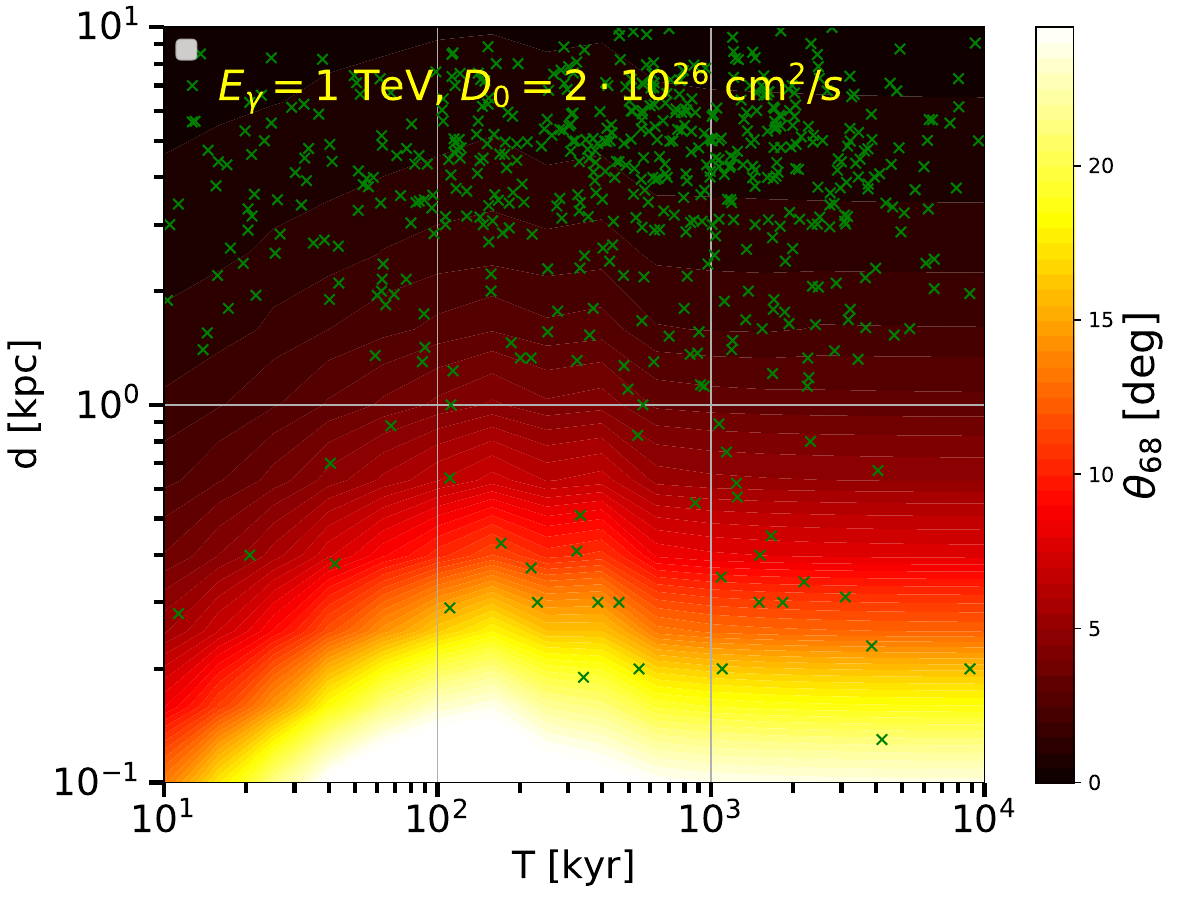}
\includegraphics[width=0.49\textwidth]{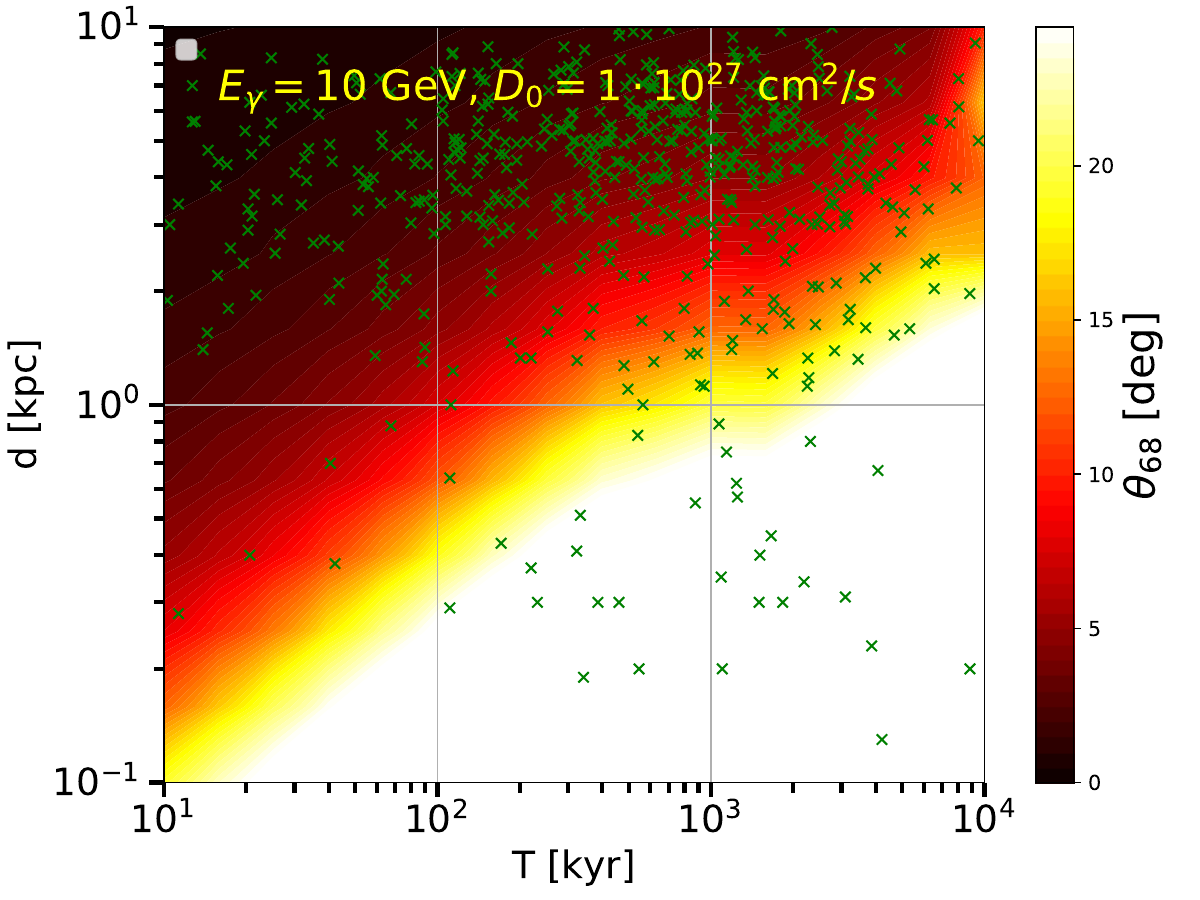}
\includegraphics[width=0.49\textwidth]{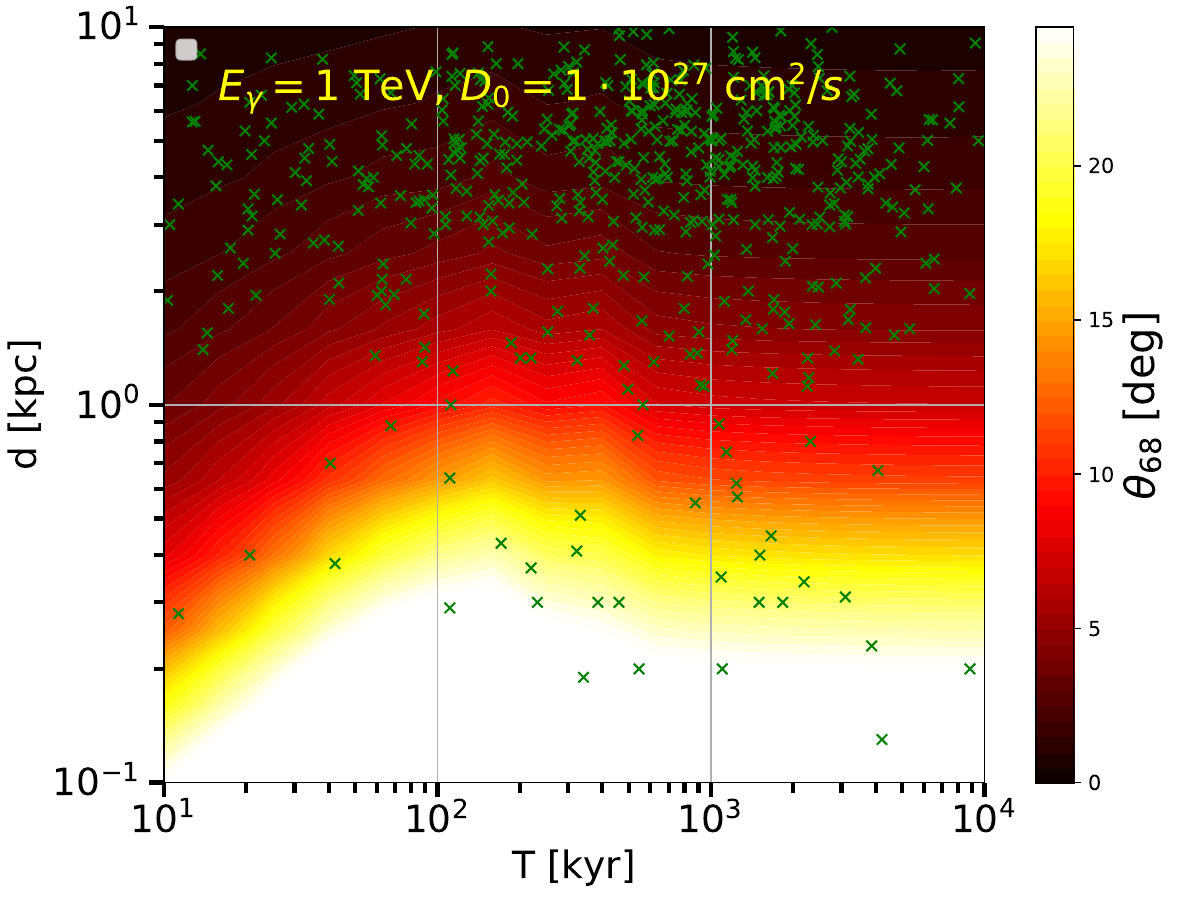}
\caption{Size of extension ($\theta_{\rm{68}}$) of the ICS halo as a function of the distance ($d$) and age ($T$) of the host pulsar. The color bar represents $\theta_{\rm{68}}$ in degrees. From top to bottom: $D_0$ = $6\cdot 10^{25}$ cm$^2$/s, $2\cdot 10^{26}$ cm$^2$/s and $1\cdot 10^{27}$ cm$^2$/s. On the left (right) side  $E_\gamma$ =10 GeV (1 TeV). The green crosses identify the ATNF catalog pulsars. }  
\label{fig:ext}
\end{figure*}
%-------------------------------

%-------------------------------
%--SIZE OF EXTENSION THETA_68
\begin{figure}
\includegraphics[width=0.49\textwidth]{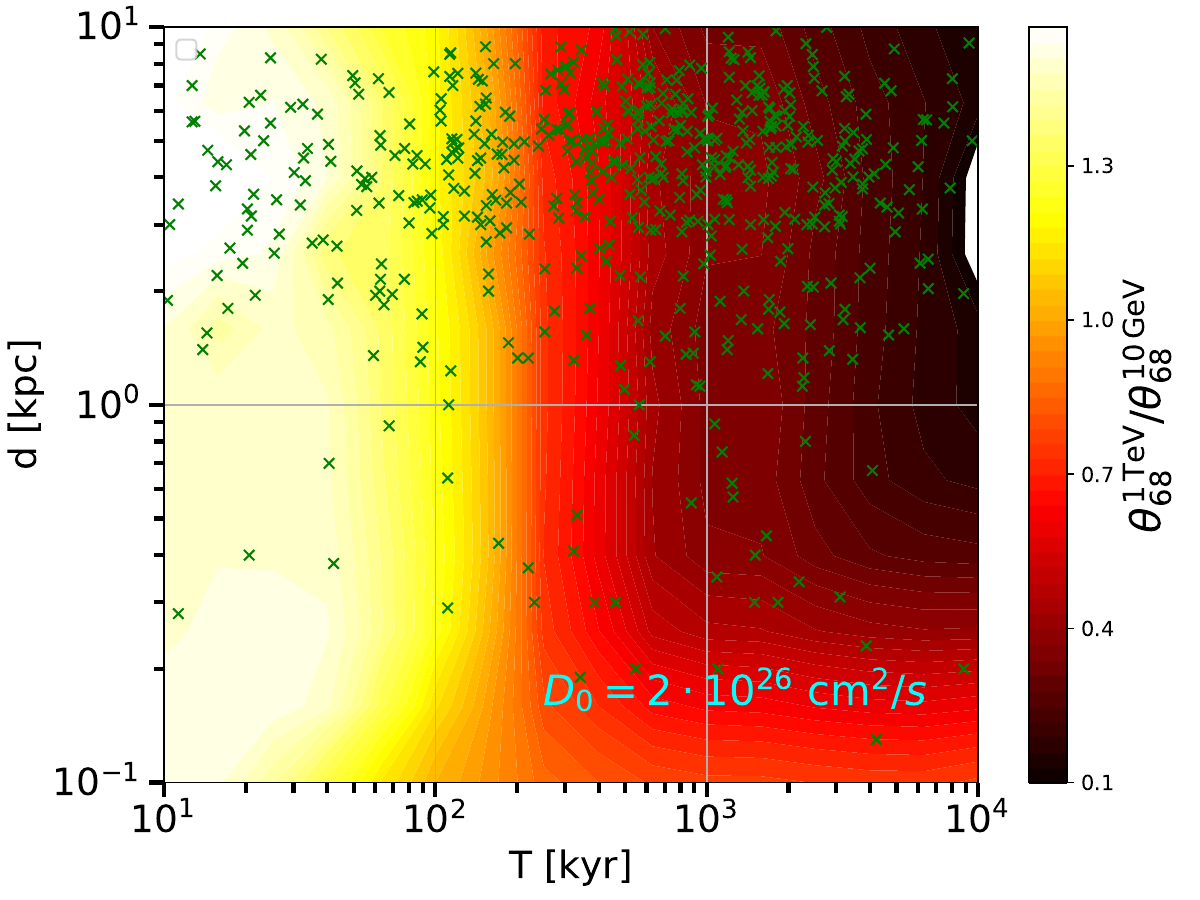}
\caption{Ratio between $\theta_{\rm{68}}$ at 1 TeV and 10 GeV as a function of the pulsar $d$ and $T$. This is calculated for $D_0$ = $2\cdot 10^{26}$ cm$^2$/s. A similar trend is present for $D_0$ = $6\cdot 10^{25}$ cm$^2$/s and $1\cdot 10^{27}$ cm$^2$/s. The green crosses identify the ATNF catalog pulsars.}  
\label{fig:extenergy}
\end{figure}
%-------------------------------

%-------------------------------
%--surface brightness ONE ZONE- TWO ZONE
\begin{figure}
\includegraphics[width=0.49\textwidth]{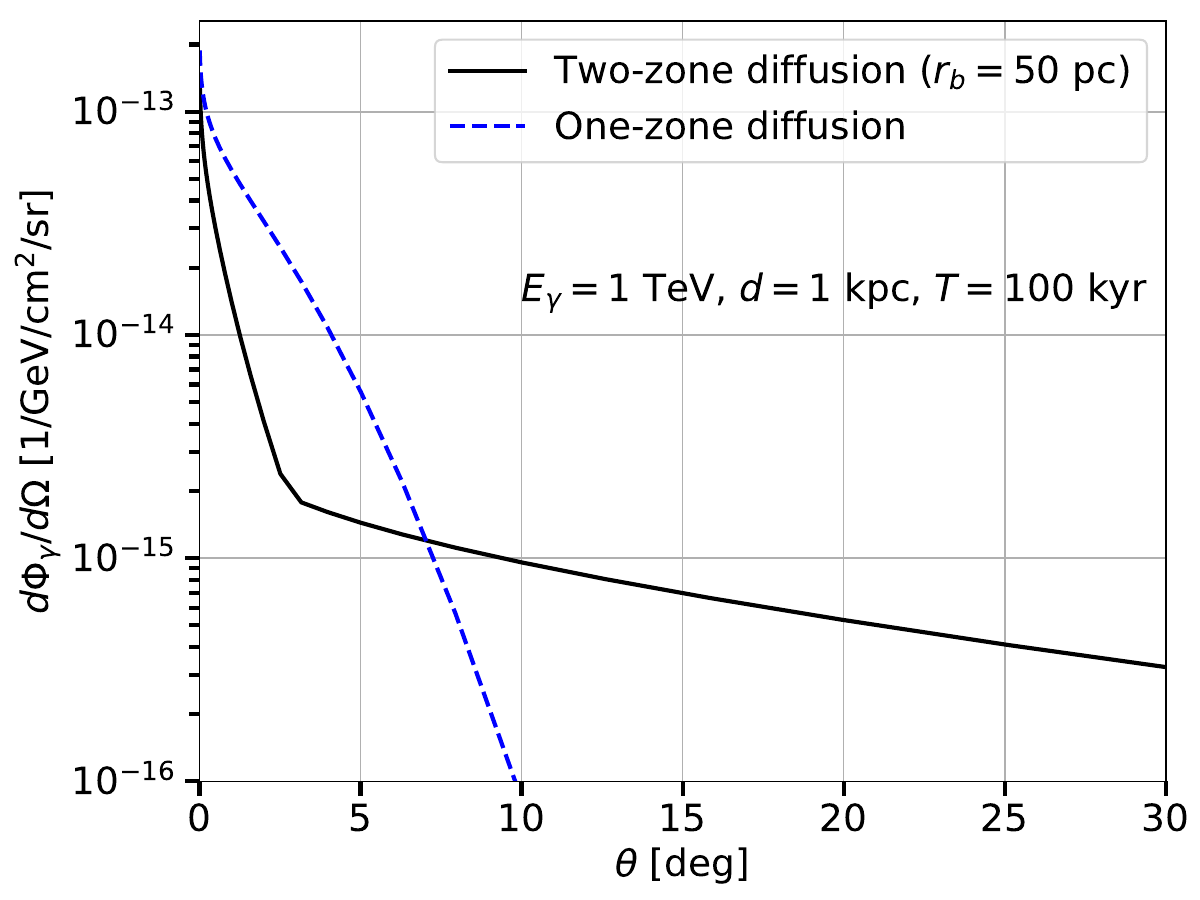}
\caption{Surface brightness for the ICS flux as a function of the angle from the central pulsar, setting $d=1$ kpc, $T=100$ kyr, $D_0=2\cdot 10^{26}$, $E_{\gamma}=1$ TeV 
 and assuming the one or two-zone diffusion models ($r_b=50$ pc, equivalent to an angular distance of $\theta = 2.86^\circ$).}  
\label{fig:SBrb}
\end{figure}
%-------------------------------

\begin{figure*}
\includegraphics[width=0.49\textwidth]{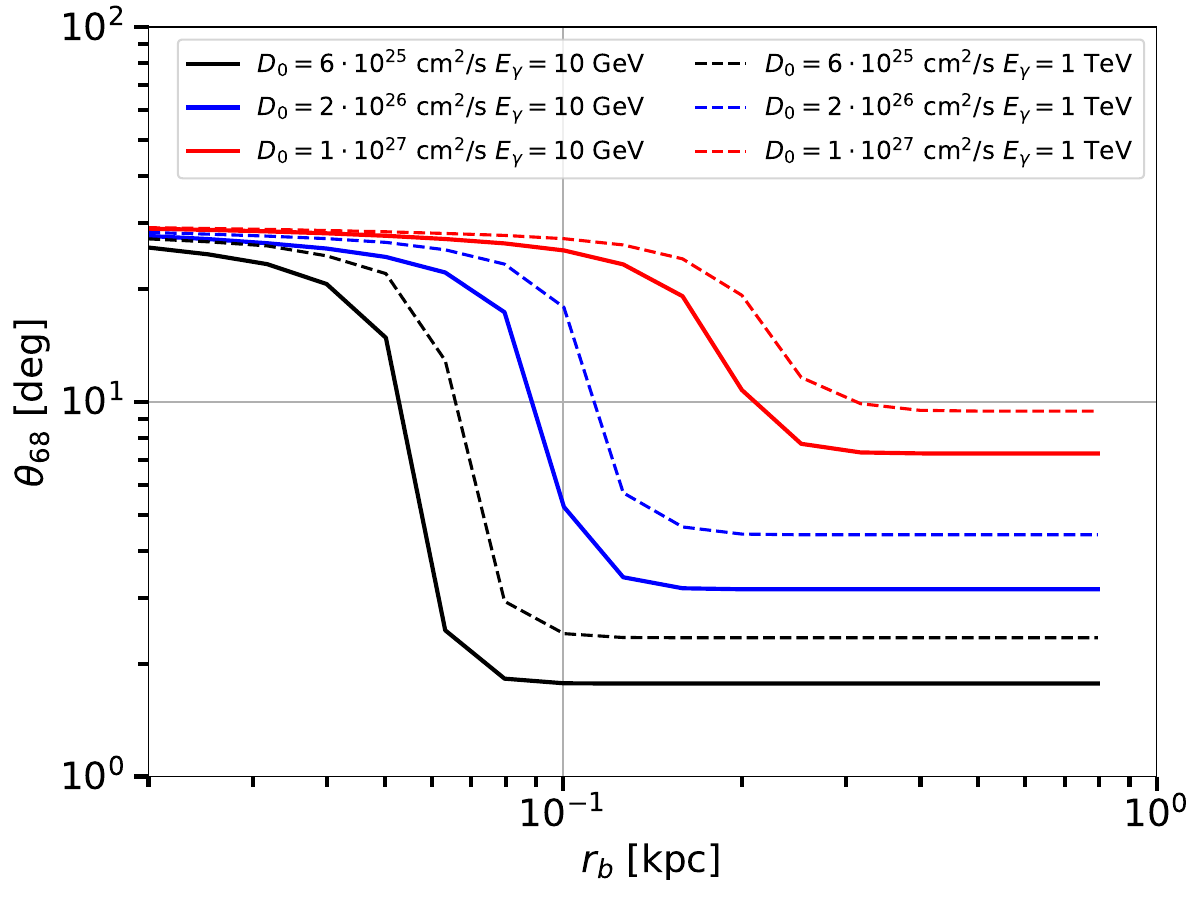}
\includegraphics[width=0.49\textwidth]{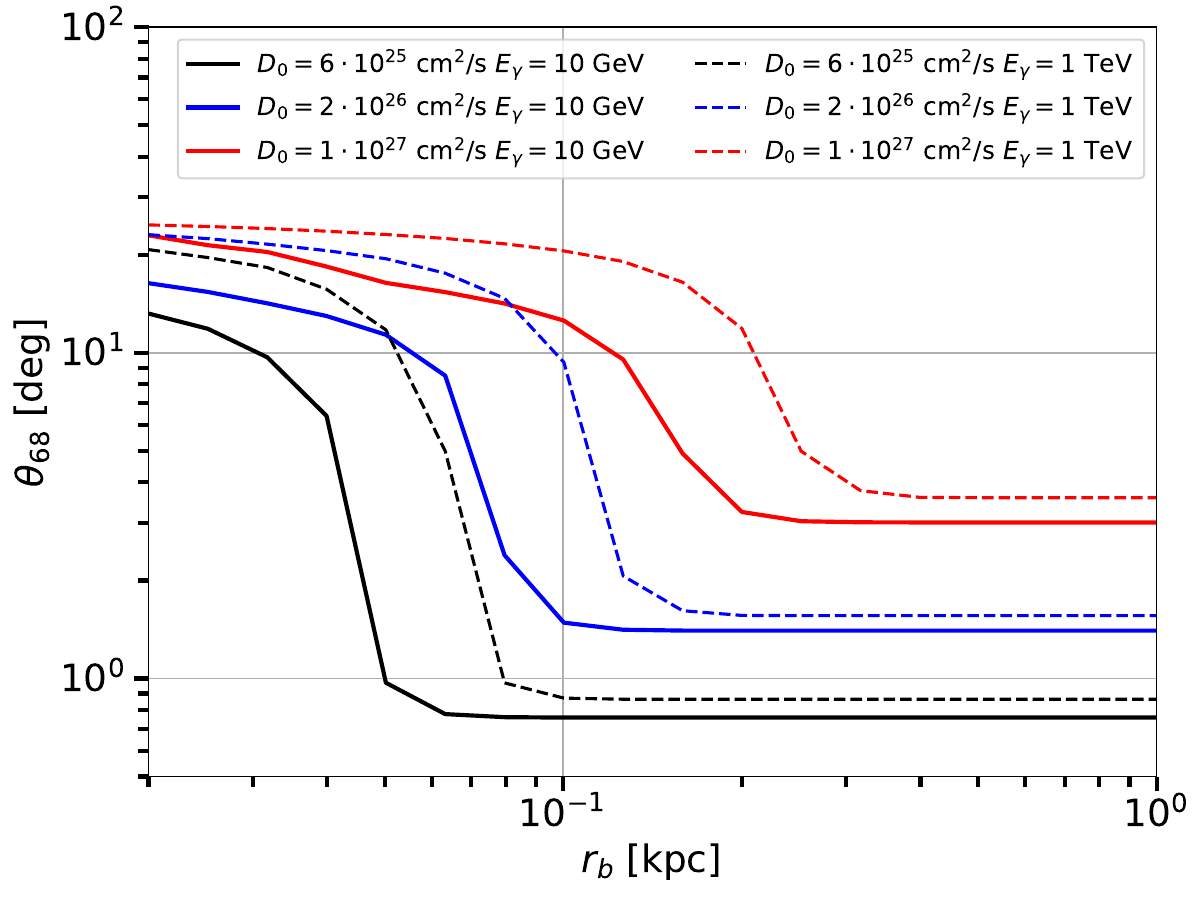}
\caption{$\theta_{\rm{68}}$ as a function of $r_b$ for $E_{\gamma}=10$ GeV (solid lines) and $E_{\gamma}=1$ TeV (dashed lines) and for a pulsar with $d=1$ kpc and $T=100$ kyr (left panel) and $d=2$ kpc and $T=60$ kyr (right panel). In each plot we show the results for $D_0 = 6\cdot 10^{25}$ cm$^2$/s, $2\cdot 10^{26}$ cm$^2$/s and $1\cdot 10^{27}$ cm$^2$/s.}
\label{fig:extrb}
\end{figure*}

The size is a key parameter for the detectability of ICS halos. 
%since IACTs, such as HESS, VERITAS and MAGIC have a limited field of view of about $5^{\circ}$.
Indeed,  IACTs have a few degrees instantaneous field of view and  a very extended halo would be  difficult to detect. 
It is also challenging to detect an halo with a size larger than about $10^{\circ}$ with {\it Fermi}-LAT data, because below 100 GeV the interstellar emission is by far the major contributor 
of the observed flux,  and an imperfect modeling of this component could produce spurious residuals and unreliable results. 
%We derive in this section size of ICS halos for different characteristics of the source (distance and age) and assuming different values for the diffusion coefficient nearby the source ($D_0$).
\\
We define the size of an ICS halo as the angle $\theta_{68}$ which contains the $68\%$ of the flux $\Phi_{\gamma}$:
\begin{equation}
 \Phi^{68\%}_{\gamma}(E_{\gamma}) = 2\pi \int_0^{\theta_{\rm{68}}} \frac{d\Phi_{\gamma}}{d\theta}(E_{\gamma}) \sin{\theta} d\theta,
  \label{eq:flux}
\end{equation}
where $d\Phi_{\gamma}/d\theta$ is the surface brightness and is computed from Eqs.~\ref{eq:phflux} and \ref{eq:M}, where  $\Delta\Omega$ depends on the angle $\theta$ from the center of the pulsar.
%We define the angle $\theta_{\rm{Ext}}$ that satisfies this relation as the size of the angular extension of the ICS halo.
This formulation of the ICS halo size follows the definition of the 68\% containment radius, used by $\gamma$-ray experiments to define the size of extended sources (see, e.g., \cite{REF:2015.3FGL}). 

We first investigate how $\theta_{\rm{68}}$ changes according to the pulsar distance and age, the diffusion coefficient and $\gamma$-ray energy.
We calculate $\theta_{\rm{68}}$ for a grid of pulsar distance and age values between $d\in[0.1,10]$ kpc and $T\in[10,10^4]$~kyr, repeated for three values of $D_0$: $6\cdot 10^{25}$~cm$^2$/s, $2\cdot 10^{26}$~cm$^2$/s and $1\cdot 10^{27}$~cm$^2$/s.
The first two values of $D_0$ are inspired to the results for Geminga found in \cite{Abeysekara:2017science,DiMauro:2019yvh} while the third one has been set to a value closer to the average Galactic diffusion.
Finally, we repeat this exercise for $E_{\gamma}=10$ GeV, which is relevant for {\it Fermi}-LAT data, and $E_{\gamma}=1$ TeV, where the IACTs have their peak of sensitivity.

We show our results for $\theta_{\rm{68}}$  in Fig.~\ref{fig:ext} and \ref{fig:extenergy}, where we superimpose the ATNF catalog pulsars.
We notice that $\theta_{\rm{68}}$ is significantly smaller at $E_{\gamma} = 1$ TeV than at 10 GeV for sources older than about 200 kyr.
Indeed, for such old sources VHE $e^{\pm}$ lose energy very quickly, so that the ICS $\gamma$-ray emission is much closer to the pulsar location.
This trend is confirmed by the recent detection of the Geminga ICS halo with a size of about $5^{\circ}$ above 5 TeV \cite{Abeysekara:2017science} and about $15^{\circ}$ at 10 GeV \cite{DiMauro:2019yvh}.
On the other hand, sources younger than about 200 kyr have extension at 1 TeV that is slightly larger than the one at 10 GeV because for these ages 1 TeV $e^{\pm}$ have a propagation length $\lambda$ (see Eq.~\ref{eq:lambda}) that is larger than the one at 10 GeV.

We also observe that the larger is $D_0$ the larger is $\theta_{\rm{68}}$. 
For example, for a source as Geminga with $d=0.25$ pc and $T=340$ kyr and at $E_{\gamma}=10$ GeV ($E_{\gamma}=1$ TeV) the size of $\theta_{\rm{68}}$ is $15^{\circ}$, $25^{\circ}$ and $30^{\circ}$ ($10^{\circ}$, $18^{\circ}$ and $25^{\circ}$) for $D_0$ equal to $6\cdot 10^{25}$~cm$^2$/s, $2\cdot 10^{26}$~cm$^2$/s and $1\cdot 10^{27}$~cm$^2$/s, respectively.
A higher diffusion coefficient makes the particle travel a larger distance in the Galaxy before losing most of its energy.

IACTs have an instantaneous field of view between $3.5-5^{\circ}$, thus if $D_0=10^{27}$ cm$^2$/s only sources farther than about 3 kpc would have a detectable ICS halo.
On the other hand, if $D_0\sim 10^{26}$ cm$^2$/s, as detected for Geminga in \cite{Abeysekara:2017science,DiMauro:2019yvh}, most of the ATNF catalog pulsars would be good targets for ICS halo searches by IACTs.
Instead, in  the {\it Fermi}-LAT energy range  most of Galactic pulsars would generate very extended halos.
More precisely, fixing $D_0=6\cdot 10^{25}$ cm$^2$/s ($D_0=2\cdot 10^{26}$ cm$^2$/s), the size of $\theta_{\rm{68}}$ would be smaller than two degrees 
only for $d\geq10^{0.58\log_{10}{(T\,[\,\rm{kyr}])}-1.2}$ kpc ($d\geq10^{0.57\log_{10}{(T\,[\,\rm{kyr}])}-0.9}$ kpc).
This means that a source with an age of 100 kyr should be farther than about 0.9 kpc (1.7 kpc) if $D_0=6\cdot 10^{25}$ cm$^2$/s ($D_0=2\cdot 10^{26}$ cm$^2$/s) to be detected with an extension smaller than two degrees.

These results may change if a two-zone diffusion model is considered. 
In this model, the pulsar is located into a bubble of low diffusion where $e^{\pm}$ are more effectively confined. 
In general, assuming a two-zone diffusion model has the effect of increasing $\theta_{\rm{68}}$.
In Fig.~\ref{fig:SBrb} we show the surface brightness $d\Phi_{\gamma}/d\theta$ 
calculated for a pulsar with $d=1$ kpc and $T=100$ kyr at $E_{\gamma}=1$ TeV, and assuming either a one or a two-zone diffusion model, where  $D_0 =2\cdot 10^{26}$ cm$^2$/s only within $r_b=50$ pc.
It is clear from the figure that the two-zone diffusion model has a much wider profile, which results into a more extended ICS flux.
This effect  depends on the value of $r_b$.

In Fig.~\ref{fig:extrb} we study  $\theta_{\rm{68}}$ as a function of $r_b$, for a pulsar with $d=1$ kpc and $T=100$ kyr, and an other one with $d=2$ kpc and $T=60$ kyr.
For these two cases we set $D_0 = 6\cdot 10^{25}$ cm$^2$/s, $2\cdot 10^{26}$ cm$^2$/s and $1\cdot 10^{27}$ cm$^2$/s.
For $r_b\geq 0.1$ kpc, $\theta_{\rm{68}}$ tends to the value obtained with the one zone model (see Fig.~\ref{fig:ext}).
Indeed, for such a large low-diffusion zone bubble most of the $e^{\pm}$ lose completely their energy before reaching the high-diffusion zone.
Therefore, they are completely trapped inside the low-diffusion bubble. This effect could be a result of the confinement of the CRs inside the PWN and/or the PWN.
For example, in the case of the pulsar with $d=1$ kpc and $T=100$ kyr, $\gamma$ rays with energies of $E_{\gamma}=1$ TeV are on average produced by $e^{\pm}$ of 10 TeV energy. 
These $e^{\pm}$ in a diffusion environment with $D_0=6\cdot 10^{25}$ cm$^2$/s have a propagation length of about 30 pc.
Therefore, if $r_b$ is larger than this length, $\theta_{\rm{68}}\approx 2^{\circ}$, similar to the value found for the one-zone model.
On the other hand, for smaller values of $r_b$, $e^{\pm}$ exit the low-diffusion zone before losing most of their energy and produce a significant ICS flux in the high-diffusion zone. 
Since outside the low-diffusion bubble $e^{\pm}$ travel significant larger distances, the ICS halo can become very extended.
We also notice  that the lower is $D_0$ the lower is the value of $r_b$ at which we observe the transition between small and large values of $\theta_{\rm{68}}$. 
This is due to the fact that with a less intense $D_0$, $e^{\pm}$ travel shorter distances before losing most of their energies. 
%Therefore, $e^{\pm}$ produce smaller ICS halos, as we have shown in Fig.~\ref{fig:ext}, and $r_b$ should be smaller in order to affect $\theta_{\rm{EXT}}$.
We conclude that for $D_0 \sim 10^{26}$ cm$^2$/s values of $r_b\geq 80$ pc do not alter significantly $\theta_{\rm{68}}$. 
In other words, diffusion coefficient values of the order of $D_0 \sim 10^{26}$ cm$^2$/s with $\theta_{\rm{68}}$ at the degree scale implies $r_b \gtrsim 80$ pc.

%MAGIC 3.5
%VERITAS 3.5
%HESS 5.0

%\begin{figure*}
%\includegraphics[width=0.49\textwidth]{plots/mapcube_Geminga_asymmetric_paperIV_1.pdf}
%\includegraphics[width=0.49\textwidth]{plots/mapcube_Geminga_asymmetric_paperIV_10.pdf}
%\caption{Intensity map in units of 1/GeV/cm$^2$/s/sr for the ICS flux from Geminga when proper motion is taken into account. The left (right) planel plot is for $E_{\gamma}=10$ GeV (1 TeV). The direction of the proper motion ($\bf{v_T}$) and the $\theta_{\rm{motion}}$ are also indicated.}  
%\label{fig:propmotmap}
%\end{figure*}

%%%%%%%%%%%%%%%%%%%%%%%%%%%%%%%%%%%%%%%%%%%%%%%%%%%%%%%%%%%%
\subsection{Pulsar proper motion}

\begin{figure*}
\includegraphics[width=0.49\textwidth]{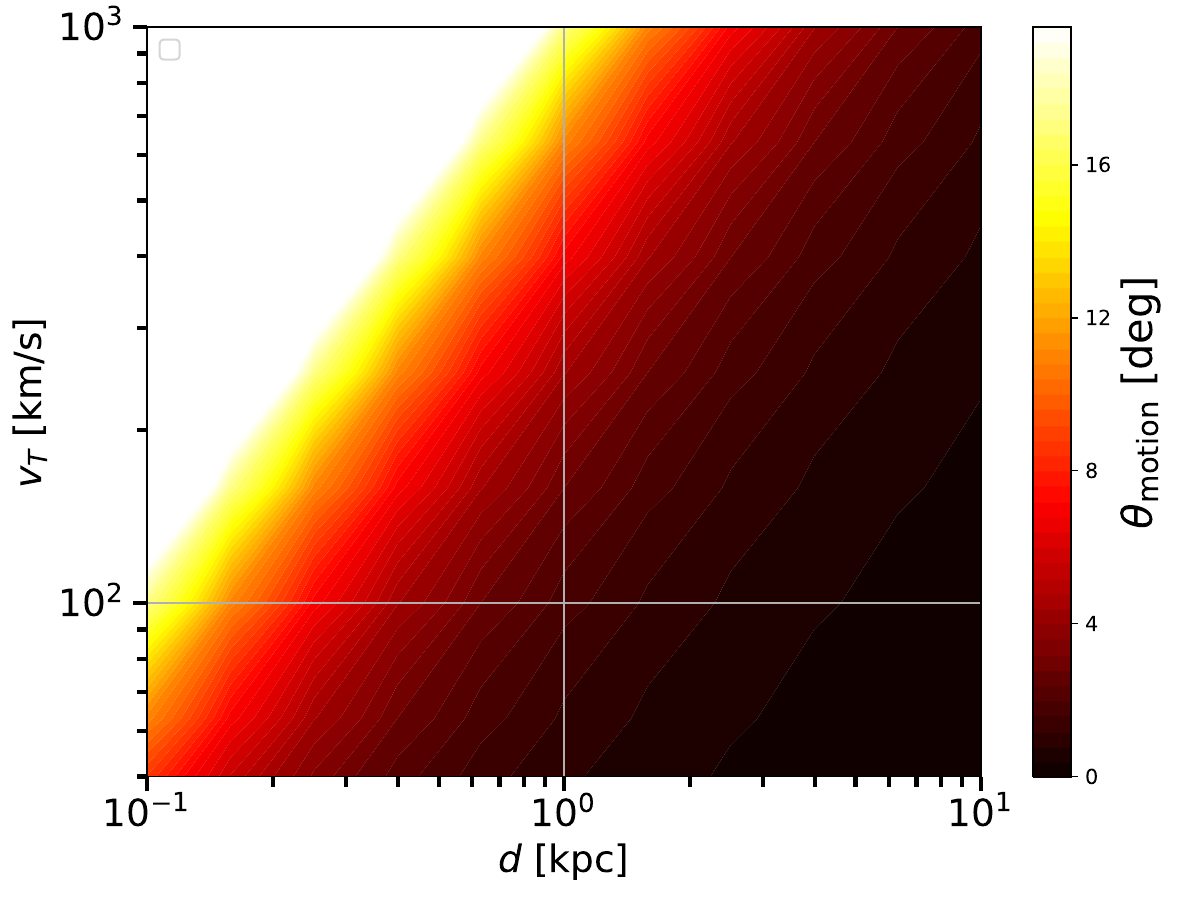}
\includegraphics[width=0.49\textwidth]{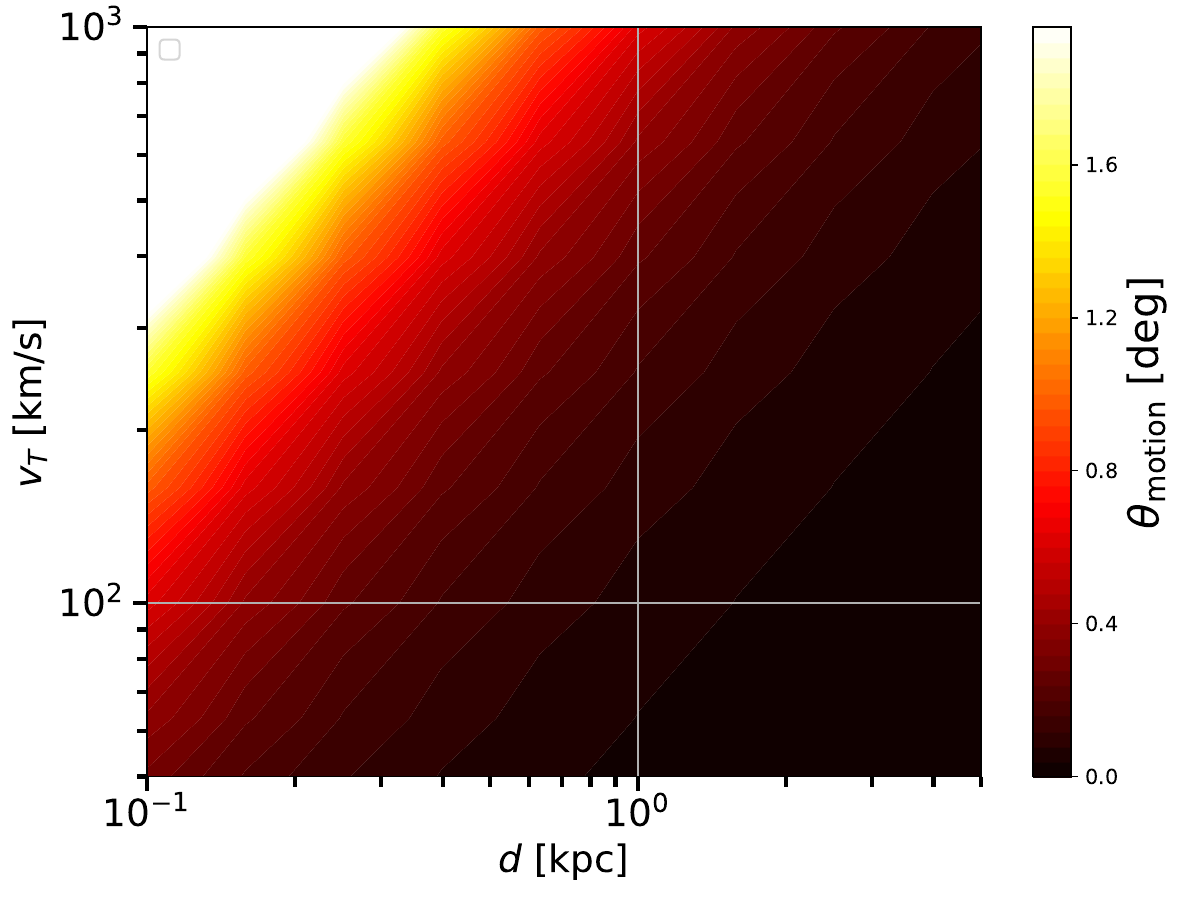}
\caption{Angular distortion $\theta_{\rm{motion}}$ as a function of the pulsar distance and transverse velocity for $E_{\gamma}=10$ GeV (left panel) and $E_{\gamma}=1$ TeV (right panel). The scale for  $\theta_{\rm{motion}}$  is different by one order of magnitude for the two energies.}  
\label{fig:extpropermotion}
\end{figure*}

An other element that can affect the spatial morphology of $\gamma$ rays produced for ICS is the pulsar proper motion.
The analysis presented in \cite{DiMauro:2019yvh} shows that the proper motion of the Geminga pulsar, which moves with a transverse velocity $v_T=$ 211 km/s  \cite{Faherty:2007}, shapes significantly the $\gamma$-ray ICS template below 100 GeV. 
In particular we have shown that at 10 GeV the ICS halo has a distortion of about $10^{\circ}$ in the opposite direction of the proper motion.
This is due to the fact that 10 GeV $\gamma$ rays are produced by $e^{\pm}$ emitted by the pulsar tens of kyr ago. 
Therefore, a significant fraction of the $\gamma$-ray flux is detected in the direction where the pulsar was in the past.  

The ICS power $\mathcal{P}^{ICS}$ has a peak at around $E_e=1.5$~TeV for $E_{\gamma}=10$ GeV and $E_e=60$ TeV for $E_{\gamma}=10$~TeV.
We use here the ISRF model as in \cite{Vernetto:2016alq}. Very similar results are found with the model presented in Refs.~\cite{2006ApJ...648L..29P,2017MNRAS.470.2539P}.
An electron of energy of 1.5 TeV (60 TeV) loses most of its energy after about 300 kyr (20 kyr).
In this time lapse the Geminga pulsar has travelled across the sky for 60 pc (4 pc).
Therefore, we expect that the size of extension of the ICS halo is distorted by about $12^{\circ}$ ($0.9^{\circ}$) in the opposite direction of the proper motion 
(see Fig.~10 in \cite{DiMauro:2019yvh}).

%If we apply our calculation for $\theta_{\rm{dist}}$ to the HAWC ({\it Fermi}-LAT) detection of Geminga, we predict that the $\gamma$-ray emission should be distorted by about $0.9^{\circ}$ ($12^{\circ}$).
%The predicted value of $\theta_{\rm{dist}}$ for the HAWC data is a very small angle with respect to the angular extension which has been measured to be about $5^{\circ}$ \cite{Abeysekara:2017science}.
%On the other hand, the value found for the {\it Fermi}-LAT energy range is measurable and indeed Ref.~\cite{DiMauro:2019yvh} found a preference for the model that includes the proper motion.
%We show in Fig.~\ref{fig:propmotmap} the intensity map at 10 GeV and 10 TeV to show how the ICS flux is distorted for the Geminga pulsar proper motion.

We generalize this calculation and derive the source distance and age values for which the proper motion is a relevant effect in the ICS flux.
The angular size $\theta_{\rm{motion}}$ by which the ICS halo is distorted   due to the pulsar proper motion can be parametrized as:
\begin{equation}
\theta_{\rm{motion}}(E_{\gamma}) = \rm{atan} \left( \frac{d_{\rm{motion}}(E_{\gamma})}{d}  \right) ,
 \label{eq:proper1}
\end{equation}
where $d$ is the actual distance of the source from Earth and $d_{\rm{motion}}$ is:
\begin{equation}
d_{\rm{motion}}(E_{\gamma}) =  \frac{v_T E_e(E_{\gamma})}{b(E_e(E_{\gamma}))}.
 \label{eq:proper2}
\end{equation}
Here $E_e(E_{\gamma})$ is the energy of the electron for which the ICS power $\mathcal{P}^{ICS}$ has its peak for a given $\gamma$-ray energy and $v_T$ is the transverse velocity of pulsar.
We can now put together Eq.~\ref{eq:proper1} and \ref{eq:proper2} finding:
%print(atan( ( (0.324/(b0*(E_e)))*vproper) /d)*180./np.pi)
\begin{equation}
\theta_{\rm{motion}}(E_{\gamma})[\rm{deg}] = \rm{atan}  \left( 0.324  \frac{ \frac{v_T [\rm{km/s}]}{b\rm{[}10^{-16} \rm{GeV/s} \rm{]} (E_e\rm{[GeV]}]) }}{d\rm{[kpc]}}  \right).
 \label{eq:properfinal}
\end{equation}

In Fig.~\ref{fig:extpropermotion} we show the value of $\theta_{\rm{motion}}$ for $\gamma$-ray energy of 10 GeV and 1 TeV.
Here we assume energy losses for ICS and synchrotron radiation parametrized as $b(E)=5 \times 10^{-17} \rm{GeV/s} \,(E_e[\,\rm{GeV}])^2$.
The angular distortion at 1~TeV is significantly smaller with respect to the 10 GeV case. 
Indeed, at 1 TeV $\theta_{\rm{motion}}>1^{\circ}$ only for pulsars with  velocities larger than about 300 km/s and  closer than a few hundred pc.
For all other $v_T$-$d$ combinations the angular distortion is not significant.

In \cite{Abdalla:2017vci} the HESS Collaboration found that the offset between the PNW $\gamma$-ray emission and the central pulsar is between $0.2-0.4^{\circ}$. 
From Fig.~\ref{fig:extpropermotion} this would be consistent with pulsar proper motion with velocities $v_T $ smaller that few hundred km/s. Indeed, most of the pulsars have velocities of the order of 100 km/s (see, e.g., \cite{Hobbs:2005yx} for a compilation of pulsar proper motion measurements).

On the other hand, at $E_{\gamma}= 10$ GeV even moderate pulsar velocities affect the morphology of the ICS emission, implying $\theta_{\rm{motion}}$ of at least a few degrees. 
This represents a limiting factor for detecting ICS halos in {\it Fermi}-LAT data, since $v_T$ is known only for a limited number of pulsars (about 230 over almost 3000 detected so far).
Indeed, performing a search for ICS emission from a pulsar with unknown $\bf{v_T}$  is challenging, since the intensity and direction of the motion can create a significant asymmetry in the morphology.
This issue is probably alleviated by the fact that the most promising pulsars for the ICS halo search are also the better observed and studied and for many of them the proper motion has been already measured.

%%%%%%%%%%%%%%%%%%%%%%%%%%%%%%%%%%%%%%%%%%%%%%%%%%%%%%%%%%%%
\section{Inverse Compton Scattering halos at TeV energies}
\label{sec:IACTs}

In this section we illustrate how the $\gamma$-ray flux selects the most promising pulsars with a detectable ICS halo.
%We consider the pulsars in the ATNF catalog\footnote{\url{http://www.atnf.csiro.au/research/pulsar/psrcat/}} to calculate the $\gamma$-ray flux for each individual source.
First, we predict the number of ICS halos that could be detected by  HAWC, HESS  and CTA as a function of the efficiency $\eta$.
The number of expected ICS halos detections with HAWC, HESS and CTA has  been recently calculated in \cite{Sudoh:2019lav}.
Their model uses different assumptions with respect to ours. In particular, instead of computing the extended ICS flux for each source, they rescale the observed Geminga gamma-ray flux to all the sources, assuming they are "Gemiga-like" systems.  Moreover, their results for the cumulative number of detections vary by about one order of magnitude according to the choice of the pulsar rotational period and the alignment of the pulsar jet. 
Therefore, their results are not easily comparable with ours.

\subsection{IACTs detectability of extended ICS halos}
\label{sensitivity}

{\bf HAWC.} 
The 2HWC catalog \cite{Abeysekara:2017hyn} reports the sensitivity for the detection of a point source as a function of the declination.
The lowest detectable flux at 7 TeV is $6\cdot 10^{-15} ({\rm TeV \, cm^2 \, s})^{-1}$ for declination angles in the range $10^{\circ}-30^{\circ}$ and a point source with a spectral slope of $-2.5$. 
However, this value is not appropriate for our scope, because we are interested in the detection of extended ICS halos with a size of a fraction of the degree (see Tab.~\ref{tab:TeVsourcesHAWC}).
We estimate the HAWC sensitivity to ICS halos by taking the publicly available data of the 2HWC Survey\footnote{\url{https://data.hawc-observatory.org/datasets/2hwc-survey/index.php}}. 
This on-line resource provides - at each direction in the HAWC field of view - the significance for the presence of a source, for different spatial and spectral assumptions.
In particular, it provides the significance, the flux measurement and the $95\%$ CL flux upper limit at 7 TeV for a point like source with a spectral index of $-2.7$, or for an extended  source sizing 0.5$^{\circ}$, 1.0$^{\circ}$ and 2.0$^{\circ}$, with a spectral index of $-2.0$.
We estimate the average flux at different sky directions for the detection at about $5\sigma$ significance to be $[8,9,10,20]\cdot 10^{-15}$ $({\rm TeV\, cm^2\, s})^{-1}$ for a point like, or extended source of size 0.5$^{\circ}$, 1.0$^{\circ}$ and 2.0$^{\circ}$, respectively.   
%The value we report here of the point source sensitivity is slightly larger than  because we average over the different declination angles.
%The value we report above for the sensitivity of a point like source is compatible with the value reported in \cite{Abeysekara:2017hyn} between declination angles $5^{\circ}-35^{\circ}$.
Most of the pulsars are predicted to have an ICS halo with an angular extension from a fraction of a degree to a few degrees (see Fig.~\ref{fig:ext}), so we fix the flux sensitivity at 7 TeV to be $1\cdot 10^{-14}\; ({\rm TeV\, cm^2\, s})^{-1}$,  that is valid for a $1^{\circ}$ extended source. 
We note that we are not including any declination dependence of the sensitivity.

\begin{figure}
\includegraphics[width=0.49\textwidth]{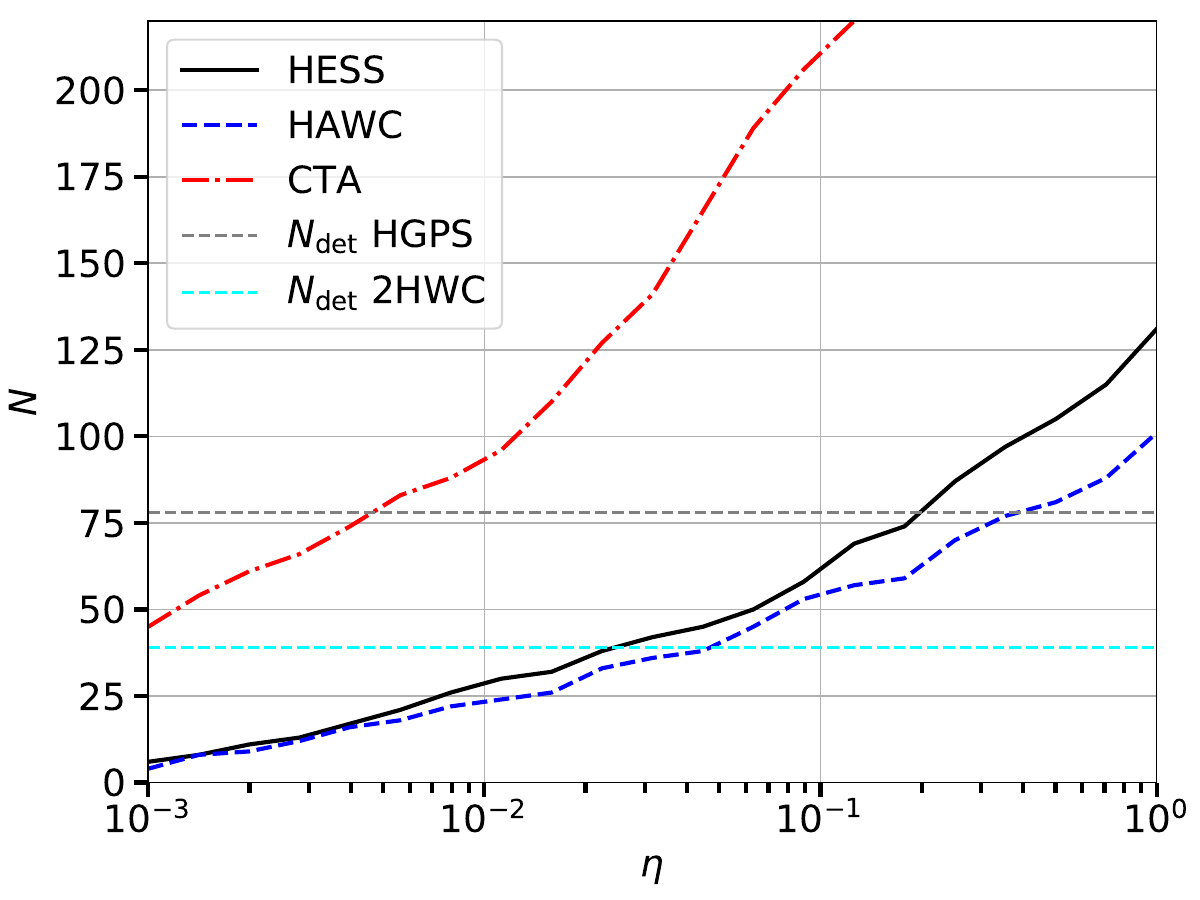}
\caption{Prediction for the number of ICS halos powered by ATNF catalog pulsars detected by HESS, HAWC and CTA as a function of the efficiency for the conversion of spin-down luminosity into $e^{\pm}$ ($\eta$). The cyan and grey horizontal lines represent the number of sources detected in the HGPS and 2HWC.}  
\label{fig:Ndet}
\end{figure}

{\bf HESS.} 
In order to estimate the flux sensitivity for HESS we use the information published in the HGPS catalog \cite{H.E.S.S.:2018zkf}.
%In \cite{H.E.S.S.:2018zkf} the flux sensitivity for a point source with a spectral index of $-2.3$ is reported as a function of longitude. 
The HESS Collaboration has calculated that the flux sensitivity for a point source with a spectral index of $-2.3$ is around $1\%$ of the Crab flux, i.e.~about $3\cdot 10^{-13}\; ({\rm TeV\, cm^2\, s})^{-1}$ at 1 TeV.
This has been calculated for the optimistic case of an isolated source, while the real sensitivity is probably higher.
They also show that there is a dependence of the flux sensitivity with the Galactic longitude. 
In the range between $l\in [40^{\circ},300^{\circ}]$ the sensitivity for point sources  is roughly constant and at its lowest level, while it increases outside these directions. 
This result cannot be directly used for extended sources.  
We select therefore the sources detected as extended with $\theta_{68} \sim 0.1^{\circ}-0.4^{\circ}$ with the faintest flux at 1 TeV.
We find a dependence for the flux of those sources with the size of extension.
For sources detected with $\theta_{68} \approx 0.1^{\circ}$ the faintest detected flux is $5\cdot 10^{-13}$ $({\rm TeV\, cm^2\, s})^{-1}$, 
for $\theta_{68} \approx 0.2$ it is $1\cdot 10^{-12}$ $({\rm TeV\, cm^2\, s})^{-1}$ and for $\theta_{68} \approx 0.4$ it is $2\cdot 10^{-12}$ $({\rm TeV\, cm^2\, s})^{-1}$.
We make the simplistic assumption of neglecting the dependence with $\theta_{\rm{68}}$, and fix the sensitivity to $1\cdot 10^{-12}$ $({\rm TeV\, cm^2\, s})^{-1}$ at 1 TeV.
Moreover, we neglect the longitude dependence which is present in a minor portion of the HESS field of view.

{\bf CTA} is the next generation ground-based observatory for $\gamma$-ray astronomy at VHE \cite{Mazin:2019ykz}. 
With more than 100 telescopes located in the northern and southern hemispheres, CTA will be the world's largest and most sensitive high-energy $\gamma$-ray observatory.
Ref.~\cite{Ambrogi:2018skq} ha calculated the flux sensitivity for the detection at the $5\sigma$ CL of an extended source with 50h observation time and different sizes of extension.
The sensitivity flux at 10 TeV is $7\cdot 10^{-16}$ $({\rm TeV\, cm^2\, s})^{-1}$ ($1.2\cdot 10^{-15}$ $({\rm TeV\, cm^2\, s})^{-1}$) for a $0.1^{\circ}$ ($0.5^{\circ}$) extension.
We will use $1\cdot 10^{-15}$ $({\rm TeV\, cm^2\, s})^{-1}$ in the rest of this section.

\medskip 

In Fig.~\ref{fig:Ndet} we show the number $N$ of ICS halos detectable by HAWC, HESS and CTA as a function of the efficiency $\eta$ (see Eq.~\ref{eq:Q_E_cont}).
We calculate $\Phi_{\gamma}$ using Eq.~\ref{eq:phflux} for all the ATNF catalog pulsars. If the flux is above the sensitivity of each experiment, it contributes to this number. 
The  design of CTA is very promising for the detection of ICS halos. Indeed, with an efficiency a slow as a few \%, 
%(as we have found for Geminga and Monogem PWNe in \cite{DiMauro:2019yvh}) 
this future experiment could detect about $100-130$ ICS halos. 
On the other hand, HAWC and HESS might have already detected around $25$ and $35$ halos, respectively.
This is a realist number, given that 2HWC and HGPS catalogs contain 39 and 78 sources, and only a fraction of them are probably associated to ICS halos.
We can revert the reasoning and use the number of sources detected in 2HWC and HGPS catalogs to find a rough upper limit for the average efficiency, which reads about 0.07 for 2HWC and 0.25 for HGPS.

%%%%%%%%%%%%%%%%%%%%%%%%%%%%%%%%%%%%%%%
\subsection{Ranking of the brightest expected ICS halos} 

\begin{table*}
\begin{center}
\begin{tabular}{|c|c|c|c|c|c|c|c|c|c|c|}
\hline
PSR   &   $l$  &  $b$  &   $d$ & $T$ & $\dot{E}$ &  $\Phi^{10\,\rm{TeV}}_{\gamma}$  &   $\theta_{68}$ & Name  & Class \\ 
\hline
  &	[deg]	 &  [deg]&   [kpc] &  [kyr]        &   [erg/s]   & [$({\rm TeV\, cm^2\, s})^{-1}$]  & [deg]  &     &   \\ 
\hline
J1826-1256   &  18.56 & -0.38 &  1.55 & 14 &  $3.6\cdot 10^{36}$  & $2.5\cdot 10^{-13}$  & 0.89  &   2HWC J1825-134  &  UNID \\
J2021+3651  &   75.22 & 0.11 &  1.80 & 17 &  $3.4\cdot 10^{36}$  & $1.6\cdot 10^{-13}$  & 0.82  &   2HWC J2019+367  &  UNID \\
J1813-1246   &   17.24 & 2.44 &  2.63 & 43 &  $6.2\cdot 10^{36}$  & $8.6\cdot 10^{-14}$  & 0.60  &   2HWC J1812-126  &  UNID \\
J1907+0602 &   40.18 & -0.89 &  2.37 & 20 &  $2.8\cdot 10^{36}$  & $6.7\cdot 10^{-14}$  & 0.64  &   2HWC J1908+063  &  UNID \\
J0633+1746  &   195.13 & 4.27 &  0.19 & 342 &  $3.3\cdot 10^{34}$  & $5.8\cdot 10^{-14}$  &  6.54      &  GEMINGA PWN  &  TEV HALO \\
B0656+14     &    201.11 & 8.26 &  0.29 & 111 &  $3.8\cdot 10^{34}$  & $3.4\cdot 10^{-14}$  & 4.71  &   2HWC J0700+143  &  TEV HALO \\
B1951+32     &    68.77 & 2.82 &  3.00 & 107 &  $3.7\cdot 10^{36}$  & $3.0\cdot 10^{-14}$  & 0.46  &   undetected  &  undetected \\
J1811-1925   &    11.18 & -0.35 &  5.00 & 23 &  $6.4\cdot 10^{36}$  & $2.8\cdot 10^{-14}$  & 0.30  &   2HWC J1809-190  &  UNID \\
B1823-13      &    18.00 & -0.69 &  3.61 & 21 &  $2.8\cdot 10^{36}$  & $2.6\cdot 10^{-14}$  & 0.41  &   2HWC J1825-134  &  UNID \\
J1935+2025  &    56.05 & -0.05 &  4.60 & 21 &  $4.7\cdot 10^{36}$  & $2.5\cdot 10^{-14}$  & 0.32  &   SNR G054.1+00.3  &  PWN \\
J1954+2836  &    65.24 & 0.38 &  1.96 & 69 &  $1.1\cdot 10^{36}$  & $2.3\cdot 10^{-14}$  & 0.77  &   2HWC J1955+285  &  UNID \\
J1809-1917   &    11.09 & 0.08 &  3.27 & 51 &  $1.8\cdot 10^{36}$  & $1.5\cdot 10^{-14}$  & 0.47  &   2HWC J1809-190  &  UNID \\
J1838-0655   &    25.25 & -0.20 &  6.60 & 23 &  $5.6\cdot 10^{36}$  & $1.3\cdot 10^{-14}$  & 0.22  &   2HWC J1837-065  &  PWN \\
J1856+0245  &    36.01 & 0.06 &  6.32 & 21 &  $4.6\cdot 10^{36}$  & $1.2\cdot 10^{-14}$  & 0.23  &   2HWC J1857+027  &  UNID \\
J1958+2846  &    65.88 & -0.35 &  1.95 & 22 &  $3.4\cdot 10^{35}$  & $1.2\cdot 10^{-14}$  & 0.79  &   2HWC J1955+285  &  UNID \\
J1740+1000  &    34.01 & 20.27 &  1.23 & 114 &  $2.3\cdot 10^{35}$  & $1.1\cdot 10^{-14}$  & 1.15  &   undetected  &  undetected \\
J1913+1011  &    44.48 & -0.17 &  4.61 & 169 &  $2.9\cdot 10^{36}$  & $9.1\cdot 10^{-15}$  & 0.27  &   2HWC J1912+099  &  SHELL \\
J1837-0604   &    25.96 & 0.27 &  4.77 & 34 &  $2.0\cdot 10^{36}$  & $8.6\cdot 10^{-15}$  & 0.32  &   2HWC J1837-065  &  UNID \\
J1907+0631  &    40.52 & -0.48 &  3.40 & 11 &  $5.3\cdot 10^{35}$  & $6.9\cdot 10^{-15}$  & 0.41  &   2HWC J1908+063  &  UNID \\
J1928+1746  &    52.93 & 0.11 &  4.34 & 83 &  $1.6\cdot 10^{36}$  & $6.5\cdot 10^{-15}$  & 0.30  &   2HWC J1928+177  &  UNID \\
J0633+0632  &    205.09 & -0.93 &  1.35 & 59 &  $1.2\cdot 10^{35}$  & $5.8\cdot 10^{-15}$  & 1.14  &   HAWC J0635+070  &  TEV HALO \\
J1831-0952   &    21.90 & -0.13 &  3.68 & 128 &  $1.1\cdot 10^{36}$  & $5.6\cdot 10^{-15}$  & 0.39  &   2HWC J1831-098  &  PWN \\
J1828-1101   &    20.50 & 0.04 &  4.77 & 77 &  $1.6\cdot 10^{36}$  & $5.3\cdot 10^{-15}$  & 0.28  &   2HWC J1831-098  &  UNID \\
\hline
\end{tabular}
\caption{List of the pulsars from the ATNF catalog in the HAWC field of view with the brightest predicted ICS halo flux at 10 TeV. We list the pulsar name, Galactic coordinates, distance, age and spin-down luminosity taken from the ATNF catalog. Then, we report the predicted extension $\theta_{68}$ and ICS flux $\Phi^{10\,\rm{TeV}}_{\gamma}$ both calculated at 10 TeV and assuming $D_0=7\cdot 10^{25}$ cm$^2$/s. Finally, we display the name as in 2HWC catalog and the classification given in TeVCat. Sources labeled as UNID are unidentified in the TeVCat catalog but are associated with potential ICS halo in our analysis since they have a pulsar within a small angular distance.}
\label{tab:TeVsourcesHAWC}
\end{center}
\end{table*}

\begin{table*}
\begin{center}
\begin{tabular}{|c|c|c|c|c|c|c|c|c|c|c|}
\hline
PSR   &   $l$  &  $b$  &   $d$ & $T$ & $\dot{E}$ & $\Phi^{10\,\rm{TeV}}_{\gamma}$   & $\theta_{\rm{68}}$ \\ 
\hline
  &	[deg]	 &  [deg]&   [kpc] &  [kyr]        &   [erg/s] & [$({\rm TeV\, cm^2\, s})^{-1}$]    & [deg]    \\ 
\hline
B1951+32   &  68.77 & 2.82 &  3.00 & 107 &  $3.7\cdot 10^{+36}$  & $3.0\cdot 10^{-14}$ & 0.46 \\
J1740+1000   &  34.01 & 20.27 &  1.23 & 114 &  $2.3\cdot 10^{+35}$  & $1.1\cdot 10^{-14}$ & 1.15 \\
J1755-0903   & 18.32 & 8.15 &  0.23 & 3870 &  $4.4\cdot 10^{+33}$  & $5.0\cdot 10^{-15}$ & 5.48 \\
J0729-1448   &  230.39 & 1.42 &  2.68 & 35 &  $2.8\cdot 10^{+35}$  & $4.0\cdot 10^{-15}$ & 0.60 \\
J0631+1036   &  201.22 & 0.45 &  2.10 & 44 &  $1.7\cdot 10^{+35}$  & $3.8\cdot 10^{-15}$ & 0.76 \\
B1929+10   &  47.38 & -3.88 &  0.31 & 3100 &  $3.9\cdot 10^{+33}$  & $2.4\cdot 10^{-15}$ & 4.11 \\
J0538+2817   &  179.72 & -1.69 &  1.30 & 618 &  $4.9\cdot 10^{+34}$  & $1.8\cdot 10^{-15}$ & 1.02 \\
J2043+2740   &  70.61 & -9.15 &  1.48 & 1200 &  $5.6\cdot 10^{+34}$  & $1.6\cdot 10^{-15}$ & 0.88 \\
J1846+0919   &  40.69 & 5.34 &  1.53 & 360 &  $3.4\cdot 10^{+34}$  & $9.2\cdot 10^{-16}$ & 0.86 \\
J1900-09    & 25.46 & 4.73 &  0.30 & 1500 &  $1.2\cdot 10^{+33}$  & $7.9\cdot 10^{-16}$ & 4.26 \\
J2055+2539   &  70.69 & -12.52 &  0.62 & 1240 &  $4.9\cdot 10^{+33}$  & $7.8\cdot 10^{-16}$ & 1.97 \\
B1702-19   &  3.19 & 13.03 &  0.75 & 1140 &  $6.1\cdot 10^{+33}$  & $6.6\cdot 10^{-16}$ & 1.75 \\
J0611+1436   &  195.38 & -2.00 &  0.89 & 1070 &  $8.0\cdot 10^{+33}$  & $6.1\cdot 10^{-16}$ & 1.47 \\
J0357+3205   &  162.76 & -16.01 &  0.83 & 540 &  $5.9\cdot 10^{+33}$  & $5.3\cdot 10^{-16}$ & 1.59 \\
B1930+22   &  57.36 & 1.55 &  10.90 & 40 &  $7.5\cdot 10^{+35}$  & $5.3\cdot 10^{-16}$ & 0.12 \\
B0450-18   &  217.08 & -34.09 &  0.40 & 1510 &  $1.4\cdot 10^{+33}$  & $5.2\cdot 10^{-16}$ & 3.21 \\
B0950+08  &   228.91 & 43.70 &  0.26 & 17500 &  $5.6\cdot 10^{+32}$  & $4.9\cdot 10^{-16}$ & 4.88 \\
J2006+3102   &  68.67 & -0.53 &  6.03 & 104 &  $2.2\cdot 10^{+35}$  & $4.5\cdot 10^{-16}$ & 0.21 \\
B0919+06   &  225.42 & 36.39 &  1.10 & 497 &  $6.8\cdot 10^{+33}$  & $3.5\cdot 10^{-16}$ & 1.22 \\
B1706-16   &  5.78 & 13.66 &  0.56 & 1640 &  $8.9\cdot 10^{+32}$  & $1.7\cdot 10^{-16}$ & 2.17 \\
J1921+0812   &  43.71 & -2.93 &  2.90 & 622 & $2.3\cdot 10^{+34}$  & $1.7\cdot 10^{-16}$ & 0.45 \\
J1816-0755   &  21.87 & 4.09 &  3.13 & 532 &  $2.5\cdot 10^{+34}$  & $1.6\cdot 10^{-16}$ & 0.41 \\
B1821-19   &  12.28 & -3.11 &  3.70 & 573 &  $3.0\cdot 10^{+34}$  & $1.4\cdot 10^{-16}$ & 0.33 \\
J1848+0647   &  38.70 & 3.65 &  1.13 & 916 &  $2.7\cdot 10^{+33}$  & $1.3\cdot 10^{-16}$ & 1.18 \\
\hline
\end{tabular}
%\caption{List of the pulsars from the ATNF catalog in the HAWC field of view with the brightest predicted ICS halo flux at 10 TeV and not included in the 2HWC catalog. We list the pulsar name, Galactic coordinates, distance, age, spin-down luminosity, predicted extension calculated using Eq.~\ref{eq:flux} and for $D_0=7\cdot 10^{25}$ cm$^2$/s. }
\caption{Same as in Tab.~\ref{tab:TeVsourcesHAWC} but for source not detected by HAWC so far.}
\label{tab:TeVsourcesHAWCfuture}
\end{center}
\end{table*}

We can also use our predictions for the ICS flux in order to outline the most promising targets among Galactic pulsars for the detection of a possible ICS halos.
We pick the distance, age and spin down energy of pulsars from the ATNF catalog\footnote{\url{http://www.atnf.csiro.au/research/pulsar/psrcat/}} and calculate, using Eq.~\ref{eq:phflux}, the ICS flux  ($\Phi_{\gamma}$) at 1 TeV, which is relevant for HESS, and at 10 TeV, where HAWC and the future CTA experiment have their peak of sensitivity.
We rank the sources according to $\Phi_{\gamma}$ assuming that all the PWNe have the same efficiency $\eta=0.01$. 
We note that the efficiency acts as a mere normalization for the ICS flux, and does not influence the relative ranking of the sources.
In Tab.~\ref{tab:TeVsourcesHAWC} we report the list of the 23 highest pulsars in the HAWC field of view ranked according to the brightest predicted ICS halo flux at 10 TeV.
We select only sources with $DEC\in [-20^{\circ},40^{\circ}]$ since this is the constrain of the HAWC field of view.
We also report the predicted extension $\theta_{68}$ at 10 TeV calculated using Eq.~\ref{eq:flux} and for $D_0=7\cdot 10^{25}$ cm$^2$/s.
$\theta_{68}$ falls in the range between $0.40^{\circ}-0.80^{\circ}$ for most of the sources,
 while for Geminga and Monogem (2HWC J0700+143), which are very close sources, $\theta_{68}$ is about $7^{\circ}$ and $5^{\circ}$.
This implies that $D_0$ should be of the order of $\sim 10^{26}$ cm$^2$/s at 1 GeV if the $\gamma$-ray emission is due to ICS.
Only two out of these 23 have not already been detected by HAWC.
These two sources are associated to the pulsars PSR B1951+32 and PSR J1740+1000 and will very likely be reported in future HAWC catalogs. 
The 2HWC Survey reports for these sources a significance of 1.3$\sigma$ and 2.3$\sigma$, respectively.  
%The reason is that these sources are outside the field of view of these experiments.
The fact that most of the sources in Tab.~\ref{tab:TeVsourcesHAWC}  have been already detected in 2HWC, demonstrates that the ICS flux is a very efficient indicator to select promising Galactic $\gamma$-ray sources.

A list of sources detectable (or already detected) by HAWC have been also presented in \cite{Linden:2017vvb}. Indeed, some of the sources  reported in this paper  are also among the most promising ones in our list in Tab.~\ref{tab:TeVsourcesHAWC}. However, the complete list in Tab.~\ref{tab:TeVsourcesHAWC} contains differences with respect to \cite{Linden:2017vvb}. This is explained by the different estimation of the ICS flux. Indeed, 
Ref.~\cite{Linden:2017vvb} uses a simplified model, that is based on a mere rescaling of the Geminga ICS flux observed by HAWC, 
defined through  the distance and spin-down luminosity of the considered sources. 
The calculation is neglecting different ingredients which can vary the ICS flux, such as the source age. Moreover, the authors focus on sources with $T>100$ kyr. Instead, we perform for each source the complete calculation of their ICS flux, including also younger sources which can still exhibit an ICS halo.

HAWC is planning to operate at least until 2023 and to upgrade the detector and the data analysis (see, e.g., \cite{Joshi:2017eou}).
These improvements and the increase of statistics will improve the sensitivity by a factor of at least 2.
Since the results presented so far in the 2HWC catalog consider only 2 years of data, we can expect that it could be able to 
detect  many more ICS halos in the near future.

According to  the ICS flux at 10 TeV, we compile a list of pulsars promisingly detectable in the direction where HAWC could reasonably have the sensitivity to detect an ICS halo. 
We list these sources in Tab.~\ref{tab:TeVsourcesHAWCfuture}, including the two non detected sources in Tab.~\ref{tab:TeVsourcesHAWC}. 
The  $\theta_{68}$ and $\Phi_{\gamma}$ are computed as in Tab.~\ref{tab:TeVsourcesHAWC}. 
The fluxes are in the range between $10^{-16}-10^{-14}$ $({\rm TeV\, cm^2\, s})^{-1}$. 
As reported above, the HAWC sensitivity for the detection of an extended source is about $1\cdot 10^{-14}$ $({\rm TeV\, cm^2\, s})^{-1}$. 
With the future HAWC improvements, the first sources of Tab.~\ref{tab:TeVsourcesHAWCfuture} could be detected by HAWC. In case our efficiency, here fixed at  $\eta=0.01$ would be underestimated, several other sources could be potentially detectable with HAWC.
\\

%%%%%%%%%%%%%%%%%%%%%%%%%%%%%%%%%%%%%%%%%%%%%%%%%%%%%%%%%%%%
\section{Derivation of $D_0$ in ICS halos}
\label{sec:D0}

The main goal of our analysis is to estimate the diffusion coefficient $D_0$ around the pulsars 
%and their efficiency $\eta$ for the conversion of spin-down energy into CR $e^{\pm}$, 
under the hypothesis that the VHE $\gamma$-ray emission is due to ICS.

\subsection{Selection of source sample}
\label{sec:sample}
In this section we build a sample of sources in order to study the physical properties ($D_0$ and $r_b$) of ICS halo candidates. 
We focus on the detected emissions around pulsars at VHE since, as we have seen in the previous section, their angular extension is much smaller 
than at lower energies and  makes the detection feasible for IACTs. 
Moreover, at these energies the pulsar proper motion does not effect significantly the ICS morphology.
%We will perform a search for ICS halos in {\it Fermi}-LAT data in future works.

\begin{table*}
\begin{center}
\begin{tabular}{|c|c|c|c|c|c|c|c|c|c|c|c|}
\hline
PSR  & $l$  &  $b$  &   $d$ & $T$ & $\dot{E}$ &  Name  &  $\theta^{\rm{HESS}}_{\rm{gauss}}$  &  $\theta_{\rm{gauss}}$ &   $\theta_{\rm{68}}$ &Type \\ 
\hline
 &	[deg]	 &  [deg]&   [kpc] &  [kyr]        &   [erg/s]   &    &     [deg]  &  [deg]     &  [deg]     &  \\ 
\hline 
J1016-5857  &  284.08 & -1.88 &  3.16 & 21 &  2.6$\cdot10^{36}$  & HESS J1018-589 B  		      & $0.15\pm0.03$    &  $0.14\pm0.03$	 & 0.14 &   PWN \\ %CHECKED
J1028-5819  &  285.06 & -0.50 &  1.42 & 90 &  8.3$\cdot10^{35}$  & HESS J1026-582 			      & $0.13\pm0.04$    & $0.18\pm0.04$	& 0.27 &   PWN \\ %CHECKED
J1459-6053 &  317.89 & -1.79 &  1.84 & 65 &  9.1$\cdot10^{35}$  & HESS J1458-608  			     & $0.37\pm0.03$    & $0.37\pm0.10$	&   0.44 &PWN \\ %CHECKED
J1632-4757  &. 336.30 & 0.08 &  4.84 & 240  & 5.0$\cdot10^{34}$& HESS J1632-478			     & $0.18\pm0.02$  & $0.25\pm0.04$	&  0.14  & PWN   \\ %CHECKED
J1718-3825  &  348.95 & -0.43 &  3.49 & 90 &  1.3$\cdot10^{36}$  & HESS J1718-385  				& $0.12\pm0.01$& $0.13\pm0.02$	&   0.09  &PWN \\ %CHECKED
J1809-1917  &  11.18 & -0.35 &  3.27 & 51.7 &  1.8$\cdot10^{36}$  & HESS J1809-193(2HWC J1809-190)& $0.40\pm0.05$ & $0.35\pm0.03$   &  0.37 &UNID \\  %CHECKED
J1813-1246  &  17.24 & 2.44 &  2.63 & 43 &  6.2$\cdot10^{36}$  & HESS J1813-126(2HWC J1812-126)  & $0.21\pm0.03$  & $0.20\pm0.09$   & 0.33    &   UNID \\ %CHECKED
B1823-13   &  18.00 & -0.69 &  3.61 & 21 &  2.8$\cdot10^{36}$  & HESS J1825-137(2HWC J1825-134)& $0.46\pm0.03$  & $0.36\pm0.02$	& 0.28 & HALO \\ %CHECKED
J1831-952  &  21.90 & -0.13 &  3.68 & 128 &  1.1$\cdot10^{36}$  & HESS J1831-098(2HWC J1831-098)& $0.15$		&  $0.19\pm0.05$  & 0.21  	&   PWN \\ %CHECKED
J1838-0655  &  25.25 & -0.20 &  6.60 & 23 &  5.6$\cdot10^{36}$  & HESS J1837-069(2HWC J1837-065)  & $0.36\pm0.03$  & $0.31\pm0.02$& 0.14 	&   PWN \\ %CHECKED
J1841-0524  & 27.02   & -0.33 &  4.12 & 30.2& 1.0$\cdot10^{36}$  &   HESS J1841-055     			&  $0.40\pm0.03$  &  $0.50\pm0.08$& 0.21	&    UNID     \\ %CHECKED
J1856+0245  &  36.01 & 0.06 &  6.32 & 21 &  4.6$\cdot10^{36}$  & HESS J1857+026(2HWC J1857+027) & $0.26\pm0.06$  & $0.23\pm0.04$	&   0.14 &UNID \\ %CHECKED
J1857+0143  &  35.17 & -0.57 &  4.57 & 71 &  4.5$\cdot10^{35}$  & HESS J1858+020			  	  & $0.08\pm0.02$  & $0.12\pm0.04$ &0.21  	&   UNID \\ %CHECKED
J1907+0602  &  40.18 & -0.89 &  2.37 & 20 &  2.8$\cdot10^{36}$  & HESS J1908+063(2HWC J1908+063) & $0.49\pm0.03$  & $0.50\pm0.10$	&  0.26 & UNID \\  %CHECKED
J1913+1011  &  44.48 & -0.17 &  4.61 & 169 &  2.9$\cdot10^{36}$  & HESS J1912+101(2HWC J1912+099)& $0.49\pm0.04$  & $0.42\pm0.06$   &0.16    &   SHELL \\ %CHECKED
\hline
\hline
B0833-45  &  263.55 & -2.79 &  0.28 & 11.3 &  6.9$\cdot10^{36}$  &  HESS J0835-455(Vela X)		& $0.58\pm0.05$  & $0.48\pm0.06$ & 0.66	  &  PWN \\  %CHECKED
J1301-6305  &  304.10 & -0.24 &  10.72 & 11 &  1.7$\cdot10^{36}$  &  HESS J1303-631				& $0.18\pm0.02$  & $0.20\pm0.02$ & 0.10   &  PWN \\  %CHECKED
J1357-6429  &  309.92 & -2.51 &  3.10 & 7.3 &  3.1$\cdot10^{36}$  &  HESS J1356-645				& $0.23\pm0.02$  & $0.22\pm0.04$  & 0.28 	  &  PWN \\  %CHECKED
J1420-6048 &  313.54 & 0.23 &  5.63 & 13 &  1.0$\cdot10^{37}$  & HESS J1420-607				  & $0.08\pm0.01$   & $0.13\pm0.02$	& 0.14     &  PWN \\ %CHECKED
J1617-5055  &  332.50 & -0.28 &  4.74 & 8.1 &  1.6$\cdot10^{36}$  &  HESS J1616-508				& $0.23\pm0.03$  & $0.22\pm0.02$ & 0.13  &  PWN \\  %CHECKED
J1640-4631  &  338.32 & -0.02 &  12.75 & 3.4 &  4.4$\cdot10^{36}$  &  HESS J1640-465				& $0.18\pm0.02$  & $0.13\pm0.01$	& 0.10   &  PWN \\  %CHECKED
B1706-44  &  343.10 & -2.69 &  2.60 & 18 &  3.4$\cdot10^{36}$  & HESS J1708-443  			        & $0.28\pm0.03$   & $0.24\pm0.07$ & 0.23 	  &  PWN \\ %CHECKED
J1813-1749  &  12.82 & -0.02 &  4.70 & 5.6 &  5.6$\cdot10^{37}$  &  HESS J1813-178(2HWC J1814-173)& $0.049\pm0.004$  & $0.019\pm0.003$   & 0.10  &  PWN \\  %CHECKED
J1826-1256  &  18.56 & -0.38 &  1.55 & 14 &  3.6$\cdot10^{36}$  & HESS J1826-130(2HWC J1825-134)& $0.15\pm0.02$  & $0.23\pm0.06$	 & 0.28   &  UNID \\  %CHECKED
J1833-1034  &  21.50 & -0.89 &  4.10 & 4.9 &  3.4$\cdot10^{37}$  & HESS J1833-105				  & $<0.05$ & $0.08\pm0.02$	& 0.30   &  PWN \\ %CHECKED
\hline
\hline
J0633+1746  &  195.13 & 4.27 &  0.19 & 342 &  3.3$\cdot10^{34}$  & GEMINGA(2HWC J0635+180)  	& $5.5\pm0.7$	  & $1.4\pm0.2$    & 6.54 & HALO \\  
B0656+14  &  201.11 & 8.26 &  0.29 & 111 &  3.8$\cdot10^{34}$  & MONOGEM(2HWC J0700+143)    	& $4.8\pm0.6$    & $2.7\pm0.4$   &  4.71 & HALO \\  
\hline
\end{tabular}
\caption{List of the pulsars considered in our analysis. See the text for more information on the criteria we use to select them. We list the pulsar name, Galactic coordinates, pulsar distance, age and spin-down luminosity, association name, extension as given in HGSP or \cite{Abeysekara:2017science} ($\theta^{\rm{HESS}}_{\rm{gauss}}$). We also show the angular size ($\theta_{\rm{gauss}}$) found by fitting with a gaussian function the source surface brightness derived with the HESS flux maps with $R_c=0.1^{\circ}$ (see the text for further details). Finally, we report predicted size of the ICS halo at 1 TeV using $D_0=7\cdot 10^{25}$ cm$^2$/s ($\theta_{68}$), and classification as in TeVcat. The first (second) block corresponds to $old$ ($young$) sources.}
\label{tab:TeVsourcesmost}
\end{center}
\end{table*}

We compute the ICS $\gamma$-ray flux for all the ATNF pulsars, and 
select the ones  with the highest predicted ICS $\gamma$-ray flux, and having an extended counterpart already detected  by HESS. 
Indeed,  we will use the flux maps, which have been released in the HGPS catalog\footnote{\url{https://www.mpi-hd.mpg.de/hfm/HESS/hgps/}}.
We also add Geminga and Monogem for which the HAWC Collaboration has released the surface brightness \cite{Abeysekara:2017science}.
We report in Tab.~\ref{tab:TeVsourcesmost} the list of pulsars corresponding to these criteria with their age, distance and position in the sky.
We also indicate the spatial extension as measured by HESS using a gaussian function ($\theta^{\rm{HESS}}_{\rm{gauss}}$).

We divide our sample in {\it old} and {\it young} pulsars fixing an age limit of 20 kyr.
Indeed, as we described in Sec.~\ref{sec:model}, $e^{\pm}$ are believed to be accelerated in PWNe to very high energies at the termination shock.
This happens in the Sedov phase, i.e.~in a time between a few up to twenty thousands of years after the supernova explosion \citep{1996ApJ...459L..83C, 2011ASSP...21..624B}.
After this stage, accelerated $e^{\pm}$  produce photons from radio, through synchrotron emission, up to VHE $\gamma$ rays by ICS. 
The size of extension thus depends on the PWN evolution.
We consider separately the {\it old} and {\it young} PWN samples to inspect  any dependence on the PWN evolution. 

The list of sources in Tab.~\ref{tab:TeVsourcesmost} exhibits an observed extended emission with $\theta^{\rm{HESS}}_{\rm{gauss}}$ 
$\sim [0.1^{\circ},0.5^{\circ}]$, which translates into a physical size of $\sim [8-35]$ pc.
This size has been calculated by HESS using a spatial gaussian function, with the size of extension as the standard deviation parameter.
We report also the predicted size of ICS emission calculated, for each source, with $\theta_{\rm{68}}$, i.e.~as the 68\% containment radius (see Eq.~\ref{eq:flux}), at 1 TeV and for $D_0 = 7 \cdot 10^{25}$ cm$^2$/s. We apply the following procedure to calculate $\theta_{\rm{68}}$.
We calculate the surface brightness for different angular distances from the source. Then we calculate, interpolating between the angle values considered, the distance that contains the 68\% of the flux following the definition in Eq.~\ref{eq:flux}. 
Overall, we find a good match between the measured and predicted size of extension, implying that the morphology of the $\gamma$-ray emission from these sources should be consistent with a diffusion environment with $D_0\sim 10^{26}$ cm$^2$/s.

Most of the sources in our sample are located in the inner 4 kpc from the Earth and are younger than 100 kyr.
7 of them are classified in the TeVCat as unidentified, since no PWN structure has been identified in radio or X rays. 
However, a very powerful pulsar is found close to them, making the presence of a PWN a viable possibility.

We add here few comments about the association of few sources in  Tab.~\ref{tab:TeVsourcesmost}. 
HESS J1858+020 is positionally compatible with the ATNF catalog pulsars PSR J1857+0143, J1857+0210 and B1855+02.
However,  assuming the same efficiency for all three, PSR J1857+0143 would have an ICS flux higher than a factor of  50 (100) with respect to J1857+0210 (B1855+02). 
For our purposes, we thus assume that HESS J1858+020 is associated to PSR J1857+0143.
HESS J1303-631 position is compatible with  PSR J1301-6305 and PSR J1301-6310. 
Computing ICS flux with the same efficiency for both, PSR J1301-6305 overclasses  PSR J1301-6310 by a factor of about 50. 
Moreover, PSR J1301-6310 has a small distance from us and is relatively old, so the ICS flux is expected much more extended than $\theta^{\rm{HESS}}_{\rm{gauss}}=0.18^{\circ}$.
Therefore, we associate HESS J1303-631 to PSR J1301-6305.
Finally, HESS J1831-098 is found to have $TS = 59$ in the main HGPS analysis, but only $TS = 17$ in the cross-check analysis made using an alternative calibration, reconstruction, and gamma-hadron separation method, and is therefore considered as a source candidate \cite{H.E.S.S.:2018zkf}.
\subsection{Analysis technique}
\label{sec:method}
%%%%%%%%%%%%%%%%%%%%%%%%%%%%%%%%%%%%%%%%%%%%%%%%%%%%%%%%%%%%
\begin{figure*}
\includegraphics[width=0.49\textwidth]{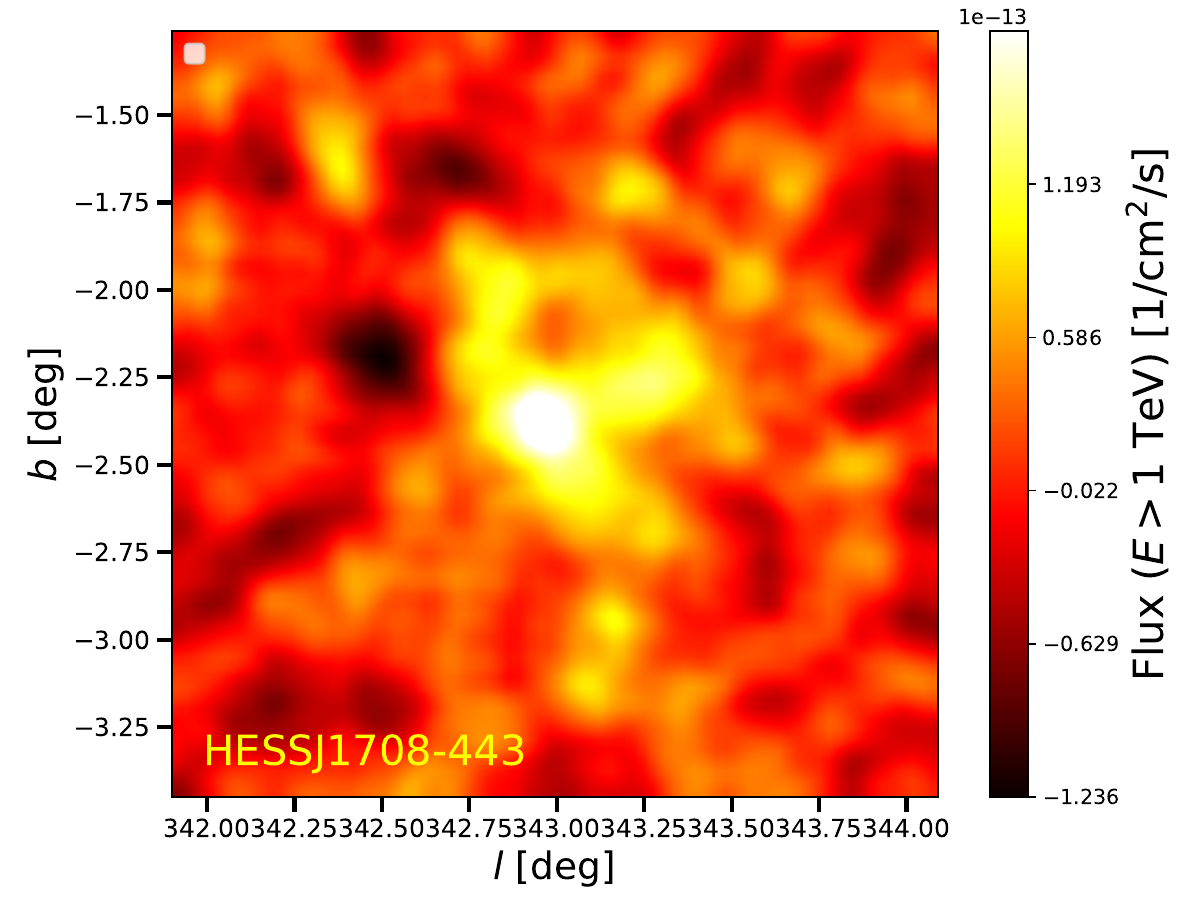}
\includegraphics[width=0.49\textwidth]{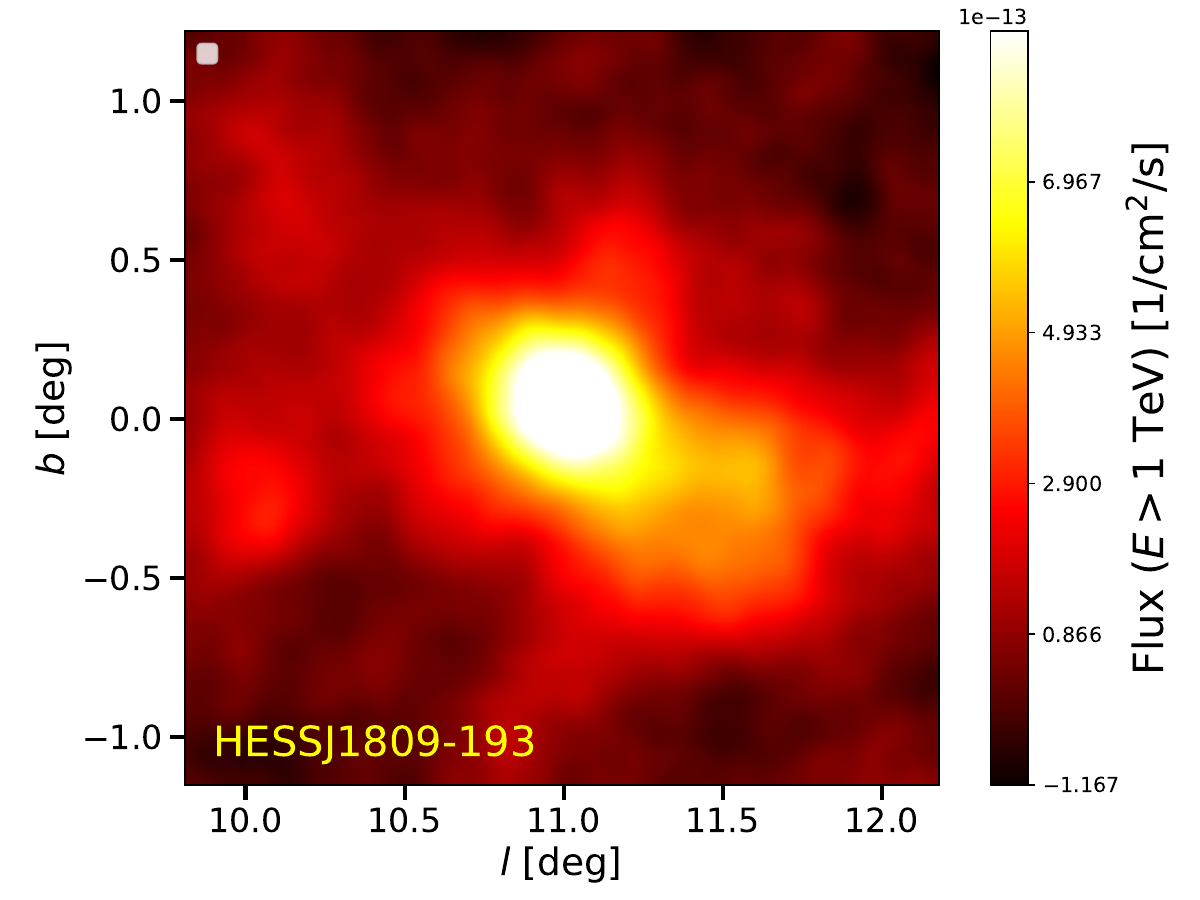}
\includegraphics[width=0.49\textwidth]{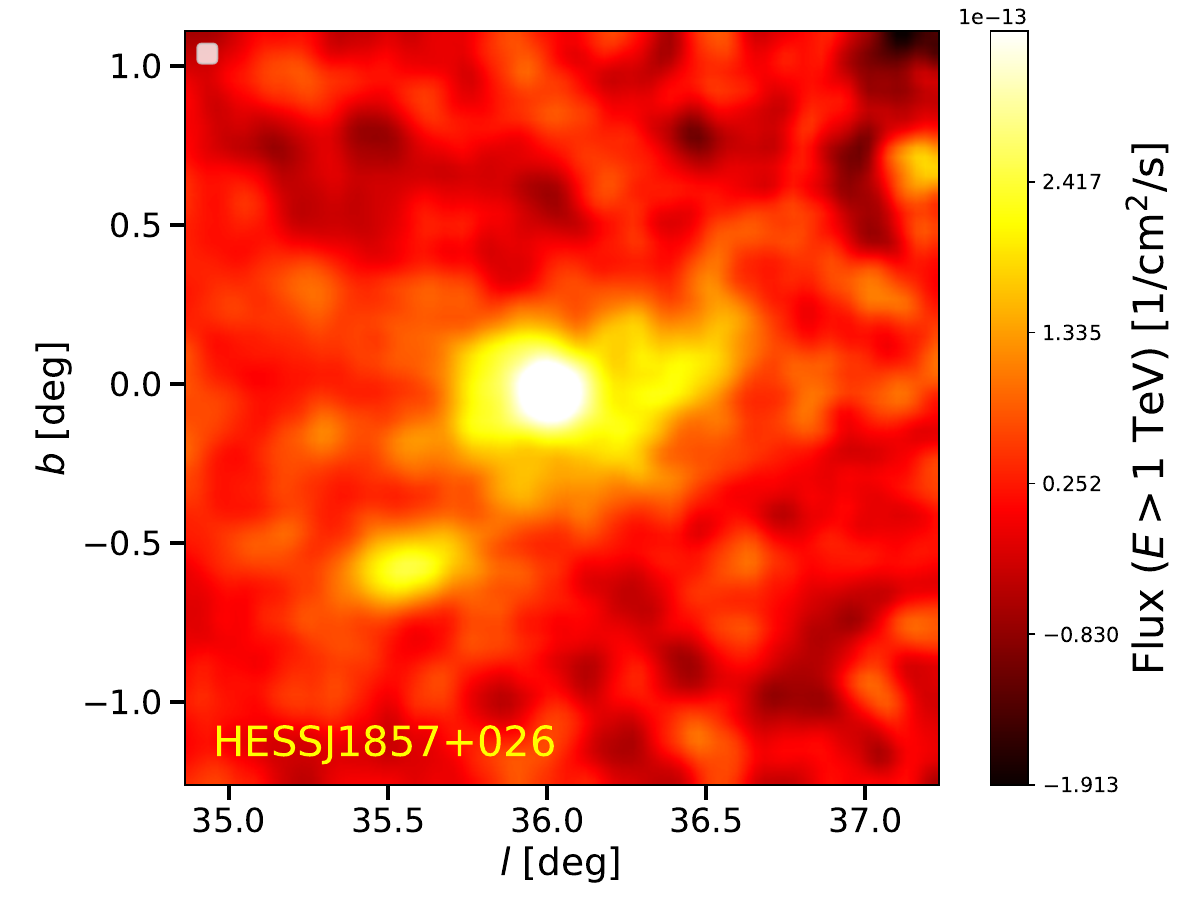}
\includegraphics[width=0.49\textwidth]{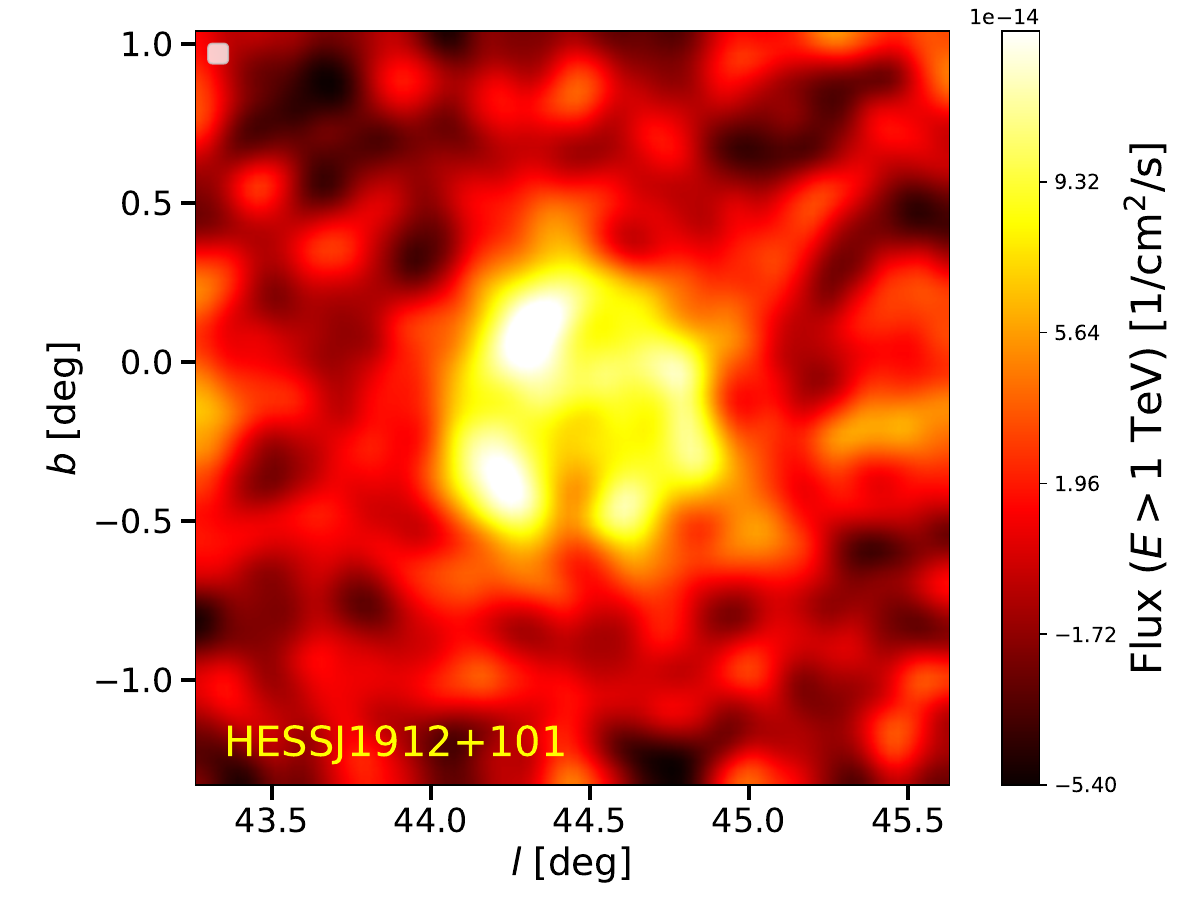}
\caption{Map of the flux integrated above 1 TeV taken from the publicly available data for the HGPS catalog. We have used the maps derived with a correlation radius of $0.1^{\circ}$.}  
\label{fig:fluxmap}
\end{figure*}

%\begin{figure*}
%\includegraphics[width=0.65\textwidth]{plots/ext_HESS_ours_Gauss_Rc0p1.pdf}
%%\includegraphics[width=0.65\textwidth]{plots/size_age_paper_hess.pdf}
%\caption{Ratio between the size of extension derived by our a fit to the surface brightness data using a gaussian template ($\theta^{\rm{HESS}}_{\rm{gauss}}$) and the same quantity published in the HGPS catalog ($\theta_{\rm{gauss}}$).}  
%\label{fig:extcomp}
%\end{figure*}

\begin{figure*}
\includegraphics[width=0.49\textwidth]{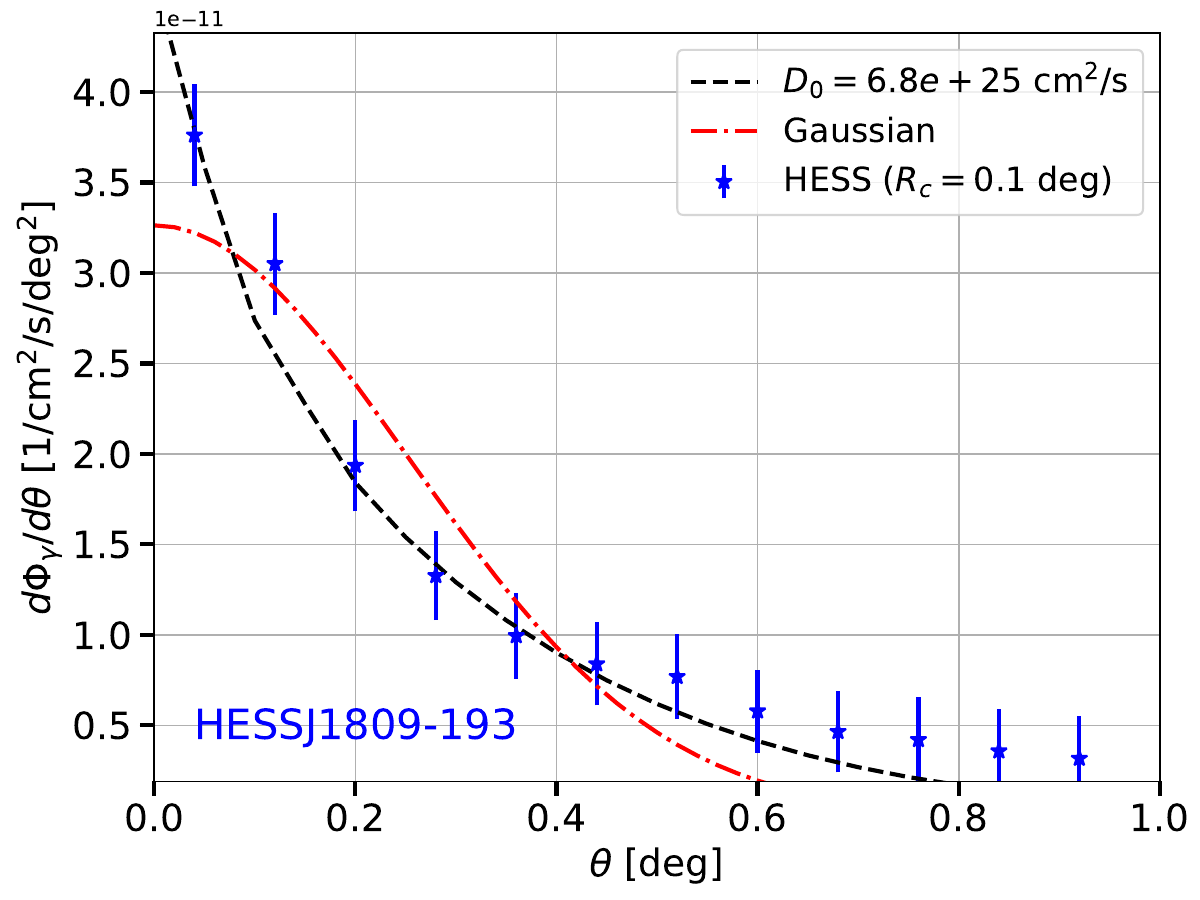}
\includegraphics[width=0.49\textwidth]{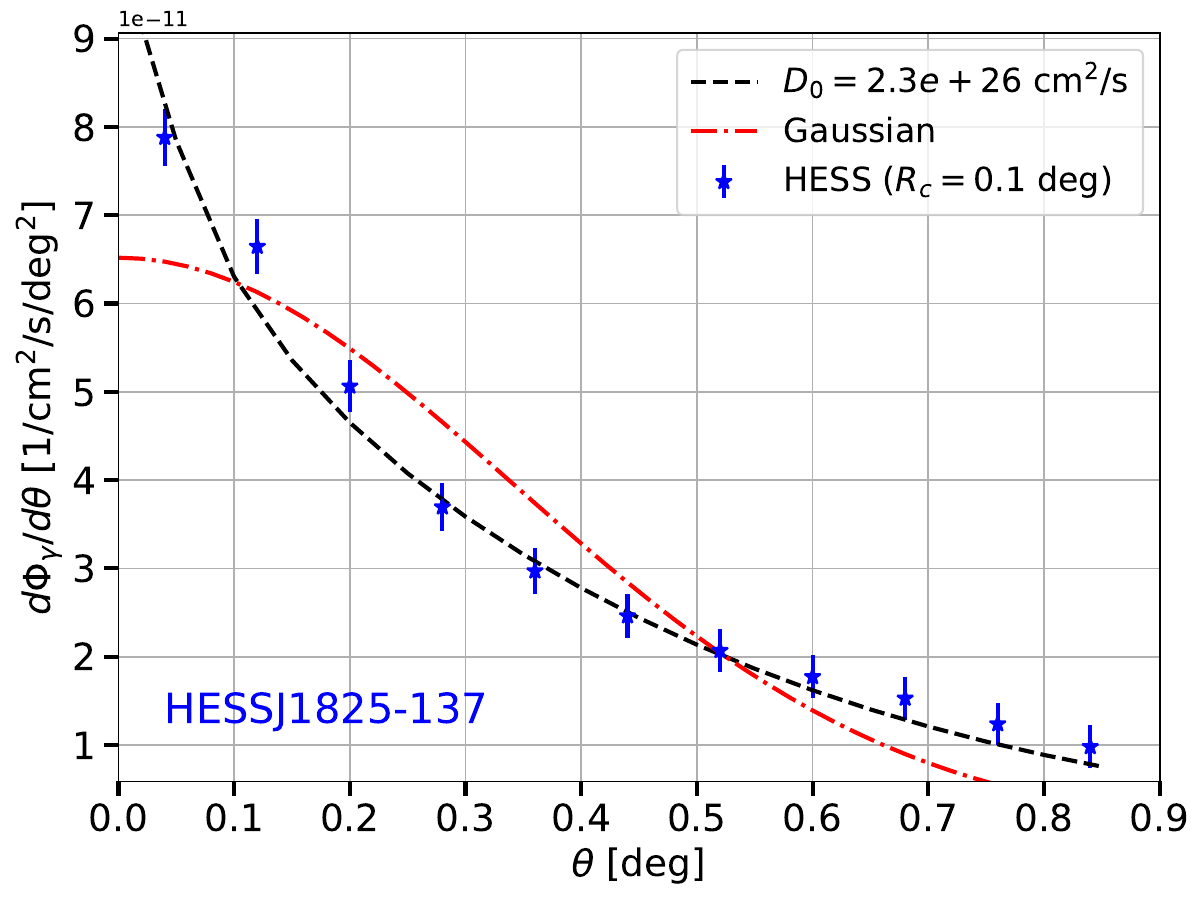}
\includegraphics[width=0.49\textwidth]{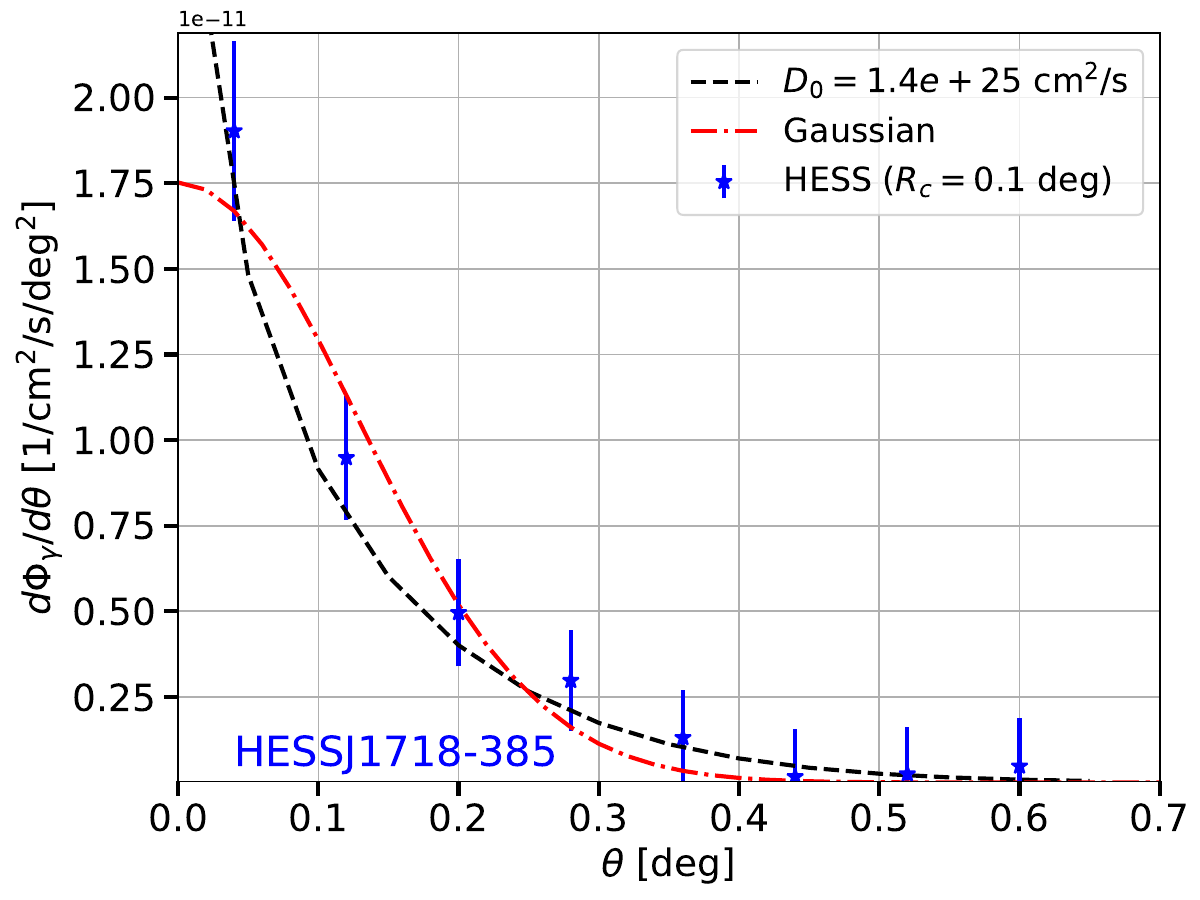}
\includegraphics[width=0.49\textwidth]{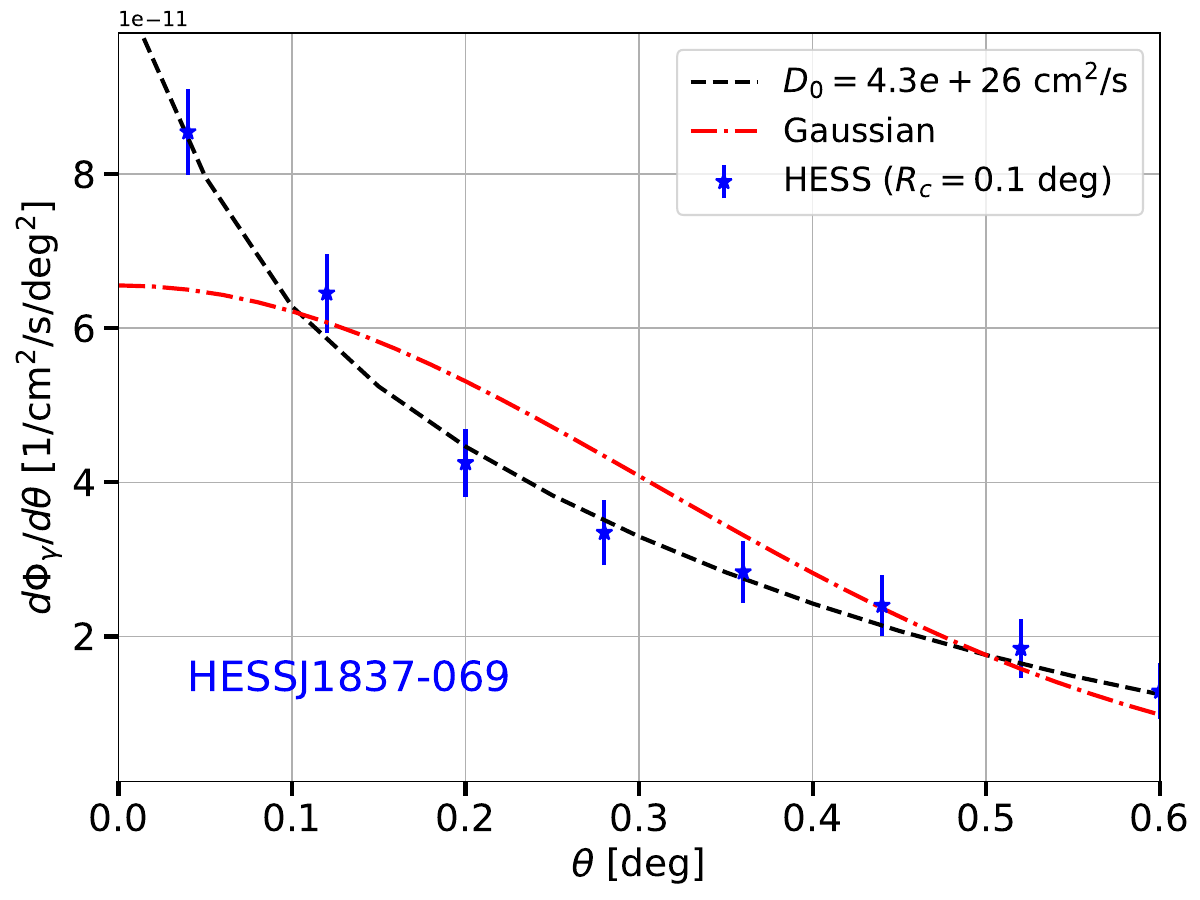}
\caption{Surface brightness above 1 TeV calculated from the flux maps publicly available for the HGPS catalog. We show in each plot the HESS data together with the best fit of our model (black line) and of simple gaussian function (red line).}  
\label{fig:SB}
\end{figure*}

$D_0$ is derived through a fit to the surface brightness, source by source. 
We use the HESS $\gamma$-ray flux maps to derive the observed spatial distribution of the $\gamma$-ray emission around HESS sources.
In order to extract the surface brightness as a function of the angular distance, we take the flux, the sensitivity and the significance maps
 from the HGPS catalog\footnote{\url{https://www.mpi-hd.mpg.de/hfm/HESS/hgps/}}.
These maps contain the flux integrated above 1 TeV, around a circular region defined by the correlation radius $R_c$. 
They are provided for $R_c = 0.1^{\circ}$ and $0.2^{\circ}$, and with a pixel size of $0.02^{\circ}$.  
Therefore, each pixel contains information partially present also in the closest pixels.
In order to limit this oversampling, we use the case with $R_c=0.1^{\circ}$ as our benchmark case, but we will comment on the results found with $R_c=0.2^{\circ}$.
We show in Fig.~\ref{fig:fluxmap} the flux maps for four sources in our sample.

We select a region of interest (ROI) around each source between $0.7^{\circ}$ and $1.1^{\circ}$ of radius depending on the extension of the source, as given in HGPS.  
%We choose the size of the ROI in order to contain most of the $\gamma$-ray emission from the source and limit the presence of the $\gamma$-ray background.
We choose the size of the ROI in order to limit the contribution of background sources and include mainly the emission of the central sources.
For example, HESS J1708-443 has a measured size of about $0.3^{\circ}$,  so we choose a ROI of $0.8^{\circ}$ which contains entirely the $\gamma$-ray flux from that source.
For sources extended $0.1^{\circ}$ ($0.4^{\circ}$) we typically select ROIs with $0.6^{\circ}$ ($1.1^{\circ}$) radius. 

We do not include in our analysis a $\gamma$-ray background component, which could be modeled with the interstellar emission and flux from faint sources.
%Indeed, we use the flux maps to find the spatial morphology of the $\gamma$-ray emission around PWNe and the presence of a uniform background is not going to affect our results.
Indeed, assuming that the background is isotropic, it should act as a mere normalization without changing significantly the angular profile of the TeV  surface brightness.
We also check whether there are other sources from the HGPS catalog in the ROI. If any other source is present in the same ROI, we remove the quadrant where this is located.
For example, the source HESS J1616-508 is located at longitude and latitude $l_S=332.48^{\circ}$ and $b_S=-0.17^{\circ}$ and is close to HESS J1614-518 ($l=331.47^{\circ}$ and $b=-0.60^{\circ}$). 
Therefore, we remove from the analysis the region given by $l<l_S$ and $b<b_S$ in order to avoid any contamination from HESS J1614-518. 
We apply the same method to the following sources: HESS J1026-582, HESS J1303-631, HESS J1420-607, HESS J1458-608, HESS J1616-508, HESS J1632-478, HESS J1718-385, HESS J1825-137, HESS J1826-130, HESS J1831-098, HESS J1833-105, HESS J1841-055, HESS J1857+026, HESS J1858+020.

We assume for the ICS flux the one-zone diffusion model since the surface brightness data is not enough precise to constrain also the $r_b$ parameter. In order to use a two-zone diffusion model one would need to fit directly the HESS data around each source, but this is not available at the moment. Moreover, a two-zone diffusion model and the specific value of $r_b$ should affect mostly the $\gamma$-ray flux in the outer part of the halo, while we fit mainly the core of the $\gamma$-ray emission for each source.

This is the procedure we use to calculate the surface brightness of each source using the HESS  flux maps.
The maps are given as the $\gamma$-ray flux integrated over the  solid angle (and  have units of GeV/cm$^2$/s/deg$^2$). 
We calculate txhe total flux in concentric annuli and we divide it by their solid angle.
We use as a benchmark case an annuli size bin of $0.08^{\circ}$. We also test larger and smaller values, finding very similar results.
%We show in Fig.~\ref{fig:SB} the surface brightness data integrated above 1 TeV for four sources in our sample.

Before using this technique to extract $D_0$ for each source, we have to verify if the flux maps extracted from the HGPS catalog represent well the flux distribution around the sources in our sample.
In order to achieve this goal, we perform a fit to the surface brightness data assuming, as in the in the HGPS catalog, a gaussian function ($\propto \exp{(-\theta^2/(2 \cdot \theta_{\rm{gauss}}^2))}$).
Then, we compare the best fit values for the size of extension ($\theta_{\rm{gauss}}$) with the ones reported in the HGPS catalog ($\theta^{\rm{HESS}}_{\rm{gauss}}$).
The best fit values and $1\sigma$ errors for $\theta^{\rm{HESS}}_{\rm{gauss}}$ and $\theta_{\rm{gauss}}$ are reported in Tab.~\ref{tab:TeVsourcesmost} for $R_c=0.1^{\circ}$ .
The source extensions we derive from the flux maps are compatible with the values reported in the HESS catalog.
We find similar results using the flux maps provided for $R_c=0.2^{\circ}$.
We are thus confident that, regardless the oversampling, the flux maps released by HESS can be used as a viable proxy to study the source spatial extension of the $\gamma$-ray flux.

%%%%%%%%%%%%%%%%%%%%%%%%%%%%%%%%%%%%%%%%%%%%%%%%%%%%%%%%%%%%
%%%%%%%%%%%%%%%%%%%%%%%%%%%%%%%%%%%%%%%%%%%%%%%%%%%%%%%%%%%%

\section{Results for the diffusion around PWNe}
\label{sec:resultsD0eta}
\begin{figure*}
\includegraphics[width=0.60\textwidth]{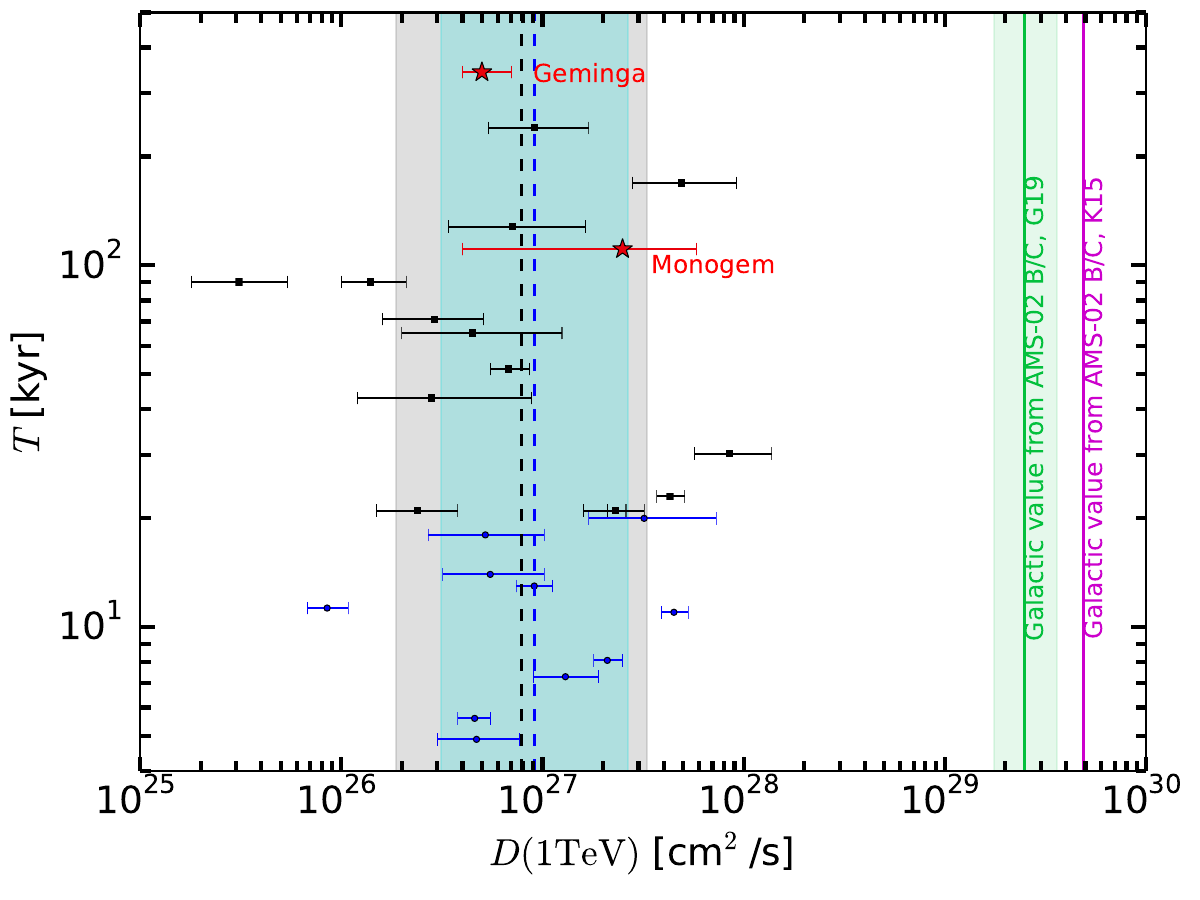}
\caption{Diffusion coefficient at 1 TeV derived for the PWNe in our sample. Blue (black) points are the results for PWNe powered by pulsars young (older) than 20 kyr. 
We also show the mean and one standard deviation for $D(1 \, \rm{TeV})$ and the results for this variable derived through fits to AMS-02 CR data in K15 and G19 \cite{Kappl:2015bqa,Genolini:2019ewc}.
The results for Monogem and Geminga PWNe derived fitting the HAWC surface brightness \cite{DiMauro:2019yvh} are outlined with red points.}  
\label{fig:D0res}
\end{figure*}

We now perform a fit to  the surface brightness data to find the diffusion coefficient around each PWN in our sample in Tab.~\ref{tab:TeVsourcesmost}.
This is performed by using the ICS flux calculation (see  Sec.~\ref{sec:model}), by leaving  $D_0$ and  $\eta$ (see Eq.~\ref{eq:Q_E_cont}) as free parameters of the fit. The efficiency $\eta$ acts as a normalization, while the diffusion coefficient at 1 GeV $D_0$ modifies the angular profile of the ICS flux.
We show in Tab.~\ref{tab:D0etta} and in Fig.~\ref{fig:D0res} our results and  the best fit and $1\sigma$ error for $D_0$. In Fig.~\ref{fig:D0res}, 
the diffusion coefficient (see Eq.~\ref{eq:Diff}) has been rescaled to 1 TeV, which is the typical energy scale of this analysis since we are considering VHE $\gamma$ rays. 
We also show in Tab.~\ref{tab:D0etta} the size of the ICS halo found implementing the empirical function \cite{Abeysekara:2017hyn}:
\begin{equation}
\frac{d\Phi_{\gamma}}{d\theta} \sim \frac{1}{\theta_{\rm{ICS}} (\theta+0.06 \cdot \theta_{\rm{ICS}})} e^{  -  \left( \frac{\theta}{\theta_{\rm{ICS}}} \right)^2},
 \label{eq:SBfunc}
\end{equation}
where $\theta_{\rm{ICS}}/2$ is the angle that contains the $80\%$ of the observed flux.
We find that this functional form indeed better describes for many sources in our sample the surface brightness data with respect to the gaussian function.
In Fig.~\ref{fig:SB} we report the surface brightness data  together with the best fit to the ICS model, found with  $D_0$  as the free
parameters. The best fit reproduces the observed surface brightness profile. Indeed, this model predicts the proper  
 angular decrease of the surface brightness through the description of leptons diffusion around the source. 
We also show the fit with a mere gaussian template which, for these and several other sources, is a worst fit than the physical ICS model. 
This does not apply to all the sources in our sample 
but for most of them the ICS model is at least as good as the gaussian template.

The best fit values for $D_0$, source by source,  are distributed in the $10^{25}-10^{27}$ cm$^2$/s range.
In particular, the mean value and the standard deviation over the entire sample are $D_0 = 9.1^{+17.4}_{-6.0}\cdot 10^{25}$ cm$^2$/s.
We find very similar values if we use a size for the annuli of $0.1^{\circ}$: $D_0 = 8.2^{+20.9}_{-5.9}\cdot 10^{25}$ cm$^2$/s.

%%%%%%%%%%%%%%%%%%%%%%%%%%%%%%%%%%%%%%%%%%%%%%%%%%%%%%%%%%%%
%%%%%%%%%%%%%%%%%%%%%%%%%%%%%%%%%%%%%%%%%%%%%%%%%%%%%%%%%%%%

The results we find for the {\it young} PWN sample could be affected by the presence of the SNR, and could thus be systematically different from the ones found for the {\it old} sample, for which on the other hand the SNR has lost its power (see discussion in Sec.~\ref{sec:model}).
Therefore, we compute $D_0$ for the {\it young} and {\it old} PWN sample separately, finding 
$D_0 = 8.9^{+17.1}_{-5.9}\cdot 10^{25}$ cm$^2$/s and $D_0 = 7.8^{+23.2}_{-5.8}\cdot 10^{25}$ cm$^2$/s, respectively. There is no clear difference for the two PWN samples. Therefore, we do not see any evolution of $D_0$ with respect to the age as predicted by \cite{Evoli:2018aza}.
This is visible in Fig.~\ref{fig:D0res}, where we show the value of the diffusion coefficient at 1 TeV ($D(1\, \rm{TeV})$) as a function of the PWN age.
We remind that we show $D(1\, \rm{TeV})$ because we use $\gamma$-ray data above hundreds GeV that are produced for ICS by $e^{\pm}$ at TeV energies.
We find for our sample $D(1\, \rm{TeV}) = 8.2^{+20.9}_{-5.9}\cdot 10^{26}$ cm$^2$/s. There is a variation in the values of $D(1\, \rm{TeV})$ of about 1 order of magnitude.
Our results for $D(1\, \rm{TeV})$ are compatible with the ones found for Geminga and Monogem with HAWC \cite{Abeysekara:2017science} and {\it Fermi}-LAT data \cite{DiMauro:2019yvh}. 
We also show in Fig.~\ref{fig:D0res} the results for the diffusion coefficient (scaled to 1 TeV considering the uncertainties on the normalization and the slope $\delta$) derived in \cite{Kappl:2015bqa,Genolini:2019ewc} from a fit to AMS-02 CR data. These numbers are representative of the average diffusion coefficient 
in the Galaxy. 
The intensity of $D(1\, \rm{TeV})$  we find with our analysis is about two orders of magnitude smaller than the one derived for the ISM. 
We also run our analysis on the HESS flux maps derived assuming $R_c=0.2^{\circ}$.
We find a diffusion coefficient at 1 GeV for the entire sample of $D_0 = 13.6^{+33.1}_{-9.6}\cdot 10^{25}$ cm$^2$/s while for the {\it young} and {\it old} PWN sample separately is $D_0 = 14.5^{+25.3}_{-9.2}\cdot 10^{25}$ cm$^2$/s and $D_0 = 13.0^{+37.8}_{-9.7}\cdot 10^{25}$ cm$^2$/s, respectively.
These values are consistent within $1\sigma$ with the ones reported above for $R_c=0.1^{\circ}$.

\begin{figure*}
\includegraphics[width=0.60\textwidth]{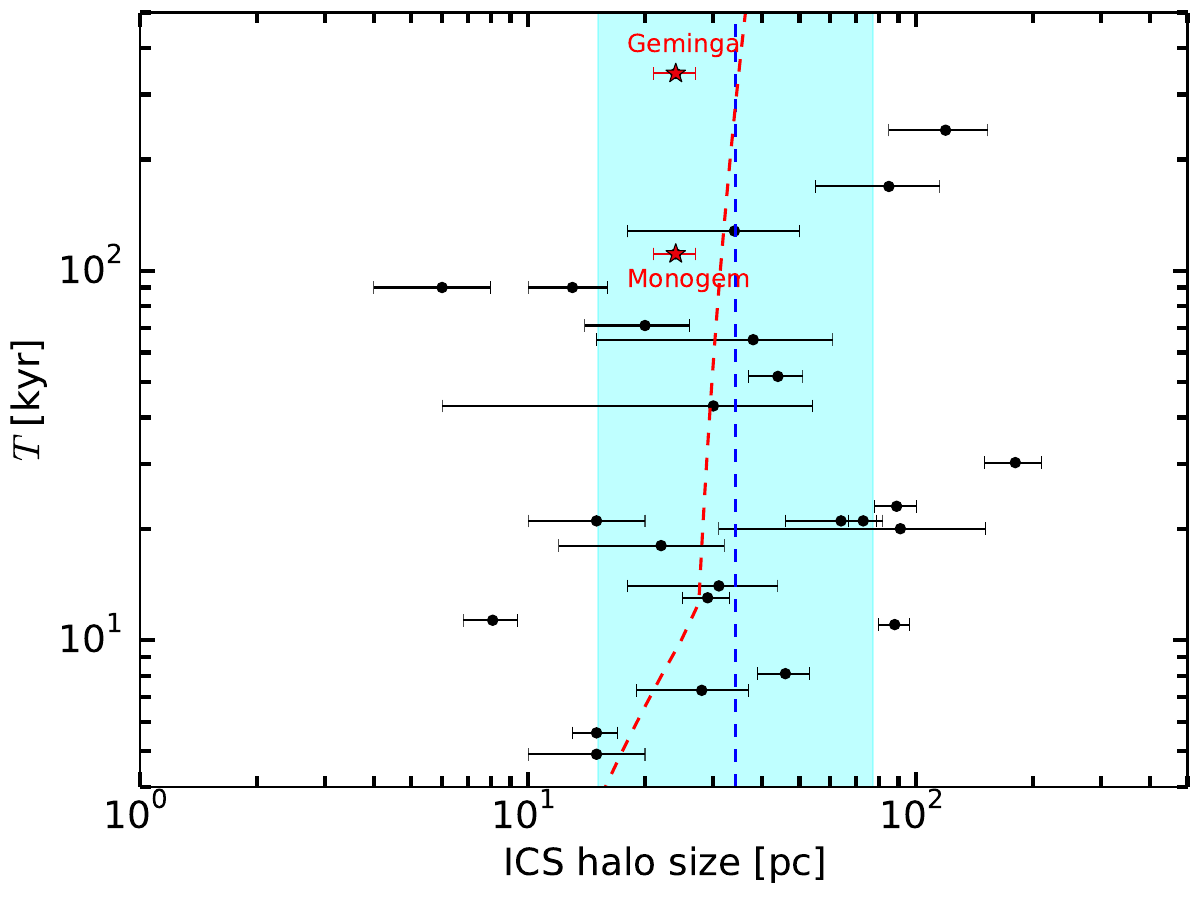}
\caption{Size of the ICS emission from the PWNe in our sample calculated using $\theta_{\rm{ICS}}/2$ (see the text for further details). 
We also show the average value (dashed blue line) and  one standard deviation variation found for the entire sample. 
The red dashed line shows the model for the PWN evolution described in the text in Sec.~\ref{sec:model}.}  
\label{fig:resext}
\end{figure*}

%%%%%%%%%%%%%%%%
An important characteristic of the ICS emission around PWNe is that their extension is connected to the size of the low-diffusion zone located around these sources (see Sec.~\ref{sec:sizeflux}). 
In particular the size of the low-diffusion bubble must be at least large as the ICS region.
We estimate the ICS halo size by considering the parameter $\theta_{\rm{ICS}}/2$ in Eq.~\ref{eq:SBfunc}. 
Then, we convert the angular scale into a physical size using $d \cdot \tan{(\theta_{\rm{ICS}}/2)}$.
The average size of the ICS halo is $34^{+43}_{-19}$~pc for the entire sample, and $29^{+30}_{-15}$ pc and $38^{+52}_{-22}$~pc for the {\it young} and {\it old} sub-samples, respectively. 
%There is a slight larger size for the {\it old} sample, that though considering the uncertainty is not significant.
We show in Fig.~\ref{fig:resext} the ICS halo size as a function of the age of the pulsar, together with the evolution model reported in Sec.~\ref{sec:model}.
In particular we use: $R \propto t^{1.2}$ for $t<3$ kyr, $R \propto t^{0.73}$ for $12<t<3$ kyr and $R \propto t^{0.3}$ for $t>12$ kyr.
This model is compatible with the observed sizes and ages, and our results are comparable with the ones released for PWN by HESS \cite{Abdalla:2017vci}.
However, there is a large scatter in the data that prevents us to refine the model for the expansion rate as a function of time.
The scatter we find is probably due to the fact that every pulsar has a different environment and a different evolution that makes the size of ICS flux significantly different for PWN with a similar age.
Since the size of the ICS halos is of the order of 35 pc for the PWNe of our sample, this implies that the low-diffusion bubble should be at least large as this size.
In particular for this average ICS halos size, $r_b$ should be at least of the order of 80 pc (see discussion in Sec.~\ref{sec:sizeflux} and Fig.~\ref{fig:extrb}).
However, some of the sources, e.g.~HESS J1632-478, HESS J1825-137, HESS J1837-069, HESS J1841-055, HESS J1912+101 and HESS J1303-631, have a much more extended ICS halos size implying that the size of the low-diffusion bubble could even exceed 100 pc.

\begin{figure}
\vspace{0.5cm}
\includegraphics[width=0.50\textwidth]{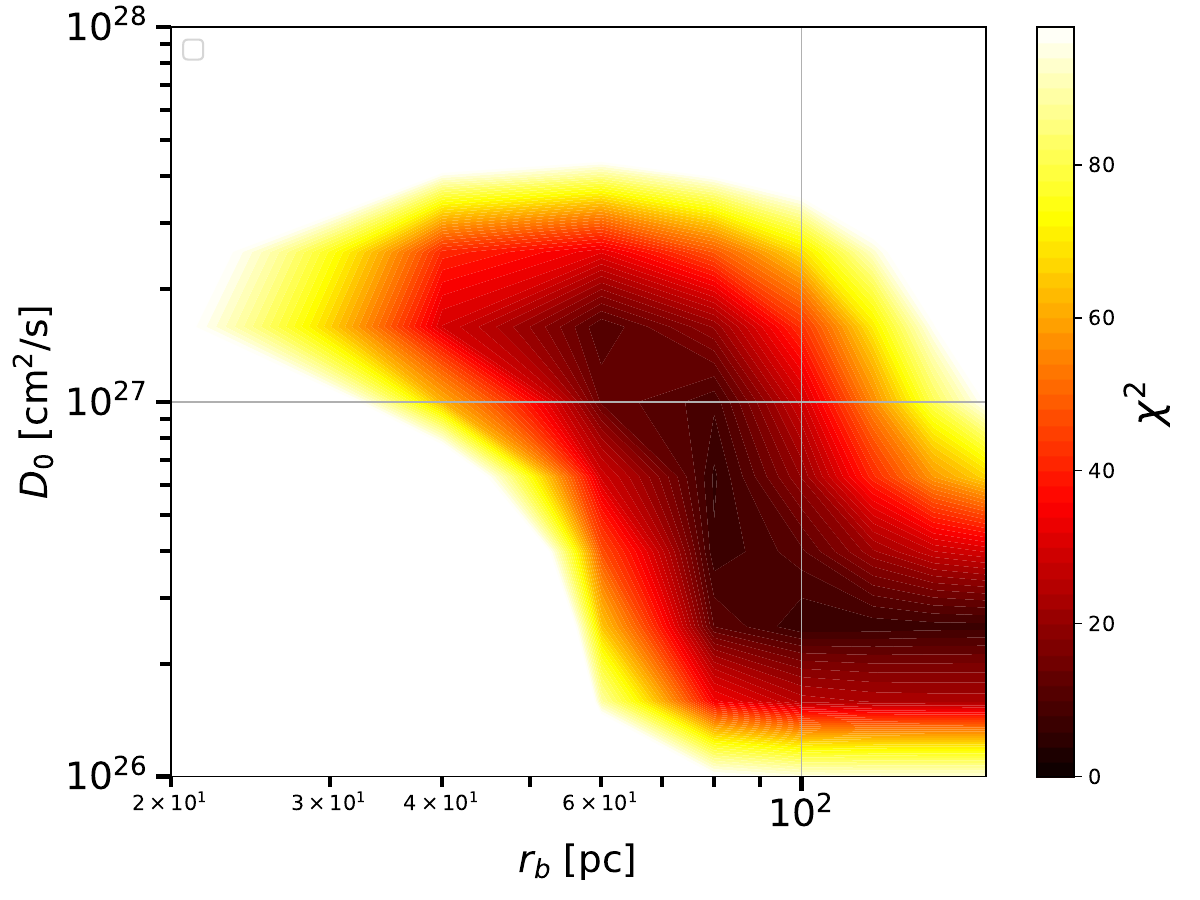}
\caption{Results of the fit to the HESS J1825-137 PWN surface brightness varying $D_0$ and $r_b$. The color bar describes the value of $\chi^2$.}  
\label{fig:D0rbHESS1825}
\end{figure}

These results have been obtained within the one-zone diffusion model (see Sect. \ref{sec:model}). 
We now explore the possibility that a low diffusive regime is present in a region around the source within a radius $r_b$, according to Eq.~\ref{eq:Diff}. 
In order to show the effect of $r_b$ on the surface brightness we consider the very bright  HESS J1825-137 source, 
 which has surface brightness data with relatively small uncertainties (see Fig.~\ref{fig:SB}). 
We calculate the best fit for $D_0$ for a two-zone diffusion model with $r_b$ variable between $10-200$ pc. 
The range of $D_0$ and $r_b$ that best represents the data are: $D_0\in[2.5,15]\cdot 10^{26}$ cm$^2$/s and $r_b>60$ pc.
The best fit is for $D_0 = 6\cdot 10^{26}$ cm$^2$/s and $r_b=80$ pc, but the $\chi^2$ distribution is flat for $r_b>70$ pc and gives comparably good fit for increasing values of $r_b$ and decreasing values of $D_0$.
In particular, for $r_b=[60,80,100,120]$ pc the best for $D_0$ is $D_0\in[15.9,6.3,2.5,2.2]\cdot 10^{26}$ cm$^2$/s.
Therefore, for $r_b>80$ pc the best fit for $D_0$ tends to the value we find with the one-zone diffusion model (see Tab.~\ref{tab:D0etta}).
We show in Fig.~\ref{fig:D0rbHESS1825} the contour plot for the $\chi^2$ values as a function of $D_0$ and $r_b$.
This exercise demonstrates that surface brightness data could be used in principle to bound the size of the low-diffusion bubble. 
However, it is prohibitive  to run this analysis for all the sources in our sample, because the surface brightness data for most of the sources have large uncertainties.

In Ref.~\cite{Giacinti:2019nbu}, a sample of PWNe and PWN candidates from the HGPS catalog have been considered to estimate the density of $e^{\pm}$ in the ICS halo.  
The $e^{\pm}$ density has been calculated with different methods finding that, for most of the sources, it is larger than the one of the ISM.  This implies that the $e^{\pm}$ that produce the ICS halos are probably confined in the PWN. 
One of the main assumption in that paper is the size of the ICS region, which is taken directly as the source extension provided in the HGPS catalog, i.e.~as the standard deviation for a Gaussian spatial distribution of $\gamma$ rays. 
These sizes are probably an underestimate of the halo size. 
Indeed for many sources the sizes they assume are much smaller than the values we report in Tab.~\ref{tab:D0etta} with $\theta_{\rm{ICS}}/2$. 
In particular, this happens for the following sources: HESS J1718-385, HESS J1809-193, HESS J1813-178, HESS J1825-137, HESS J1858+020, HESS J1908+063, HESS J1303-631, HESS J1356-645, HESS J1420-607 and HESS J1833-105. The differences in the halo size is for most of the sources about a factor of 2 thus bringing a difference in halo volume of almost 1 order of magnitude. 
If this factor is considered in their calculation, many of their sources would have a $e^{\pm}$ density comparable to the one of the ISM.
This would change significantly their conclusion because they could not exclude that, for most of their sources, the $e^{\pm}$ are probably not confined in the PWN and actually are traveling in the ISM. 
For example the source HESS J1825-137 has a $e^{\pm}$ density of about 0.2 eV/cm$^3$ in \cite{Giacinti:2019nbu}, about twice the one of the ISM, calculated using a size of the halo of 48.3 pc. 
On the other hand, we find for the same source that the size is about 73 pc. Using this number, the $e^{\pm}$ density becomes 0.06 eV/cm$^3$, i.e.~smaller than the ISM one.

We can now estimate the efficiency $\eta$ and the power-law index 
$\gamma_e$ using the measurement of the differential flux at 1 TeV and of the $\gamma$-ray flux spectral index published in the HGPS catalog.
Specifically, we fit the value of $\gamma_e$ to the observed $\gamma$-ray slope and then we find the efficiency which reproduces the flux data at 1 TeV.
The $\gamma_e$ values are derived assuming Eq.~\ref{eq:Q_E_cont} for the $e^{\pm}$ injection spectrum.
We report in Tab.~\ref{tab:D0etta} the results for  $\gamma_e$, together with the size of the halo and the diffusion coefficient. 
Indeed, $\gamma_e$ turns out to be well determined for each single source, but showing a great spread over the whole sample. 
Most of the values of $\gamma_e$ are very soft and in the range $2.5-3.0$. Only 7 of them are harder, with values between $1.2-1.9$.
Given that our study is devoted to energies well above the TeV, 
we do not introduce any further break at lower energies as instead assumed to model low-energy data from {\it Fermi}-LAT and X-ray telescopes \cite{Torres:2014iua}.

We calculate the efficiency for the conversion of pulsar spin-down energy into $e^{\pm}$ using Eq.~\ref{eq:Q_E_cont}. 
We assume for each source the $\gamma_e$ values reported in Tab.~\ref{tab:D0etta}. 
We find very high values of $\eta$, sometimes even exceeding 1.
These high values for $\eta$ are likely due to the choice not to set a break into Eq.~\ref{eq:Q_E_cont}. 
The $e^{\pm}$ injection spectrum is indeed usually modeled with a broken power law with a break around hundreds GeV and with an index above and below the break of about 1.4 and 2.2, respectively (see, e.g., \cite{Torres:2014iua}.)
The bias here is that we are extrapolating very soft indexes below the energy break where actually $\gamma_e$ hardens.
Indeed, we remind that the efficiency is calculated from an $e^{\pm}$ energy of 0.1 GeV while this analysis is constraining the injection spectrum for TeV energies. 
We can revert the sentence inferring that the $e^{\pm}$ injection spectrum is probably harder below the energy range covered by HESS.
For all the reasons reported above we decide to not show the values of $\eta$ that we have found.
In order to constrain more efficiently the efficiency and $e^{\pm}$ injection spectrum $\gamma$-ray data at GeV energies must be considered.
{\it Fermi}-LAT data are ideal to this scope, as we have already shown in \cite{DiMauro:2019yvh}. 
We are planning to perform, in a followup paper, a combined analysis of HESS and {\it Fermi}-LAT data from the sources considered in this paper in order to derive $\eta$ and $\gamma_e$.

\begin{table*}
\begin{center}
\begin{tabular}{|c|c|c|c|c|c|}
\hline
 Name &   $\theta_{\rm{ICS}}/2$   &   Size   &   $D_0$   &   $\gamma_e$   \\
\hline
           &    [deg]      &          pc           &    $10^{25}$ [cm$^2$/s]        &   \\
\hline
 HESS J1018-589B				& $0.27\pm0.10$	&   $15\pm5$     &  $2.2^{+1.5}_{-0.9}$($2.4^{+1.4}_{-0.9}$)    & $2.5\pm0.1$   \\
 HESS J1026-582				& $0.25\pm0.09$	&   $6\pm2$  	 &  $0.48^{+0.38}_{-0.20}$($0.31^{+0.23}_{-0.13}$)       &  $1.6\pm0.1$  \\
 HESS J1458-608				& $1.20\pm0.73$	&   $38\pm23$   &  $4.7^{+12.3}_{-2.8}$($4.5^{+8.0}_{-2.5}$)    &  $2.7\pm0.1$ \\
 HESS J1632-478				& $1.4\pm0.4$		&   $119\pm34$ &   $8.7^{+9.1}_{-3.9}$($9.1^{+7.9}_{-3.7}$)     &  $1.9\pm0.1$  \\ 
 HESS J1718-385				& $0.21\pm0.05$	&   $13\pm3$     &  $1.4^{+0.8}_{-0.5}$($1.4^{+0.7}_{-0.4}$)      &  $1.2\pm0.1$  \\ 
 HESS J1809-193  				& $0.76\pm0.12$      &   $44\pm7$     &  $7.3^{+2.1}_{-1.5}$($6.8^{+1.8}_{-1.3}$)      &  $2.3\pm0.1$  \\ 
 HESS J1813-126  				& $0.63\pm0.40$      &   $30\pm24$     &  $2.4^{+6.5}_{-1.7}$($2.8^{+6.0}_{-1.6}$)      &  $1.9\pm0.2$  \\ 
 HESS J1825-137				& $1.15\pm0.09$	&   $73\pm6$     &  $21^{+3}_{-3}$($23^{+3}_{-2}$)   &   $2.8\pm0.1$  \\
 HESS J1831-098				& $0.52\pm0.26$  	&   $34\pm16$   &  $6.0^{+10.1}_{-3.2}$($7.1^{+9.3}_{-3.7}$)    &  $1.2\pm0.1$ \\
 HESS J1837-069				& $0.77\pm0.09$	&   $89\pm11$ &  $41^{+8}_{-6}$($43^{+8}_{-6}$)    &  $2.6\pm0.1$ \\
 HESS J1841-055				& $2.50\pm0.42$	&   $180\pm30$     &  $93^{+84}_{-35}$($85^{+53}_{-28}$)        & $2.8\pm0.1$   \\  
 HESS J1857+026				& $0.58\pm0.16$	&   $64\pm18$     &  $23^{+11}_{-7}$($23^{+9}_{-7}$)     &  $2.9\pm0.1$  \\ 
 HESS J1858+020				& $0.25\pm0.08$      &   $20\pm6$	  &	$2.8^{+2.7}_{-1.4}$($2.9^{+2.2}_{-1.3}$)      & $1.8\pm0.2$   \\ 
 HESS J1908+063				& $2.2\pm1.7$		&   $91\pm60$     &  $32^{+56}_{-16}$($32^{+41}_{-15}$)    & $2.7\pm0.1$  \\ 
 HESS J1912+101				& $1.05\pm0.38$      &   $85\pm30$     &  $43^{+46}_{-20}$($49^{+43}_{-21}$)    & $1.8\pm0.1$  \\ 
\hline
\hline
 HESS J0835-455				& $1.65\pm0.27$	&   $8.1\pm1.3$   &  $0.84^{+0.27}_{-0.19}$($0.85^{+0.23}_{-0.17}$)      &  $2.4\pm0.1$  \\  
 HESS J1303-631				& $0.47\pm0.04$	&   $88\pm8$     &  $48^{+9}_{-7}$($45^{+8}_{-6}$)     &  $2.4\pm0.1$  \\  
 HESS J1356-645				& $0.52\pm0.17$	&   $28\pm9$     &  $12^{+7}_{-4}$($13^{+6}_{-4}$)       &  $2.8\pm0.1$  \\ 
 HESS J1420-607				& $0.30\pm0.04$	&   $29\pm4$       &  $8.7^{+2.0}_{-1.6}$($9.1^{+2.1}_{-1.7}$)      &  $2.5\pm0.1$  \\ 
 HESS J1616-508				& $0.55\pm0.09$	&   $46\pm7$       &  $19^{+5}_{-4}$($21^{+4}_{-3}$)      &  $2.9\pm0.1$  \\ 
 HESS J1640-465				& $0.17\pm0.01$	&   $39\pm3$         &  $18^{+3}_{-2}$($19^{+2}_{-1}$)      &  $2.9\pm0.1$  \\ 
 HESS J1708-443				& $0.49\pm0.23$	&   $22\pm10$     &  $5.4^{+6.1}_{-2.9}$($5.2^{+5.0}_{-2.5}$)    &  $2.6\pm0.1$  \\  
 HESS J1813-178				& $0.19\pm0.02$      &   $15\pm2$       &  $5.0^{+1.0}_{-0.9}$($4.6^{+0.9}_{-0.8}$) 	            & $2.6\pm0.1$   \\ 
 HESS J1826-130				& $1.13\pm0.46$	&   $31\pm13$         &  $4.9^{+4.8}_{-2.2}$($5.5^{+4.7}_{-2.3}$) 	          & $2.4\pm0.2$   \\ 
 HESS J1833-105				& $0.21\pm0.07$	&   $15\pm5$               &  $4.6^{+3.4}_{-1.2}$($4.7^{+3.0}_{-1.7}$)     &  $3.0\pm0.2$  \\
%gammae = np.array([2.5,1.4,1.5,1.3,1.3,2.3,2.1,2.5,1.3,2.1,2.4,2.2,1.9,2.5,1.8])
\hline
\hline
 Geminga            			        & $5.5 \pm 0.7$   	&   $24$   &  $5.0^{+2.0}_{-1.0}$($2.1^{+1.0}_{-0.7}$)     &  2.3  \\ 
 Monogem              				&   $4.8 \pm 0.6$ 	&   $24$   &  $25^{+3.3}_{-2.1}$    &  2.1 \\ 
\hline
\hline
\end{tabular}
\caption{Results of our analysis for $D_0$ and $\eta$. We report the source name, the size of extension of the ICS halo found using the function in Eq.~\ref{eq:SBfunc} ($\theta_{\rm{ICS}}/2$), half of the the size of the ICS halo calculated using $\theta_{\rm{ICS}}/2$, the best fit and $1\sigma$ error for $D_0$ and the $e^{\pm}$ spectral index.}
\label{tab:D0etta}
\end{center}
\end{table*}

\section{Conclusions}
\label{sec:conclusions}

The detection of low-diffusion regions, few tens of pc wide, found around Geminga and Monogem pulsars analyzing {\it Fermi}-LAT  \cite{DiMauro:2019yvh} and HAWC \cite{Abeysekara:2017science} $\gamma$-ray data raises the question if this is a peculiarity or a general property of Galactic pulsars. 

In this paper, trying to answer this question, we have studied the physical properties of these halos, believed to be generated by $e^\pm$ accelerated 
by PWN and ICS with the ISRF. 
We have studied the size of ICS halos as a function of the strength and size of the low-diffusion bubble, the age and distance of the host pulsar, and of its proper motion.
We find that current IACTs  are able to probe diffusion coefficients $\leq 10^{27}$ cm$^2$/s around most of the pulsars closer than 10 kpc from the Earth. We show that, at VHE, the pulsar proper motion has a limited effect on the ICS spatial morphology.

We then rank ATNF pulsars according to the ICS flux and demonstrate that this parameter is very efficient to indicate the detectability of each source. 
Indeed, out of 23 pulsars in the HAWC field of view and predicted by our model to have the brightest ICS halo fluxes, 21 have been included in the 2HWC catalog.
We provide in Tab.~\ref{tab:TeVsourcesHAWCfuture} the list of sources not yet detected by HAWC, and ranked by their 
ICS $\gamma$-ray flux. Given the ICS emission is the process producing the VHE photons similarly with Geminga and Monogem, 
we predict these sources to be the next-to-be-discovered as ICS halos in HAWC data. 
As a further  prediction, we also list the angular size of the ICS halo of each source. 
The number of ICS halos potentially already detected by HAWC and HESS ranges between 25-50 assuming a conversion efficiency $\eta$ at the \% level. 
As for CTA, an efficiency as low as 0.01 could lead to the detection of at least one hundred ICS halos. 

We employ the flux maps provided in the HGPS catalog and the Geminga and Monogem surface brightness published by the HAWC Collaboration in order to derive the diffusion around a sample of 27 PWNe and PWN candidates. 
We demonstrate that the $e^{\pm}$, released from the sources in our sample, propagate in a low-diffusion Galactic environment with a diffusion coefficient which is about two orders of magnitude lower than the value recently derived for the entire Galaxy through a fit to AMS-02 CR data. 
The mean value and the standard deviation over the entire sample are at 1 GeV $D_0 = 9.1^{+17.4}_{-6.0}\cdot 10^{25}$~cm$^2$/s. 
We do not register any dependence of this numbers on the age of pulsar, meaning that probably the effect of confinement of the SNR and PWN is not very strong even for the younger sources in our sample.
The characterization of the pulsar environment by a low diffusion region turns out to be a general trend for all the analyzed sources.
The size of the ICS halos have been found to be on average 35 pc implying that the low-diffusion bubbles should be larger than this size.
For some of the sources in our sample, e.g.~HESS J1632-478, HESS J1825-137, HESS J1837-069, HESS J1841-055, HESS J1912+101 and HESS J1303-631, the low-diffusion bubble size could exceed 100 pc.
These numbers should be used as an estimate for $r_b$ in the two-zone diffusion model employed to propagate $e^\pm$ from the pulsar to the Earth. 
Since, as we have explained in Sec.~\ref{sec:sample}, the PWNe consider in this paper are also the highest ranked according to the ICS flux at TeV energies, we do not believe our results are biased towards objects that have smaller $D_0$ and so have a more concentrated $\gamma$-ray emission.
The consequences of the present results for the interpretation of the $e^+$ flux data in terms of Galactic PWNe and for the propagation of cosmic rays will be investigating in a forthcoming paper.

\begin{acknowledgments}
The authors thank Luigi Tibaldo, Ke Fang, Andrew James Smith, Regina Caputo and Roger Romani for insightful discussions.
MDM acknowledges support by the NASA Fermi Guest Investigator Program 2019 Cycle 12 through the Fermi Program N. 121119 (P.I. MDM) entitled ``Detecting $\gamma$-ray halos around PWNe and interpretation of the positron excess''.
The work of FD and SM  is supported by the "Departments of Excellence 2018 - 2022" Grant awarded by
the Italian Ministry of Education, University and Research (MIUR) (L. 232/2016).
FD and SM acknowledge financial contribution from the agreement ASI-INAF
n.2017-14-H.0 and the Fondazione CRT for the grant 2017/58675.
\end{acknowledgments}

\bibliography{paper}

\begin{thebibliography}{52}
\expandafter\ifx\csname natexlab\endcsname\relax\def\natexlab#1{#1}\fi
\expandafter\ifx\csname bibnamefont\endcsname\relax
  \def\bibnamefont#1{#1}\fi
\expandafter\ifx\csname bibfnamefont\endcsname\relax
  \def\bibfnamefont#1{#1}\fi
\expandafter\ifx\csname citenamefont\endcsname\relax
  \def\citenamefont#1{#1}\fi
\expandafter\ifx\csname url\endcsname\relax
  \def\url#1{\texttt{#1}}\fi
\expandafter\ifx\csname urlprefix\endcsname\relax\def\urlprefix{URL }\fi
\providecommand{\bibinfo}[2]{#2}
\providecommand{\eprint}[2][]{\url{#2}}

\bibitem[{\citenamefont{Abeysekara
  et~al.}(2017{\natexlab{a}})}]{Abeysekara:2017science}
\bibinfo{author}{\bibfnamefont{A.~U.} \bibnamefont{Abeysekara}}
  \bibnamefont{et~al.} (\bibinfo{collaboration}{HAWC}),
  \bibinfo{journal}{Science} \textbf{\bibinfo{volume}{358}},
  \bibinfo{pages}{911} (\bibinfo{year}{2017}{\natexlab{a}}),
  \eprint{1711.06223}.

\bibitem[{\citenamefont{{Abdo} et~al.}(2009)\citenamefont{{Abdo}, {Allen},
  {Aune} et~al.}}]{2009ApJ...700L.127A}
\bibinfo{author}{\bibfnamefont{A.~A.} \bibnamefont{{Abdo}}},
  \bibinfo{author}{\bibfnamefont{B.~T.} \bibnamefont{{Allen}}},
  \bibinfo{author}{\bibfnamefont{T.}~\bibnamefont{{Aune}}},
  \bibnamefont{et~al.}, \bibinfo{journal}{\apjl}
  \textbf{\bibinfo{volume}{700}}, \bibinfo{pages}{L127} (\bibinfo{year}{2009}),
  \eprint{0904.1018}.

\bibitem[{\citenamefont{Di~Mauro et~al.}(2019)\citenamefont{Di~Mauro, Manconi,
  and Donato}}]{DiMauro:2019yvh}
\bibinfo{author}{\bibfnamefont{M.}~\bibnamefont{Di~Mauro}},
  \bibinfo{author}{\bibfnamefont{S.}~\bibnamefont{Manconi}}, \bibnamefont{and}
  \bibinfo{author}{\bibfnamefont{F.}~\bibnamefont{Donato}}
  (\bibinfo{year}{2019}), \eprint{1903.05647}.

\bibitem[{\citenamefont{Adriani et~al.}(2013)}]{Adriani:2013uda}
\bibinfo{author}{\bibfnamefont{O.}~\bibnamefont{Adriani}} \bibnamefont{et~al.}
  (\bibinfo{collaboration}{PAMELA}), \bibinfo{journal}{Phys. Rev. Lett.}
  \textbf{\bibinfo{volume}{111}}, \bibinfo{pages}{081102}
  (\bibinfo{year}{2013}), \eprint{1308.0133}.

\bibitem[{\citenamefont{{Ackermann} et~al.}(2012)\citenamefont{{Ackermann},
  {Ajello}, {Allafort} et~al.}}]{2012PhRvL.108a1103A}
\bibinfo{author}{\bibfnamefont{M.}~\bibnamefont{{Ackermann}}},
  \bibinfo{author}{\bibfnamefont{M.}~\bibnamefont{{Ajello}}},
  \bibinfo{author}{\bibnamefont{{Allafort}}}, \bibnamefont{et~al.},
  \bibinfo{journal}{Physical Review Letters} \textbf{\bibinfo{volume}{108}},
  \bibinfo{eid}{011103} (\bibinfo{year}{2012}), \eprint{1109.0521}.

\bibitem[{\citenamefont{Aguilar et~al.}(2019)\citenamefont{Aguilar,
  Ali~Cavasonza, Ambrosi et~al.}}]{PhysRevLett.122.041102}
\bibinfo{author}{\bibfnamefont{M.}~\bibnamefont{Aguilar}},
  \bibinfo{author}{\bibfnamefont{L.}~\bibnamefont{Ali~Cavasonza}},
  \bibinfo{author}{\bibfnamefont{G.}~\bibnamefont{Ambrosi}},
  \bibnamefont{et~al.} (\bibinfo{collaboration}{AMS Collaboration}),
  \bibinfo{journal}{Phys. Rev. Lett.} \textbf{\bibinfo{volume}{122}},
  \bibinfo{pages}{041102} (\bibinfo{year}{2019}),
  \urlprefix\url{https://link.aps.org/doi/10.1103/PhysRevLett.122.041102}.

\bibitem[{\citenamefont{Kappl et~al.}(2015)\citenamefont{Kappl, Reinert, and
  Winkler}}]{Kappl:2015bqa}
\bibinfo{author}{\bibfnamefont{R.}~\bibnamefont{Kappl}},
  \bibinfo{author}{\bibfnamefont{A.}~\bibnamefont{Reinert}}, \bibnamefont{and}
  \bibinfo{author}{\bibfnamefont{M.~W.} \bibnamefont{Winkler}},
  \bibinfo{journal}{JCAP} \textbf{\bibinfo{volume}{1510}}, \bibinfo{pages}{034}
  (\bibinfo{year}{2015}), \eprint{1506.04145}.

\bibitem[{\citenamefont{Genolini et~al.}(2015)\citenamefont{Genolini, Putze,
  Salati, and Serpico}}]{Genolini:2015cta}
\bibinfo{author}{\bibfnamefont{Y.}~\bibnamefont{Genolini}},
  \bibinfo{author}{\bibfnamefont{A.}~\bibnamefont{Putze}},
  \bibinfo{author}{\bibfnamefont{P.}~\bibnamefont{Salati}}, \bibnamefont{and}
  \bibinfo{author}{\bibfnamefont{P.~D.} \bibnamefont{Serpico}},
  \bibinfo{journal}{Astron. Astrophys.} \textbf{\bibinfo{volume}{580}},
  \bibinfo{pages}{A9} (\bibinfo{year}{2015}), \eprint{1504.03134}.

\bibitem[{\citenamefont{Genolini et~al.}(2019)}]{Genolini:2019ewc}
\bibinfo{author}{\bibfnamefont{Y.}~\bibnamefont{Genolini}} \bibnamefont{et~al.}
  (\bibinfo{year}{2019}), \eprint{1904.08917}.

\bibitem[{\citenamefont{Linden et~al.}(2017)\citenamefont{Linden, Auchettl,
  Bramante, Cholis, Fang, Hooper, Karwal, and Li}}]{Linden:2017vvb}
\bibinfo{author}{\bibfnamefont{T.}~\bibnamefont{Linden}},
  \bibinfo{author}{\bibfnamefont{K.}~\bibnamefont{Auchettl}},
  \bibinfo{author}{\bibfnamefont{J.}~\bibnamefont{Bramante}},
  \bibinfo{author}{\bibfnamefont{I.}~\bibnamefont{Cholis}},
  \bibinfo{author}{\bibfnamefont{K.}~\bibnamefont{Fang}},
  \bibinfo{author}{\bibfnamefont{D.}~\bibnamefont{Hooper}},
  \bibinfo{author}{\bibfnamefont{T.}~\bibnamefont{Karwal}}, \bibnamefont{and}
  \bibinfo{author}{\bibfnamefont{S.~W.} \bibnamefont{Li}},
  \bibinfo{journal}{Phys. Rev.} \textbf{\bibinfo{volume}{D96}},
  \bibinfo{pages}{103016} (\bibinfo{year}{2017}), \eprint{1703.09704}.

\bibitem[{\citenamefont{Giacinti et~al.}(2019)\citenamefont{Giacinti, Mitchell,
  L\'opez-Coto, Joshi, Parsons, and Hinton}}]{Giacinti:2019nbu}
\bibinfo{author}{\bibfnamefont{G.}~\bibnamefont{Giacinti}},
  \bibinfo{author}{\bibfnamefont{A.~M.~W.} \bibnamefont{Mitchell}},
  \bibinfo{author}{\bibfnamefont{R.}~\bibnamefont{L\'opez-Coto}},
  \bibinfo{author}{\bibfnamefont{V.}~\bibnamefont{Joshi}},
  \bibinfo{author}{\bibfnamefont{R.~D.} \bibnamefont{Parsons}},
  \bibnamefont{and} \bibinfo{author}{\bibfnamefont{J.~A.} \bibnamefont{Hinton}}
  (\bibinfo{year}{2019}), \eprint{1907.12121}.

\bibitem[{\citenamefont{Abdalla et~al.}(2018{\natexlab{a}})}]{H.E.S.S.:2018zkf}
\bibinfo{author}{\bibfnamefont{H.}~\bibnamefont{Abdalla}} \bibnamefont{et~al.}
  (\bibinfo{collaboration}{HESS}), \bibinfo{journal}{Astron. Astrophys.}
  \textbf{\bibinfo{volume}{612}}, \bibinfo{pages}{A1}
  (\bibinfo{year}{2018}{\natexlab{a}}), \eprint{1804.02432}.

\bibitem[{\citenamefont{Hooper et~al.}(2017)\citenamefont{Hooper, Cholis,
  Linden, and Fang}}]{Hooper:2017gtd}
\bibinfo{author}{\bibfnamefont{D.}~\bibnamefont{Hooper}},
  \bibinfo{author}{\bibfnamefont{I.}~\bibnamefont{Cholis}},
  \bibinfo{author}{\bibfnamefont{T.}~\bibnamefont{Linden}}, \bibnamefont{and}
  \bibinfo{author}{\bibfnamefont{K.}~\bibnamefont{Fang}},
  \bibinfo{journal}{Phys. Rev.} \textbf{\bibinfo{volume}{D96}},
  \bibinfo{pages}{103013} (\bibinfo{year}{2017}), \eprint{1702.08436}.

\bibitem[{\citenamefont{Xi et~al.}(2018)\citenamefont{Xi, Liu, Huang, Fang,
  Yan, and Wang}}]{Shao-Qiang:2018zla}
\bibinfo{author}{\bibfnamefont{S.-Q.} \bibnamefont{Xi}},
  \bibinfo{author}{\bibfnamefont{R.-Y.} \bibnamefont{Liu}},
  \bibinfo{author}{\bibfnamefont{Z.-Q.} \bibnamefont{Huang}},
  \bibinfo{author}{\bibfnamefont{K.}~\bibnamefont{Fang}},
  \bibinfo{author}{\bibfnamefont{H.}~\bibnamefont{Yan}}, \bibnamefont{and}
  \bibinfo{author}{\bibfnamefont{X.-Y.} \bibnamefont{Wang}}
  (\bibinfo{year}{2018}), \eprint{1810.10928}.

\bibitem[{\citenamefont{Tang and Piran}(2018)}]{Tang:2018wyr}
\bibinfo{author}{\bibfnamefont{X.}~\bibnamefont{Tang}} \bibnamefont{and}
  \bibinfo{author}{\bibfnamefont{T.}~\bibnamefont{Piran}}
  (\bibinfo{year}{2018}), \eprint{1808.02445}.

\bibitem[{\citenamefont{Fang et~al.}(2018)\citenamefont{Fang, Bi, Yin, and
  Yuan}}]{Fang:2018qco}
\bibinfo{author}{\bibfnamefont{K.}~\bibnamefont{Fang}},
  \bibinfo{author}{\bibfnamefont{X.-J.} \bibnamefont{Bi}},
  \bibinfo{author}{\bibfnamefont{P.-F.} \bibnamefont{Yin}}, \bibnamefont{and}
  \bibinfo{author}{\bibfnamefont{Q.}~\bibnamefont{Yuan}},
  \bibinfo{journal}{Astrophys. J.} \textbf{\bibinfo{volume}{863}},
  \bibinfo{pages}{30} (\bibinfo{year}{2018}), \eprint{1803.02640}.

\bibitem[{\citenamefont{Abeysekara
  et~al.}(2017{\natexlab{b}})}]{Abeysekara:2017hyn}
\bibinfo{author}{\bibfnamefont{A.~U.} \bibnamefont{Abeysekara}}
  \bibnamefont{et~al.}, \bibinfo{journal}{Astrophys. J.}
  \textbf{\bibinfo{volume}{843}}, \bibinfo{pages}{40}
  (\bibinfo{year}{2017}{\natexlab{b}}), \eprint{1702.02992}.

\bibitem[{\citenamefont{Sudoh et~al.}(2019)\citenamefont{Sudoh, Linden, and
  Beacom}}]{Sudoh:2019lav}
\bibinfo{author}{\bibfnamefont{T.}~\bibnamefont{Sudoh}},
  \bibinfo{author}{\bibfnamefont{T.}~\bibnamefont{Linden}}, \bibnamefont{and}
  \bibinfo{author}{\bibfnamefont{J.~F.} \bibnamefont{Beacom}},
  \bibinfo{journal}{Phys. Rev.} \textbf{\bibinfo{volume}{D100}},
  \bibinfo{pages}{043016} (\bibinfo{year}{2019}), \eprint{1902.08203}.

\bibitem[{\citenamefont{Hobbs et~al.}(2005)\citenamefont{Hobbs, Lorimer, Lyne,
  and Kramer}}]{Hobbs:2005yx}
\bibinfo{author}{\bibfnamefont{G.}~\bibnamefont{Hobbs}},
  \bibinfo{author}{\bibfnamefont{D.~R.} \bibnamefont{Lorimer}},
  \bibinfo{author}{\bibfnamefont{A.~G.} \bibnamefont{Lyne}}, \bibnamefont{and}
  \bibinfo{author}{\bibfnamefont{M.}~\bibnamefont{Kramer}},
  \bibinfo{journal}{Mon. Not. Roy. Astron. Soc.}
  \textbf{\bibinfo{volume}{360}}, \bibinfo{pages}{974} (\bibinfo{year}{2005}),
  \eprint{astro-ph/0504584}.

\bibitem[{\citenamefont{Acharya et~al.}(2017)}]{Acharya:2017ttl}
\bibinfo{author}{\bibfnamefont{B.~S.} \bibnamefont{Acharya}}
  \bibnamefont{et~al.} (\bibinfo{collaboration}{Cherenkov Telescope Array
  Consortium}) (\bibinfo{year}{2017}), \eprint{1709.07997}.

\bibitem[{\citenamefont{{Manchester} et~al.}(2005)\citenamefont{{Manchester},
  {Hobbs}, {Teoh}, and {Hobbs}}}]{2005AJ....129.1993M}
\bibinfo{author}{\bibfnamefont{R.~N.} \bibnamefont{{Manchester}}},
  \bibinfo{author}{\bibfnamefont{G.~B.} \bibnamefont{{Hobbs}}},
  \bibinfo{author}{\bibfnamefont{A.}~\bibnamefont{{Teoh}}}, \bibnamefont{and}
  \bibinfo{author}{\bibfnamefont{M.}~\bibnamefont{{Hobbs}}},
  \bibinfo{journal}{\aj} \textbf{\bibinfo{volume}{129}}, \bibinfo{pages}{1993}
  (\bibinfo{year}{2005}), \eprint{astro-ph/0412641}.

\bibitem[{\citenamefont{{Chi} et~al.}(1996)\citenamefont{{Chi}, {Cheng}, and
  {Young}}}]{1996ApJ...459L..83C}
\bibinfo{author}{\bibfnamefont{X.}~\bibnamefont{{Chi}}},
  \bibinfo{author}{\bibfnamefont{K.~S.} \bibnamefont{{Cheng}}},
  \bibnamefont{and} \bibinfo{author}{\bibfnamefont{E.~C.~M.}
  \bibnamefont{{Young}}}, \bibinfo{journal}{\apjl}
  \textbf{\bibinfo{volume}{459}}, \bibinfo{pages}{L83} (\bibinfo{year}{1996}).

\bibitem[{\citenamefont{{Amato}}(2014)}]{Amato:2013fua}
\bibinfo{author}{\bibfnamefont{E.}~\bibnamefont{{Amato}}},
  \bibinfo{journal}{International Journal of Modern Physics Conference Series}
  \textbf{\bibinfo{volume}{28}}, \bibinfo{eid}{1460160} (\bibinfo{year}{2014}),
  \eprint{1312.5945}.

\bibitem[{\citenamefont{Gaensler and Slane}(2006)}]{Gaensler:2006ua}
\bibinfo{author}{\bibfnamefont{B.~M.} \bibnamefont{Gaensler}} \bibnamefont{and}
  \bibinfo{author}{\bibfnamefont{P.~O.} \bibnamefont{Slane}},
  \bibinfo{journal}{Ann. Rev. Astron. Astrophys.}
  \textbf{\bibinfo{volume}{44}}, \bibinfo{pages}{17} (\bibinfo{year}{2006}),
  \eprint{astro-ph/0601081}.

\bibitem[{\citenamefont{{Chevalier}}(1977)}]{1977ASSL...66...53C}
\bibinfo{author}{\bibfnamefont{R.~A.} \bibnamefont{{Chevalier}}}, in
  \emph{\bibinfo{booktitle}{Supernovae}}, edited by
  \bibinfo{editor}{\bibfnamefont{D.~N.} \bibnamefont{{Schramm}}}
  (\bibinfo{year}{1977}), vol.~\bibinfo{volume}{66} of
  \emph{\bibinfo{series}{Astrophysics and Space Science Library}},
  p.~\bibinfo{pages}{53}.

\bibitem[{\citenamefont{{Gelfand} et~al.}(2009)\citenamefont{{Gelfand},
  {Slane}, and {Zhang}}}]{2009ApJ...703.2051G}
\bibinfo{author}{\bibfnamefont{J.~D.} \bibnamefont{{Gelfand}}},
  \bibinfo{author}{\bibfnamefont{P.~O.} \bibnamefont{{Slane}}},
  \bibnamefont{and} \bibinfo{author}{\bibfnamefont{W.}~\bibnamefont{{Zhang}}},
  \bibinfo{journal}{\apj} \textbf{\bibinfo{volume}{703}}, \bibinfo{pages}{2051}
  (\bibinfo{year}{2009}), \eprint{0904.4053}.

\bibitem[{\citenamefont{{Reynolds} and
  {Chevalier}}(1984)}]{1984ApJ...278..630R}
\bibinfo{author}{\bibfnamefont{S.~P.} \bibnamefont{{Reynolds}}}
  \bibnamefont{and} \bibinfo{author}{\bibfnamefont{R.~A.}
  \bibnamefont{{Chevalier}}}, \bibinfo{journal}{\apj}
  \textbf{\bibinfo{volume}{278}}, \bibinfo{pages}{630} (\bibinfo{year}{1984}).

\bibitem[{\citenamefont{{van der Swaluw} et~al.}(2001)\citenamefont{{van der
  Swaluw}, {Achterberg}, {Gallant}, and {T{\'o}th}}}]{2001A&A...380..309V}
\bibinfo{author}{\bibfnamefont{E.}~\bibnamefont{{van der Swaluw}}},
  \bibinfo{author}{\bibfnamefont{A.}~\bibnamefont{{Achterberg}}},
  \bibinfo{author}{\bibfnamefont{Y.~A.} \bibnamefont{{Gallant}}},
  \bibnamefont{and}
  \bibinfo{author}{\bibfnamefont{G.}~\bibnamefont{{T{\'o}th}}},
  \bibinfo{journal}{\aap} \textbf{\bibinfo{volume}{380}}, \bibinfo{pages}{309}
  (\bibinfo{year}{2001}).

\bibitem[{\citenamefont{Yuksel et~al.}(2009)\citenamefont{Yuksel, Kistler, and
  Stanev}}]{Yuksel:2008rf}
\bibinfo{author}{\bibfnamefont{H.}~\bibnamefont{Yuksel}},
  \bibinfo{author}{\bibfnamefont{M.~D.} \bibnamefont{Kistler}},
  \bibnamefont{and} \bibinfo{author}{\bibfnamefont{T.}~\bibnamefont{Stanev}},
  \bibinfo{journal}{Phys. Rev. Lett.} \textbf{\bibinfo{volume}{103}},
  \bibinfo{pages}{051101} (\bibinfo{year}{2009}), \eprint{0810.2784}.

\bibitem[{\citenamefont{{Aharonian} et~al.}(1995)\citenamefont{{Aharonian},
  {Atoyan}, and {Voelk}}}]{1995A&A...294L..41A}
\bibinfo{author}{\bibfnamefont{F.~A.} \bibnamefont{{Aharonian}}},
  \bibinfo{author}{\bibfnamefont{A.~M.} \bibnamefont{{Atoyan}}},
  \bibnamefont{and} \bibinfo{author}{\bibfnamefont{H.~J.}
  \bibnamefont{{Voelk}}}, \bibinfo{journal}{\aap}
  \textbf{\bibinfo{volume}{294}}, \bibinfo{pages}{L41} (\bibinfo{year}{1995}).

\bibitem[{\citenamefont{Malyshev et~al.}(2009)\citenamefont{Malyshev, Cholis,
  and Gelfand}}]{Malyshev:2009tw}
\bibinfo{author}{\bibfnamefont{D.}~\bibnamefont{Malyshev}},
  \bibinfo{author}{\bibfnamefont{I.}~\bibnamefont{Cholis}}, \bibnamefont{and}
  \bibinfo{author}{\bibfnamefont{J.}~\bibnamefont{Gelfand}},
  \bibinfo{journal}{Phys. Rev.} \textbf{\bibinfo{volume}{D80}},
  \bibinfo{pages}{063005} (\bibinfo{year}{2009}), \eprint{0903.1310}.

\bibitem[{\citenamefont{Torres et~al.}(2014)\citenamefont{Torres, Cillis,
  Martín, and de~Oña~Wilhelmi}}]{Torres:2014iua}
\bibinfo{author}{\bibfnamefont{D.~F.} \bibnamefont{Torres}},
  \bibinfo{author}{\bibfnamefont{A.}~\bibnamefont{Cillis}},
  \bibinfo{author}{\bibfnamefont{J.}~\bibnamefont{Martín}}, \bibnamefont{and}
  \bibinfo{author}{\bibfnamefont{E.}~\bibnamefont{de~Oña~Wilhelmi}},
  \bibinfo{journal}{JHEAp} \textbf{\bibinfo{volume}{1-2}}, \bibinfo{pages}{31}
  (\bibinfo{year}{2014}), \eprint{1402.5485}.

\bibitem[{\citenamefont{Buesching et~al.}(2008)\citenamefont{Buesching,
  de~Jager, Potgieter, and Venter}}]{Buesching:2008hr}
\bibinfo{author}{\bibfnamefont{I.}~\bibnamefont{Buesching}},
  \bibinfo{author}{\bibfnamefont{O.~C.} \bibnamefont{de~Jager}},
  \bibinfo{author}{\bibfnamefont{M.~S.} \bibnamefont{Potgieter}},
  \bibnamefont{and} \bibinfo{author}{\bibfnamefont{C.}~\bibnamefont{Venter}},
  \bibinfo{journal}{Astrophys. J.} \textbf{\bibinfo{volume}{678}},
  \bibinfo{pages}{L39} (\bibinfo{year}{2008}), \eprint{0804.0220}.

\bibitem[{\citenamefont{Sushch and Hnatyk}(2014)}]{Sushch:2013tna}
\bibinfo{author}{\bibfnamefont{I.}~\bibnamefont{Sushch}} \bibnamefont{and}
  \bibinfo{author}{\bibfnamefont{B.}~\bibnamefont{Hnatyk}},
  \bibinfo{journal}{Astron. Astrophys.} \textbf{\bibinfo{volume}{561}},
  \bibinfo{pages}{A139} (\bibinfo{year}{2014}), \eprint{1312.0777}.

\bibitem[{\citenamefont{Vernetto and Lipari}(2016)}]{Vernetto:2016alq}
\bibinfo{author}{\bibfnamefont{S.}~\bibnamefont{Vernetto}} \bibnamefont{and}
  \bibinfo{author}{\bibfnamefont{P.}~\bibnamefont{Lipari}},
  \bibinfo{journal}{Phys. Rev.} \textbf{\bibinfo{volume}{D94}},
  \bibinfo{pages}{063009} (\bibinfo{year}{2016}), \eprint{1608.01587}.

\bibitem[{\citenamefont{{Sun} et~al.}(2007)\citenamefont{{Sun}, {Han}, {Reich},
  {Reich}, {Shi}, {Wielebinski}, and {F{\"u}rst}}}]{2007A&A...463..993S}
\bibinfo{author}{\bibfnamefont{X.~H.} \bibnamefont{{Sun}}},
  \bibinfo{author}{\bibfnamefont{J.~L.} \bibnamefont{{Han}}},
  \bibinfo{author}{\bibfnamefont{W.}~\bibnamefont{{Reich}}},
  \bibinfo{author}{\bibfnamefont{P.}~\bibnamefont{{Reich}}},
  \bibinfo{author}{\bibfnamefont{W.~B.} \bibnamefont{{Shi}}},
  \bibinfo{author}{\bibfnamefont{R.}~\bibnamefont{{Wielebinski}}},
  \bibnamefont{and}
  \bibinfo{author}{\bibfnamefont{E.}~\bibnamefont{{F{\"u}rst}}},
  \bibinfo{journal}{\aap} \textbf{\bibinfo{volume}{463}}, \bibinfo{pages}{993}
  (\bibinfo{year}{2007}), \eprint{astro-ph/0611622}.

\bibitem[{\citenamefont{Profumo et~al.}(2018)\citenamefont{Profumo,
  Reynoso-Cordova, Kaaz, and Silverman}}]{Profumo:2018fmz}
\bibinfo{author}{\bibfnamefont{S.}~\bibnamefont{Profumo}},
  \bibinfo{author}{\bibfnamefont{J.}~\bibnamefont{Reynoso-Cordova}},
  \bibinfo{author}{\bibfnamefont{N.}~\bibnamefont{Kaaz}}, \bibnamefont{and}
  \bibinfo{author}{\bibfnamefont{M.}~\bibnamefont{Silverman}},
  \bibinfo{journal}{Phys. Rev.} \textbf{\bibinfo{volume}{D97}},
  \bibinfo{pages}{123008} (\bibinfo{year}{2018}), \eprint{1803.09731}.

\bibitem[{\citenamefont{Evoli et~al.}(2018)\citenamefont{Evoli, Linden, and
  Morlino}}]{Evoli:2018aza}
\bibinfo{author}{\bibfnamefont{C.}~\bibnamefont{Evoli}},
  \bibinfo{author}{\bibfnamefont{T.}~\bibnamefont{Linden}}, \bibnamefont{and}
  \bibinfo{author}{\bibfnamefont{G.}~\bibnamefont{Morlino}},
  \bibinfo{journal}{Phys. Rev.} \textbf{\bibinfo{volume}{D98}},
  \bibinfo{pages}{063017} (\bibinfo{year}{2018}), \eprint{1807.09263}.

\bibitem[{\citenamefont{{Slane}}(2017)}]{2017hsn..book.2159S}
\bibinfo{author}{\bibfnamefont{P.}~\bibnamefont{{Slane}}}
  (\bibinfo{year}{2017}), \eprint{1703.09311}.

\bibitem[{\citenamefont{{Posselt} et~al.}(2017)\citenamefont{{Posselt},
  {Pavlov}, {Slane}, {Romani}, {Bucciantini}, {Bykov}, {Kargaltsev},
  {Weisskopf}, and {Ng}}}]{2017ApJ...835...66P}
\bibinfo{author}{\bibfnamefont{B.}~\bibnamefont{{Posselt}}},
  \bibinfo{author}{\bibfnamefont{G.~G.} \bibnamefont{{Pavlov}}},
  \bibinfo{author}{\bibfnamefont{P.~O.} \bibnamefont{{Slane}}},
  \bibinfo{author}{\bibfnamefont{R.}~\bibnamefont{{Romani}}},
  \bibinfo{author}{\bibfnamefont{N.}~\bibnamefont{{Bucciantini}}},
  \bibinfo{author}{\bibfnamefont{A.~M.} \bibnamefont{{Bykov}}},
  \bibinfo{author}{\bibfnamefont{O.}~\bibnamefont{{Kargaltsev}}},
  \bibinfo{author}{\bibfnamefont{M.~C.} \bibnamefont{{Weisskopf}}},
  \bibnamefont{and} \bibinfo{author}{\bibfnamefont{C.-Y.} \bibnamefont{{Ng}}},
  \bibinfo{journal}{\apj} \textbf{\bibinfo{volume}{835}}, \bibinfo{eid}{66}
  (\bibinfo{year}{2017}), \eprint{1611.03496}.

\bibitem[{\citenamefont{{Blumenthal} and {Gould}}(1970)}]{1970RvMP...42..237B}
\bibinfo{author}{\bibfnamefont{G.~R.} \bibnamefont{{Blumenthal}}}
  \bibnamefont{and} \bibinfo{author}{\bibfnamefont{R.~J.}
  \bibnamefont{{Gould}}}, \bibinfo{journal}{Reviews of Modern Physics}
  \textbf{\bibinfo{volume}{42}}, \bibinfo{pages}{237} (\bibinfo{year}{1970}).

\bibitem[{\citenamefont{Cirelli et~al.}(2011)\citenamefont{Cirelli, Corcella,
  Hektor, Hutsi, Kadastik, Panci, Raidal, Sala, and Strumia}}]{Cirelli:2010xx}
\bibinfo{author}{\bibfnamefont{M.}~\bibnamefont{Cirelli}},
  \bibinfo{author}{\bibfnamefont{G.}~\bibnamefont{Corcella}},
  \bibinfo{author}{\bibfnamefont{A.}~\bibnamefont{Hektor}},
  \bibinfo{author}{\bibfnamefont{G.}~\bibnamefont{Hutsi}},
  \bibinfo{author}{\bibfnamefont{M.}~\bibnamefont{Kadastik}},
  \bibinfo{author}{\bibfnamefont{P.}~\bibnamefont{Panci}},
  \bibinfo{author}{\bibfnamefont{M.}~\bibnamefont{Raidal}},
  \bibinfo{author}{\bibfnamefont{F.}~\bibnamefont{Sala}}, \bibnamefont{and}
  \bibinfo{author}{\bibfnamefont{A.}~\bibnamefont{Strumia}},
  \bibinfo{journal}{JCAP} \textbf{\bibinfo{volume}{1103}}, \bibinfo{pages}{051}
  (\bibinfo{year}{2011}), \bibinfo{note}{[Erratum: JCAP1210,E01(2012)]},
  \eprint{1012.4515}.

\bibitem[{\citenamefont{{Delahaye} et~al.}(2010)\citenamefont{{Delahaye},
  {Lavalle}, {Lineros}, {Donato}, and {Fornengo}}}]{2010A&A...524A..51D}
\bibinfo{author}{\bibfnamefont{T.}~\bibnamefont{{Delahaye}}},
  \bibinfo{author}{\bibfnamefont{J.}~\bibnamefont{{Lavalle}}},
  \bibinfo{author}{\bibfnamefont{R.}~\bibnamefont{{Lineros}}},
  \bibinfo{author}{\bibfnamefont{F.}~\bibnamefont{{Donato}}}, \bibnamefont{and}
  \bibinfo{author}{\bibfnamefont{N.}~\bibnamefont{{Fornengo}}},
  \bibinfo{journal}{\aap} \textbf{\bibinfo{volume}{524}}, \bibinfo{eid}{A51}
  (\bibinfo{year}{2010}), \eprint{1002.1910}.

\bibitem[{\citenamefont{{Porter} et~al.}(2006)\citenamefont{{Porter},
  {Moskalenko}, and {Strong}}}]{2006ApJ...648L..29P}
\bibinfo{author}{\bibfnamefont{T.~A.} \bibnamefont{{Porter}}},
  \bibinfo{author}{\bibfnamefont{I.~V.} \bibnamefont{{Moskalenko}}},
  \bibnamefont{and} \bibinfo{author}{\bibfnamefont{A.~W.}
  \bibnamefont{{Strong}}}, \bibinfo{journal}{\apjl}
  \textbf{\bibinfo{volume}{648}}, \bibinfo{pages}{L29} (\bibinfo{year}{2006}),
  \eprint{astro-ph/0607344}.

\bibitem[{\citenamefont{{Popescu} et~al.}(2017)\citenamefont{{Popescu}, {Yang},
  {Tuffs}, {Natale}, {Rushton}, and {Aharonian}}}]{2017MNRAS.470.2539P}
\bibinfo{author}{\bibfnamefont{C.~C.} \bibnamefont{{Popescu}}},
  \bibinfo{author}{\bibfnamefont{R.}~\bibnamefont{{Yang}}},
  \bibinfo{author}{\bibfnamefont{R.~J.} \bibnamefont{{Tuffs}}},
  \bibinfo{author}{\bibfnamefont{G.}~\bibnamefont{{Natale}}},
  \bibinfo{author}{\bibfnamefont{M.}~\bibnamefont{{Rushton}}},
  \bibnamefont{and}
  \bibinfo{author}{\bibfnamefont{F.}~\bibnamefont{{Aharonian}}},
  \bibinfo{journal}{\mnras} \textbf{\bibinfo{volume}{470}},
  \bibinfo{pages}{2539} (\bibinfo{year}{2017}), \eprint{1705.06652}.

\bibitem[{\citenamefont{Faherty et~al.}(2007)\citenamefont{Faherty, Walter, and
  Anderson}}]{Faherty:2007}
\bibinfo{author}{\bibfnamefont{J.}~\bibnamefont{Faherty}},
  \bibinfo{author}{\bibfnamefont{F.}~\bibnamefont{Walter}}, \bibnamefont{and}
  \bibinfo{author}{\bibnamefont{Anderson}}, \bibinfo{journal}{Astrophysics and
  Space Science} \textbf{\bibinfo{volume}{308}}, \bibinfo{pages}{225–230}
  (\bibinfo{year}{2007}), \eprint{astro-ph/0504584}.

\bibitem[{\citenamefont{{Acero} et~al.}(2015)}]{REF:2015.3FGL}
\bibinfo{author}{\bibfnamefont{F.}~\bibnamefont{{Acero}}} \bibnamefont{et~al.}
  (\bibinfo{collaboration}{{Fermi-LAT Collaboration}}),
  \bibinfo{journal}{\apjs} \textbf{\bibinfo{volume}{218}}, \bibinfo{eid}{23}
  (\bibinfo{year}{2015}), \eprint{1501.02003}.

\bibitem[{\citenamefont{Abdalla et~al.}(2018{\natexlab{b}})}]{Abdalla:2017vci}
\bibinfo{author}{\bibfnamefont{H.}~\bibnamefont{Abdalla}} \bibnamefont{et~al.}
  (\bibinfo{collaboration}{HESS}), \bibinfo{journal}{Astron. Astrophys.}
  \textbf{\bibinfo{volume}{612}}, \bibinfo{pages}{A2}
  (\bibinfo{year}{2018}{\natexlab{b}}), \eprint{1702.08280}.

\bibitem[{\citenamefont{Mazin}(2019)}]{Mazin:2019ykz}
\bibinfo{author}{\bibfnamefont{D.}~\bibnamefont{Mazin}}
  (\bibinfo{collaboration}{CTA Consortium}), in \emph{\bibinfo{booktitle}{{36th
  ICRC 2019}}} (\bibinfo{year}{2019}), \eprint{1907.08530}.

\bibitem[{\citenamefont{Ambrogi et~al.}(2018)\citenamefont{Ambrogi, Celli, and
  Aharonian}}]{Ambrogi:2018skq}
\bibinfo{author}{\bibfnamefont{L.}~\bibnamefont{Ambrogi}},
  \bibinfo{author}{\bibfnamefont{S.}~\bibnamefont{Celli}}, \bibnamefont{and}
  \bibinfo{author}{\bibfnamefont{F.}~\bibnamefont{Aharonian}},
  \bibinfo{journal}{Astropart. Phys.} \textbf{\bibinfo{volume}{100}},
  \bibinfo{pages}{69} (\bibinfo{year}{2018}), \eprint{1803.03565}.

\bibitem[{\citenamefont{Joshi and Jardin-Blicq}(2018)}]{Joshi:2017eou}
\bibinfo{author}{\bibfnamefont{V.}~\bibnamefont{Joshi}} \bibnamefont{and}
  \bibinfo{author}{\bibfnamefont{A.}~\bibnamefont{Jardin-Blicq}}
  (\bibinfo{collaboration}{HAWC}), \bibinfo{journal}{PoS}
  \textbf{\bibinfo{volume}{ICRC2017}}, \bibinfo{pages}{806}
  (\bibinfo{year}{2018}), \bibinfo{note}{[35,806(2017)]}, \eprint{1708.04032}.

\bibitem[{\citenamefont{{Blasi} and {Amato}}(2011)}]{2011ASSP...21..624B}
\bibinfo{author}{\bibfnamefont{P.}~\bibnamefont{{Blasi}}} \bibnamefont{and}
  \bibinfo{author}{\bibfnamefont{E.}~\bibnamefont{{Amato}}},
  \bibinfo{journal}{Astrophysics and Space Science Proceedings}
  \textbf{\bibinfo{volume}{21}}, \bibinfo{pages}{624} (\bibinfo{year}{2011}),
  \eprint{1007.4745}.

\end{thebibliography}

\end{document}